\documentclass[11pt]{article}

\usepackage{latexsym,graphicx,amssymb,pstcol,epsfig} 

\renewcommand{\baselinestretch}{1.125}

\newcommand{\punt}[1]{}

\newcommand{\ang}[1]{\langle #1 \rangle}

\newcommand{\myforall}[3]{\ang{\forall \: #1 : #2 : #3 }}
\newcommand{\myexists}[3]{\ang{\exists \: #1 : #2 : #3 }}

\newcommand{\defined}{\;\;\triangleq\;\;}

\newcommand{\reflexive}[1]{\underline{#1}}

\renewcommand{\leq}{\leqslant}
\renewcommand{\geq}{\geqslant}

\newcommand{\hb}{\rightarrow}

\newcommand{\nhb}{\!\not\hb}

\newcommand{\po}[1]{\arrows{#1}{P}}
\newcommand{\poeq}[1]{\arrows{\reflexive{#1}}{P}}

\newcommand{\arrows}[2]{\stackrel{\mbox{\tiny $#2$}}{#1}}

\newcommand{\implies}{\Rightarrow}  
\newcommand{\follows}{\Leftarrow}

\newcommand{\myequiv}[1][]{\stackrel{{#1}}{\cong}}

\newcommand{\scc}[1]{\mbox{\sf scc}({#1})}
\newcommand{\height}[1]{\mbox{\sf height}({#1})}

\newcommand{\join}{\, \sqcup \,}
\newcommand{\meet}{\, \sqcap \,}

\newcommand{\possibly}[1]{{possibly\!:}\:{#1}}

\newcommand{\definitely}[1]{{de\!f\!initely\!:}\:{#1}}

\newcommand{\controllable}[1]{{contr\!ollable\!:}\:{#1}}

\newcommand{\invariant}[1]{{invariant\!:}\:{#1}}

\newcommand{\possiblyz}{possibly}
\newcommand{\definitelyz}{de\!f\!initely}
\newcommand{\controllablez}{contr\!ollable}
\newcommand{\invariantz}{invariant}

\newcommand{\comptn}[3][]{\ang{{#2},{#3}}_{#1}}

\newcommand{\true}{\ensuremath {\sf true }}
\newcommand{\false}{\ensuremath {\sf false }}

\newcommand{\reglrz}[1]{reg \, ({#1})}

\newcommand{\proc}[1]{proc(#1)}
\renewcommand{\succ}[1]{succ(#1)}
\newcommand{\pred}[1]{pred(#1)}

\newcommand{\frontier}[1]{f\!rontier(#1)}

\newcommand{\vertices}[1]{\mbox{\sf V}({#1})}
\newcommand{\edges}[1]{\mbox{\sf E}({#1})}
\newcommand{\cuts}[2][]{\mathcal{C}_{#1}({#2})}
\newcommand{\cutsz}[1][]{\mathcal{C}_{#1}}
\newcommand{\paths}[1]{\mathcal{P}({#1})}
\newcommand{\ji}[1]{\mathcal{JI}({#1})}
\newcommand{\joins}[2][]{\mathcal{J}_{#1}({#2})}
\newcommand{\skeletal}[2][]{\mathcal{S}_{#1}(#2)}
\newcommand{\canonical}[2][]{\mathcal{G}_{#1}(#2)}

\newcommand{\fvector}[2][]{F_{#1}(#2)}
\newcommand{\jvector}[2][]{J_{#1}(#2)}
\newcommand{\kvector}[2][]{K_{#1}(#2)}

\newcommand{\fvectorz}[1][]{F_{#1}}
\newcommand{\jvectorz}[1][]{J_{#1}}
\newcommand{\kvectorz}[1][]{K_{#1}}

\newcommand{\hgraph}[2][]{\mathcal{H}_{#1}(#2)}

\newcommand{\satisfies}[2]{#1 \: \models \: #2}
\newcommand{\nsatisfies}[2]{#1 \: \not\models \: #2}

\newcommand{\snd}[3]{snd\ang{{#1},{#2}}({#3})}
\newcommand{\rcv}[3]{rcv\ang{{#1},{#2}}({#3})}

\newcommand{\prevents}[2]{{prevents({#1},{#2})}}

\newcommand{\projectE}[2]{{#1}({#2})}
\newcommand{\projectHB}[2]{{#1}({#2})}
\newcommand{\projectC}[4][]{\comptn[#1]{{#2}({#4})}{{#3}({#4})}}

\newcommand{\projectFF}[4][]{{#4}_{#1}(#2,#3)}

\newcommand{\reachable}[1]{\mathcal{RCH}}

\newcommand{\forbiddenC}[2]{f\!orbidden(#1,#2)}
\newcommand{\forbidden}[1]{f\!orbidden\,_{#1}}

\newcommand{\computej}{{\sf ComputeJ}}
\newcommand{\computef}{{\sf ComputeF}}
\newcommand{\computes}{{\sf SliceForRegular}}
\newcommand{\computeklr}{{\sf SliceForKLocalRegular}}

\newcommand{\supportQ}[1][]{{Q}^{#1}}

\newcommand{\halflinespacing}{\vspace*{0.5em}}
\newcommand{\greaterlinespacing}{\vspace*{0.75em}}

\newtheorem{theorem}{Theorem}

\newtheorem{lemma}[theorem]{Lemma}
\newtheorem{corollary}[theorem]{Corollary}

\newtheorem{definition}{Definition}

\newtheorem{observation}{Observation}

\newtheorem{example}{Example}

\newcommand{\secref}[1]{Section~\ref{sec:#1}}
\newcommand{\figref}[1]{Figure~\ref{fig:#1}}
\newcommand{\tabref}[1]{Table~\ref{tab:#1}}

\newcommand{\thmref}[1]{Theorem~\ref{thm:#1}}
\newcommand{\lemref}[1]{Lemma~\ref{lem:#1}}

\newcommand{\eqref}[1]{(\ref{eq:#1})}

\newcommand{\obsref}[1]{Observation~\ref{obs:#1}}
\newcommand{\lineref}[1]{line~\ref{lin:#1}}

\newcommand{\Secref}[1]{Section~\ref{sec:#1}}
\newcommand{\Figref}[1]{Figure~\ref{fig:#1}}

\newcommand{\Thmref}[1]{Theorem~\ref{thm:#1}}

\newcommand{\Obsref}[1]{Observation~\ref{obs:#1}}

\newcommand{\theqed}{$\Box$}

\newcommand{\qed}{\hspace*{\fill}\theqed\\\vspace*{-0.5em}}

\newcommand{\nsqed}{\hspace*{\fill} \theqed}

\newcommand{\eoe}{\qed}

\newcommand{\nseoe}{\nsqed}

\newcommand{\comment}[1]{// {\em #1}}

\newtheorem{linenum}{\hspace*{-1ex}}[figure]

\newcommand{\algoline}[1]{\begin{minipage}{3em}{\begin{linenum}
\label{lin:#1}\end{linenum}}\end{minipage}} 

\newtheorem{algonum}{\hspace*{-1ex}}

\newcommand{\pseudocode}[1]
{\begin{center}
   \noindent \framebox[0.98\textwidth][l]{
   \begin{minipage}[c]{0.96\textwidth}
   \fontsize{10}{11} \selectfont
   \renewcommand{\baselinestretch}{1.25}
   \sf
	
   \begin{tabbing}
   aaaa\=aaaa\=aaaa\=aaaa\=aaaa\=aaaa\=aaaa\=aaaa\=aaaaaaaaa\=aaa\=  \kill
   #1
   \end{tabbing}
   \end{minipage}
   }
\end{center}}

\newcommand{\statement}[1]{\> #1 \\}
\newcommand{\reason}[2]{#1 \> \{ #2 \} \\}
\newcommand{\conclusion}[1]{\> #1}

\newcommand{\raw}[1]{#1}

\newenvironment{formal}
{
\renewcommand{\baselinestretch}{1.4}
\selectfont
\begin{tabbing} 
\hspace{\parindent}\=$\equiv$aa\=\hspace{30em}\=a\+ \kill
}
{\end{tabbing} 
}

\graphicspath{{Figures/}}

\title{\bf Techniques and Applications of Computation
	   Slicing\footnote{Parts of this paper have appeared earlier in
	   conference proceedings
	   \cite{GarMit:2001:ICDCS,MitGar:2001:DISC,MitGar:2003:ICDCS}.}}

\author{Neeraj Mittal \\
Dept.
of Computer Science \\
The University of Texas at Dallas \\
Richardson, TX 75083, USA \\
{\sf neerajm@utdallas.edu}
\and  
Vijay K. Garg\thanks{Supported in part by the NSF Grants 
ECS-9907213, CCR-9988225, Texas Education Board Grant ARP-320, an
Engineering Foundation Fellowship, and an IBM grant.} \\
Dept. of 
Electrical and Computer Engineering  \\
The University of Texas at Austin \\
Austin, TX 78712, USA\\
{\sf garg@ece.utexas.edu}}

\date{}

\begin{document}

\maketitle

\begin{abstract}

Writing correct distributed programs is hard. In spite of extensive
testing and debugging, software faults persist even in commercial
grade software. Many distributed systems, especially those employed in
safety-critical environments, should be able to operate properly even
in the presence of software faults.  Monitoring the execution of a
distributed system, and, on detecting a fault, initiating the
appropriate corrective action is an important way to tolerate such
faults. This gives rise to the predicate detection problem which
requires finding a consistent cut of a given computation exists that 
satisfies a given global predicate, if it exists.

Detecting a predicate in a computation is, however, an NP-complete
problem in general. In order to ameliorate the associated
combinatorial explosion problem, we introduce the notion of
computation slice. Formally, the slice of a computation with respect
to a predicate is a (sub)computation with the least number of
consistent cuts that contains all consistent cuts of the computation
satisfying the predicate. Intuitively, slice is a concise
representation of those consistent cuts of a computation that satisfy
a certain condition. To detect a predicate, rather than searching the
state-space of the computation, it is much more efficient to search
the state-space of the slice.

We prove that the slice exists and is uniquely defined for all
predicates. We present \mbox{efficient} algorithms for computing the
slice for several useful classes of predicates. We establish that the
problem of computing the slice for an arbitrary predicate is
NP-complete in general.  We develop efficient heuristic algorithms for
computing an approximate slice for such predicates for which computing
the slice is otherwise provably intractable. Our experimental results
demonstrate that slicing can lead to an exponential improvement over
existing techniques for predicate detection in terms of time and
space.

\greaterlinespacing
                                                   
\end{abstract}

\section{Introduction}

Writing distributed programs is an error prone activity; it is hard to
reason about them because they suffer from the combinatorial explosion
problem.
Software faults (bugs), in particular global faults, are caused by
subtle interactions between various components of the system. As such,
they may occur only for specific combinations of inputs and certain
interleavings of events. This makes it difficult to eliminate them
entirely using testing and debugging.
In fact, in spite of extensive testing and debugging, software faults
may persist even in commercial grade software. Many distributed
systems, especially those employed in safety-critical environments,
should be able to operate properly even in the presence of software
faults.  Monitoring the execution of a distributed system, and, on
detecting a fault, initiating the appropriate corrective action is an
important way to tolerate such bugs.

A system for tolerating global faults will, in general, consist of
three components: {\em program tracing module}, {\em fault detection
module}, and {\em fault recovery module}. The program tracing module
is responsible for recording the values of variables or objects being
monitored (that is, on which the predicate depends) whenever they
change. The fault detection module analyzes the trace to check for the
possible occurrence of a fault. On detecting a fault, the fault
recovery module takes the necessary corrective measure to recover from
the fault.  It could involve halting the program execution, or
resetting the values of variables, or rolling back the execution of
the program to a consistent cut before the fault followed by replay
(or retry), possibly under control.
The ability to detect global faults is therefore an important step in
tolerating them. In this paper, we focus on detecting those faults
that can be expressed as predicates on variables of processes. For
example, ``no process has the token'' can be written as $no\_token_1
\wedge no\_token_2 \wedge \cdots \wedge no\_token_n$, where
$no\_token_i$ denotes the absence of token on process $p_i$.  This
gives rise to the {\em predicate detection problem}, which involves
finding a consistent cut of a distributed computation, if it exists,
that satisfies the given global predicate. (This problem is also
referred to as detecting a predicate under {\em possibly} modality in
the literature.)  Predicate detection problem also arises in other
areas in distributed systems such as testing and debugging, for
example, to set conditional breakpoints.

Detecting a predicate in a computation is a hard problem in general
\cite{Gar:2002:Book,StoSch:1995:WDAG,MitGar:2001:ICDCS}.  The reason
is the combinatorial explosion in the number of possible consistent
cuts. Given $n$ processes each with at most $k$ local states, the
number of possible consistent cuts in the computation could be as
large as $O(k^n)$.  Finding a consistent cut that satisfies the given
predicate may, therefore, require looking at a large number of
consistent cuts. In fact, we prove in \cite{MitGar:2001:ICDCS} that
detecting a predicate in \mbox{2-CNF} (conjunctive normal form), even
when no two clauses contain variables from the same process, is an
NP-complete problem, in general.  An example of such a predicate is:
$(x_1 \vee x_2) \wedge (x_3 \vee x_4) \wedge \cdots \wedge (x_{n-1}
\vee x_n)$, where each $x_i$ is a boolean variable on process $p_i$.

The approaches for solving the predicate detection problem
can be divided into three categories.
The first approach involves repeatedly computing global snapshots
of the computation until the given predicate becomes true
\cite{ChaLam:1985:TrCS,Bou:1987:TCS,SpeKea:1986:ICDCS}. This approach
works only for stable predicates, that is, predicates that stay true
once they become true.  Some examples of stable predicates are
termination and deadlock.  The given predicate may not be stable and
may turn true only between two successive snapshots.
The second approach is based on searching the state-space of the
computation.  This approach involves incrementally building the
lattice corresponding to the computation until the desired predicate
turns true
\cite{CooMar:1991:WPDD,JegMed+:1995:WSCT,StoUnn+:2000:CAV,AlaVen:2001:TSE}.
Unlike the first approach, this approach can be used to detect
unstable predicates.  However, the algorithms based on this approach
may have exponential running time.
The third approach exploits the structure of the predicate itself---by
imposing restrictions---to evaluate its value efficiently for a given
computation. Polynomial-time algorithms have been developed for
several useful classes of predicates including conjunctive predicates
\cite{Gar:2002:Book,HurMiz+:1996:SPDP}, linear and semi-linear
predicates \cite{ChaGar:1998:DC}, and relational predicates
\cite{ChaGar:1995:WDAG}.

We develop the {\em computation slicing\/} technique for reducing the
size of the computation and therefore the number of consistent cuts to be
analyzed for detecting a predicate. The {\em slice\/} of a computation
with respect to a predicate is the (sub)computation satisfying the
following two conditions. First, it contains {\em all\/} consistent
cuts for which the predicate evaluates to true.  Second, among all
computations that fulfill the first condition, it contains the {\em
least\/} number of consistent cuts.  Intuitively, slice is a {\em
concise representation\/} of consistent cuts satisfying a given
property.  We establish that the slice exists and is uniquely defined
for all predicates.  Since we expect global faults to be relatively
rare, their slice will be much smaller---exponentially in many
cases---than the computation itself.  Therefore, in order to detect a
global fault, rather than searching the state-space of the
computation, it is much more efficient to search the state-space of
the slice.

\begin{figure}[t]
\centerline{\resizebox{4.5in}{!}{\input{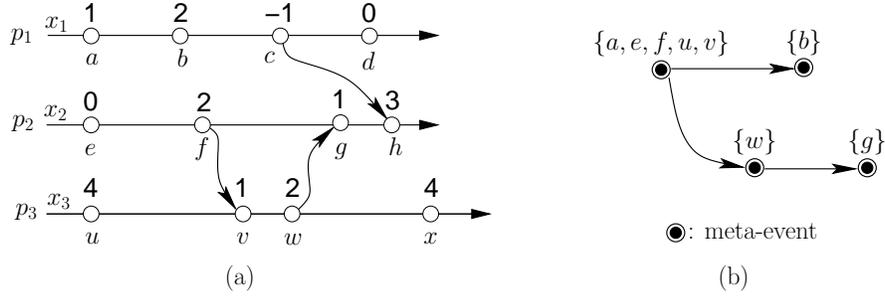}}}
\caption{\label{fig:motivation} (a) A computation and (b) its slice with
respect to $(x_1 \geq 1) \wedge (x_3 \leq 3)$.}
\end{figure}

As an illustration, suppose we want to detect the predicate $(x_1 *
x_2 + x_3 < 5)$ $\wedge (x_1 \geq 1) \wedge (x_3 \leq 3)$ in the
computation shown in \figref{motivation}(a).  The computation consists
of three processes $p_1$, $p_2$ and $p_3$ hosting integer variables
$x_1$, $x_2$ and $x_3$, respectively. The events are represented by
circles. Each event is labeled with the value of the respective
variable immediately after the event is executed. For example, the
value of variable $x_1$ immediately after executing the event $c$ is
$-1$. The first event on each process initializes the state of the
process and every consistent cut contains these initial
events. Without computation slicing, we are forced to examine all
consistent cuts of the computation, twenty eight in total, to
ascertain whether some consistent cut satisfies the
predicate. Alternatively, we can compute the slice of the computation
with respect to the predicate $(x_1 \geq 1) \wedge (x_3 \leq 3)$ as
portrayed in \figref{motivation}(b). A slice is modeled by a directed
graph. Each vertex of the graph corresponds to a {\em meta-event},
which is a subset of events. If a vertex is contained in a consistent
cut, it means that all events corresponding to the vertex are
contained in the cut. Moreover, a vertex belongs to a consistent cut
only if all its incoming neighbours are also present in the cut. We
can now restrict our search to the consistent cuts of the slice which
are only six in number, namely $\{ a, e, f, u, v \}$, $\{ a, e, f, u,
v, b \}$, $\{ a, e, f, u, v, w \}$, $\{ a, e, f, u, v, b, w \}$, $\{
a, e, f, u, v, w, g \}$ and $\{ a, e, f, u, v, b, w, g \}$. The slice
has much fewer consistent cuts than the computation
itself---exponentially smaller in many cases---resulting in
substantial savings.

The slice for a predicate may contain consistent cuts that do not
satisfy the predicate.  We identify a class of predicates called {\em
regular predicates\/} for which the slice is {\em lean}. In other
words, the slice for a regular predicate contains precisely those
consistent cuts for which the predicate evaluates to true.  The set of
consistent cuts satisfying a regular predicate forms a sublattice (of
the lattice of consistent cuts).
Some examples of regular predicates are: {\em conjunctive predicates},
which can be expressed as conjunction of local predicates, like ``all
processes are in red state'' 
\cite{Gar:2002:Book}, 
and {\em monotonic channel predicates\/} such as ``all control
messages have been received'' 
\cite{Gar:2002:Book}.
We prove that the class of regular predicates is closed under
conjunction, that is, the conjunction of two regular predicates is
also a regular predicate. 
We devise an efficient algorithm to compute the slice for a regular
predicate.  The time-complexity of the algorithm is $O(n^2 |E|)$,
where $n$ is the number of processes and $E$ is the set of events.  In
case the regular predicate can be decomposed into a conjunction of
clauses, where each clause itself is a regular predicate but 
depends on variables of only a small subset of processes, a faster
algorithm for computing the slice can be provided. Also, for special
cases of regular predicates such as conjunctive predicates and certain
monotonic channel predicates, we provide {\em optimal\/} algorithms
for computing the slice, which have $O(|E|)$ time-complexity.

In addition to regular predicates, we also design efficient algorithms
to compute the slice for many classes of non-regular predicates such
as {\em linear predicates} and {\em post-linear
predicates} 
\cite{Gar:2002:Book}. 
Our algorithms have time-complexity of $O(n^2 |E|)$.
We prove that it is intractable in general to compute the slice for an
arbitrary predicate. Nonetheless, it is still useful to be able to
compute an {\em approximate slice\/} for such a predicate
efficiently. An approximate slice may be bigger than the actual slice
but will be much smaller than the computation itself.  To that end, we
develop efficient algorithms to compose two slices. Specifically,
given two slices, {\em composition}\footnote{Composition was called
grafting in our earlier paper \cite{MitGar:2001:DISC}} involves
computing either (1)~the smallest slice that contains all consistent
cuts common to both the slices, or (2)~the smallest slice that
contains all consistent cuts that belong to at least one of the
slices.  We use slice composition to efficiently compute the slice for
a {\em co-regular predicate}---the complement of a regular
predicate---and a {\em $k$-local predicate}---depends on variables of
at most $k$ processes---for constant $k$ \cite{StoSch:1995:WDAG}.  The
algorithms have time-complexities of $O(n^2 |E|^2)$ and $O(n m^{k-1}
|E|)$, respectively, where $m$ is the maximum number of events on a
process.
More importantly, we use slice composition to compute an approximate
slice---in polynomial-time---for a predicate derived from regular and
co-regular predicates, linear and post-linear predicates, and
$k$-local predicates for constant $k$, using $\neg$, $\wedge$ and
$\vee$ operators.  Example of such a predicate is: $(x_1 \vee \neg
x_2) \wedge (x_3 \vee \neg x_1) \wedge (x_2 \vee x_3)$, where each
$x_i$ is a linear predicate. Finally, we conduct simulation tests to
experimentally measure the effectiveness of computation slicing in
pruning the search space when detecting a global fault.  Our results
indicate that slicing can lead to an {\em exponential\/} improvement
over existing techniques in terms of time and space. Furthermore,
other techniques for reducing the time-complexity
\cite{StoUnn+:2000:CAV} and/or the space-complexity
\cite{AlaVen:2001:TSE} are orthogonal to slicing, and as such can
actually be used in conjunction with slicing. For instance, Alagar and
Venkatesan's polynomial space algorithm \cite{AlaVen:2001:TSE} for
searching the state-space of a computation can also be used for searching
the state-space of a slice.

Although, in this paper, we focus on application of slicing to
predicate detection, slicing can be employed to reduce the
search-space when monitoring a predicate under other modalities as well
including $\definitelyz$,
$\invariantz$
and $\controllablez$
\cite{CooMar:1991:WPDD,Gar:2002:Book,BabFro+:1996:JSS,MitGar:2000:PODC}.
We also show that many results pertaining to consistent global
checkpoints can be derived as special cases of slicing.  In
particular, we furnish an alternate characterization of the condition
under which individual local checkpoints can be combined with others
to form a consistent global checkpoint (consistency theorem by Netzer
and Xu \cite{NetXu:1995:TPDS}): a set of local checkpoints can belong
to the same consistent global snapshot if and only if the local
checkpoints in the set are mutually consistent (including with itself)
in the slice. Moreover, the \mbox{R-graph} (rollback-dependency graph)
defined by Wang \cite{Wan:1997:TC} is a special case of the slice.
The minimum and maximum consistent global checkpoints that contain a
set of local checkpoints \cite{Wan:1997:TC} can also be easily
obtained using the slice.
We have recently applied slicing to solve several problems in
combinatorics as well \cite{Gar:2002:FSTTCS}.


Parts of this paper have appeared earlier in conference proceedings
\cite{GarMit:2001:ICDCS,MitGar:2001:DISC,MitGar:2003:ICDCS}.  This
paper combines results of the above papers in a single uniform
framework.  Preliminary versions of \secref{regular},
\secref{establish:regular} and \secref{regular:efficient} first
appeared in \cite{GarMit:2001:ICDCS}.  In this paper, we have
generalized the model of slice used in \cite{GarMit:2001:ICDCS} from a
partially ordered set (poset) on meta-events to a general directed
graph on events.  Also, parts of \secref{regular} including
\lemref{non-decreasing|regular} and \thmref{monotonic|regular} are new
and did not appear in \cite{GarMit:2001:ICDCS}. Preliminary versions
of \secref{establish:general}, \secref{skeletal},
Sections \ref{sec:regular:conjunctive} and \ref{sec:regular:applications},
and Sections \ref{sec:composing}, \ref{sec:co-regular:algorithm} and
\ref{sec:general:approximate} first appeared in
\cite{MitGar:2001:DISC}. Part of \secref{co-regular:algorithm}
involving time-complexity analysis---which follows
\thmref{co-regular:disjunction}---is new and did not appear in
\cite{MitGar:2001:DISC}. Further, we have rewritten proofs of
various lemmas and theorems including 
\lemref{b:min}, 
\thmref{min:cuteqv:b} and \thmref{co-regular:disjunction}. Finally,
preliminary versions of
Sections \ref{sec:slice|linear}, \ref{sec:klocal} and
\ref{sec:general:experimental} first appeared in
\cite{MitGar:2003:ICDCS}. We also gave a brief sketch of the algorithm
described in \secref{regular:decomposable} in
\cite{MitGar:2003:ICDCS}. However, details of the algorithm and its
proof of correctness were not provided due to lack of space.

The paper is organized as follows.
\Secref{model} describes our model of distributed system and the
notation we use in this paper.
We formally define the notion of computation slice in
\secref{slice}. 
In \secref{regular}, we introduce the class of regular predicates,
using which we establish the existence and uniqueness of slice for all
predicates in \secref{uniqueness}.
\Secref{skeletal} describes an efficient representation for slice. 
In \secref{slicing|regular} and \Secref{slicing|general}, we discuss
our results pertaining to regular predicates and general
\mbox{predicates}, respectively.
Finally, in \secref{discussion}, we describe our recent results in
slicing and applications of slicing to solving problems in
combinatorics.


\section{Model and Notation}
\label{sec:model}

Traditionally, a distributed computation is modeled as a partial order
on a set of events \cite{Lam:1978:CACM}. In this paper, we relax the
restriction that the order on events must be a partial order. Instead,
we use directed graphs to model distributed computations as well as
slices. Directed graphs allow us to handle both of them in a uniform
and convenient manner.

Given a directed graph $G$, let $\vertices{G}$ and $\edges{G}$ denote
its set of vertices and edges, respectively.  A subset of vertices of
a directed graph forms a {\em consistent cut\/} if the subset
contains a vertex only if it also contains all its incoming
neighbours. Formally,
\[ C \mbox{ is a consistent cut of } G \defined \ang{\forall e,f \in
\vertices{G} : (e,f) \in \edges{G} : f \in C \: \implies \: e \in C}
\]

Observe that a consistent cut either contains all vertices in a cycle
or none of them.  This observation can be generalized to a strongly
connected component.  Traditionally, the notion of consistent cut
({\em down-set\/} or {\em order ideal\/}) is defined for partially
ordered sets \cite{DavPri:1990:CUP}.  Here, we extend the notion to
sets with arbitrary orders.  Let $\cuts{G}$ denote the set of
consistent cuts of a directed graph $G$. Observe that the empty set
$\emptyset$ and the set of vertices $\vertices{G}$ trivially belong to
$\cuts{G}$. We call them {\em trivial\/} consistent cuts.  Let
$\paths{G}$ denote the set of pairs of vertices $(u,v)$ such that
there is a path from $u$ to $v$ in $G$.  We assume that each vertex
has a path to itself.


\subsection{Directed Graph: Path- and Cut-Equivalence}

A directed graph $G$ is {\em cut-equivalent\/} to a directed graph
$H$, denoted by $G \myequiv[C] H$, if they have the same set of
consistent cuts. Formally, $G \myequiv[C] H \defined \cuts{G} =
\cuts{H}$. 

Likewise, a directed graph $G$ is {\em path-equivalent\/} to a
directed graph $H$, denoted by $G \myequiv[P] H$, if a path from
vertex $u$ to vertex $v$ in $G$ implies a path from vertex $u$ to
vertex $v$ in $H$ and vice versa. Formally, $G \myequiv[P] H
\defined \paths{G} = \paths{H}$. 
%
The next lemma explores the relation between the two notions.

\begin{lemma}
\label{lem:graph:paths:cuts}
Let $G$ and $H$ be directed graphs with the same set of vertices. Then,
\[ \paths{G} \subseteq \paths{H} \; \equiv \; \cuts{G} \supseteq
\cuts{H} \]
\end{lemma}


Evidently, \lemref{graph:paths:cuts} implies that two directed graphs
are cut-equivalent if and only if they are path-equivalent. In other
words, to determine whether two directed graphs are cut-equivalent, it
is {\em necessary and sufficient\/} to ascertain that they are
path-equivalent. This is significant \mbox{because}, whereas
path-equivalence can be verified in polynomial-time ($|\paths{G}| =
O(|\vertices{G}|^2)$), cut-equivalence is computationally expensive to
ascertain in general ($|\cuts{G}| = O(2^{|\mbox{\scriptsize
$\vertices{G}$}|})$). In the rest of the paper, we use $\myequiv$ to
denote both $\myequiv[C]$ and $\myequiv[P]$.


\subsection{Distributed Computation as Directed Graph}

A {\em distributed computation} (or simply a {\em computation})
$\comptn{E}{\hb}$ is a directed graph with vertices as the set of
events $E$ and edges as $\hb$.  To limit our attention to only those
consistent cuts that can actually occur during an execution, we assume
that $\paths{\comptn{E}{\hb}}$ contains at least the Lamport's
happened-before relation \cite{Lam:1978:CACM}.  A distributed
computation in our model can contain cycles. This is because whereas a
computation in the traditional or happened-before model captures the
{\em observable} order of execution of events, a computation in our
model captures the set of possible consistent cuts.  Intuitively, each
strongly connected component of a computation can be viewed as a
{\em meta-event}; all events in a meta-event should be executed
atomically.

We denote the set of processes in the system by $P = \{ p_1, p_2,
\ldots, p_n \}$. For an event $e$, let $\proc{e}$ denote the process
on which $e$ occurs.  The predecessor and successor events of $e$ on
$\proc{e}$ are denoted by $\pred{e}$ and $\succ{e}$, respectively, if
they exist.  When events $e$ and $f$ occur on the same process and $e$
occurs before $f$ in real-time, then we write $e \po{\hb} f$. Let
$\poeq{\hb}$ denote the reflexive closure of $\po{\hb}$.

We assume the presence of fictitious initial and final events on each
process. The initial event on process $p_i$, denoted by $\bot_i$,
occurs before any other event on $p_i$. Likewise, the final event on
process $p_i$, denoted by $\top_i$, occurs after all other events on
$p_i$.
For convenience, let $\bot$ and $\top$ denote the set of all
initial events and final events, respectively.
We assume that all initial events belong to the same strongly
connected component. Similarly, all final events belong to the same
strongly connected component. This ensures that any non-trivial
consistent cut will contain all initial events and none of the final
events. As a result, every consistent cut of a computation in the
traditional model is a non-trivial consistent cut of the corresponding
computation in our model and vice versa.  Only non-trivial consistent
cuts are of real interest to us. As we will see later, the extended
model allows us to capture empty slices in a very convenient fashion.

The {\em frontier} of a consistent cut $C$, denoted by
$\frontier{C}$, is defined as the set of those events in $C$ whose
successors are not in $C$. Formally,
\[\frontier{C} \defined \{\: e \in C \:|\: e \not\in \top \: \implies \:
\succ{e} \not\in C \:\}\]

A consistent cut {\em passes through\/} an event if the event belongs
to the frontier of the cut.
Two events are said to be {\em consistent\/} if they are contained in
the frontier of some consistent cut, otherwise they are {\em inconsistent}.
It can be verified that events $e$ and $f$ are consistent if and only
if there is no path in the computation from $\succ{e}$, if it exists,
to $f$ and from $\succ{f}$, if it exists, to $e$. Note that, in the
extended model, in contrast to the traditional model, an event can be
inconsistent with itself.


\subsection{Global Predicate}

A {\em global predicate\/} (or simply a {\em predicate\/}) is defined
as a boolean-valued function on variables of processes. Given a
consistent cut, a predicate is evaluated with respect to the values of
variables resulting after executing all events in the cut. If a
predicate $b$ evaluates to true for a consistent cut $C$, we say that
``$C$ satisfies $b$''.
We leave the predicate undefined for the trivial consistent cuts.

A global predicate is {\em local\/} if it depends on variables of a
single process. Note that it is possible to evaluate a local predicate
with respect to an event on the appropriate process. In case the
predicate evaluates to true, the event is called a {\em true event};
otherwise, it is called a {\em false event}.  
Further, a predicate is said to be {\em $k$-local\/} if it depends on
variables of at most $k$ processes \cite{StoSch:1995:WDAG}. For
example, suppose $x_i$ is an integer variable on process $p_i$ for
each $i \in [1 \ldots n]$. Then, $x_1 + x_2 < 3$ is an example of
2-local predicate, and $x_1 * x_2 + x_3 < 6$ an example of 3-local
predicate.
%


\section{Computation Slice}
\label{sec:slice}

Informally, a {\em computation slice\/} (or simply a {\em slice\/}) is a
%
%
concise representation of all those consistent cuts of the
computation that satisfy the predicate. 
Formally,

\begin{definition}[slice]
The {\em slice} of a computation with respect to a predicate is the
smallest directed graph---with the least number of consistent
cuts---that contains all \mbox{consistent} cuts of the given
computation for which the predicate evaluates to true.
\end{definition}

We will later show that the notion of {\em smallest directed graph} in
the definition is well-defined for every predicate.  The slice of
computation $\comptn{E}{\hb}$ with respect to a predicate $b$ is
denoted by $\comptn[b]{E}{\hb}$.  Note that $\comptn{E}{\hb} =
\comptn[\true]{E}{\hb}$. In the rest of the paper, we use the terms
``computation'', ``slice'' and ``directed graph'' interchangeably.

Note that every slice derived from the computation $\comptn{E}{\hb}$
will have the \mbox{trivial} consistent cuts ($\emptyset$ and $E$)
among its set of consistent cuts.  Thus a slice is {\em empty} if it
has no non-trivial consistent cuts. In the rest of the paper, unless
otherwise stated, a consistent cut refers to a non-trivial consistent
cut. In general, a slice will contain consistent cuts that do not
satisfy the predicate (besides trivial consistent cuts). In case a
slice does not contain any such cut, it is called
{\em lean}. Formally,
 
\begin{definition}[lean slice]
The slice of a computation with respect to a predicate is {\em lean}
if every consistent cut of the slice satisfies the
predicate.
\end{definition}

An interesting question to ask is: ``For what class of predicates is the
slice always lean?'' To answer the question, we introduce the class of
regular predicates next.


\section{Regular Predicate}
\label{sec:regular}

Informally, the set of consistent cuts that satisfy a regular
predicate is closed under set intersection and set union. Formally,

\begin{definition}[regular predicate]
A global predicate is called {\em regular} if, given two consistent
cuts that satisfy the predicate, the consistent cuts given by their
set intersection and set union also satisfy the
predicate. Mathematically, given a regular predicate $b$ and
consistent cuts $C_1$ and $C_2$,
\[
(C_1 \mbox{ satisfies } b) \wedge (C_2 \mbox{ satisfies } b) \; \implies \;
(C_1 \cap \: C_2 \mbox{ satisfies } b) \wedge (C_1 \cup \:
C_2 \mbox{ satisfies } b)
\]
\end{definition}

It can be verified that a local predicate is regular.
Hence the following predicates are regular.

\begin{itemize}

\item process $p_i$ is in ``red'' state

\item the leader has sent all ``prepare to commit'' messages

\end{itemize}

We now provide more examples of regular predicates.
Consider a function $f(x,y)$ with two arguments such that it is
monotonic in its first argument $x$ but anti-monotonic in its second
argument $y$.  Some examples of the function $f$ are: $x - y$, $3x
-5y$, $x / y$ when $x, y > 0$, and $\log_y x$ when $x, y \geq 1$.  We
establish that the predicates of the form $f(x,y) < c$ and $f(x,y)
\leq c$, where $c$ is some constant, are regular when either both $x$
and $y$ are monotonically non-decreasing variables or both $x$ and $y$
are monotonically non-increasing variables.

\begin{lemma}
\label{lem:non-decreasing|regular}
Let $x$ and $y$ be monotonically non-decreasing variables. Then the
predicates $f(x,y) < c$ and $f(x,y) \leq c$ are regular predicates.
\end{lemma}
\begin{proof} We show that the predicate $f(x,y) < c$ is regular. The
proof for the other predicate is similar and has been omitted.
For a consistent $C$, let $x(C)$ and $y(C)$ denote the values of
variables $x$ and $y$, respectively, immediately after all events in
$C$ are executed. Consider consistent cuts $C_1$ and $C_2$ that
satisfy the predicate $f(x,y) < c$. Note that, by definition of $C_1
\cap C_2$, $y(C_1 \cap C_2)$ is either $y(C_1)$ or $y(C_2)$. Without
loss of generality, assume that $y(C_1 \cap C_2) = y(C_1)$. Then,

\begin{formal}
\statement{$f(x(C_1 \cap C_2), y(C_1 \cap C_2))$}
\reason{$=$}{assumption}
\raw{\> $f(x(C_1 \cap C_2), y(C_1))$ \\[0.5em]}
\raw{$\leq$ \> $\left \{ \begin{array}{@{}l@{}l@{}} 
&  x \mbox{ is monotonically non-decreasing implies } x(C_1 \cap
C_2) \leq x(C_1), \\ & \mbox{and } f \mbox{ is monotonic in } x
\end{array} \right \} $ \\[0.25em]}
\statement{$f(x(C_1),y(C_1))$}
\reason{$<$}{$C_1$ satisfies the predicate $f(x,y) < c$}
\conclusion{$c$}
\end{formal}

Thus $C_1 \cap C_2$ satisfies the predicate $f(x, y) < c$. 
Likewise, it can be proved that $C_1 \cup C_2$ satisfies the predicate
$f(x,y) < c$.
%
\qed
\end{proof}

It can be established that \lemref{non-decreasing|regular} holds even
when both $x$ and $y$ are \mbox{monotonically} non-increasing
variables.  Similar results can be proved for the case when $<$ and $\leq$ are
replaced by $>$ and $\geq$, respectively.
The following theorem combines all the above-mentioned results.

\begin{theorem}
\label{thm:monotonic|regular}
Let $f$ be a function with two arguments such that it is monotonic in
its first argument and anti-monotonic in its second argument.  Then
the predicate of the form $f(x,y) \mbox{{\em~\texttt{relop}~}} c$, where
{\em \texttt{relop}} $\in \{ <, \leq, >, \geq \}$ and $c$ is some constant,
is regular when either both $x$ and $y$ are monotonically
non-decreasing variables or both $x$ and $y$ are monotonically
non-increasing variables.
\end{theorem}

By substituting $f(x,y)$ with $x - y$, $x$ with ``the number of
messages that process $p_i$ has sent to process $p_j$ so far'' and $y$
with ``the number of messages sent by process $p_i$ that process $p_j$
has received so far'', it can be verified that the following
predicates are regular.

\begin{itemize}

\item no outstanding message in the channel from process
$p_i$ to process $p_j$


\item at most $k$ messages in transit from process $p_i$ to
process $p_j$

\item at least $k$ messages in transit from process $p_i$ to
process $p_j$

\end{itemize}

We next show that the conjunction of two regular predicates is also
a regular predicate.

\begin{theorem}
\label{thm:regular:closed}
The class of regular predicates is closed under conjunction.
\end{theorem}

The proof is given in the appendix.
The closure under conjunction implies that the following predicates
are also regular.

\begin{itemize}

\item any conjunction of local predicates


\item no process has the token and no channel has the token



\item every ``request'' message has been ``acknowledged'' in the
system

\end{itemize}



\section{Establishing the Existence and Uniqueness of Slice}
\label{sec:uniqueness}

In this section, we show that the slice exists and is uniquely defined
for all predicates. Our approach is to first prove that the slice not
only exists for a regular predicate, but is also lean. Using this fact
we next establish that the slice exists even for a predicate that is
not regular.


\subsection{Regular Predicate}
\label{sec:establish:regular}

It is well known in distributed systems that the set of all consistent
cuts of a computation forms a lattice under the subset relation
\cite{JohZwa:1988:PODC,Mat:1989:WDAG}.  We ask the question does the
lattice of consistent cuts satisfy any additional property? The answer
to this question is in affirmative. Specifically, we show that the set
of consistent cuts of a directed graph not only forms a lattice but
that the lattice is {\em distributive}. A lattice is said to be {\em
distributive\/} if meet distributes over join
\cite{DavPri:1990:CUP}. Formally,

\[ a \meet (b \join c) \: \equiv \: (a \meet b) \join (a \meet c) \]
where $\meet$ and $\join$ denote the meet (infimum) and join
(supremum) operators, respectively. (It can be proved that meet
distributes over join if and only if join distributes over meet.)

\begin{theorem}
\label{thm:graph:cuts:lattice}
Given a directed graph $G$, $\ang{\cuts{G};\subseteq}$ forms a
distributive lattice.
\end{theorem}

\begin{proof}
Let $C_1$ and $C_2$ be consistent cuts of $G$. We define their meet and
join as follows:
\begin{eqnarray*}
C_1 \meet C_2 \! & \! \defined \! & \! C_1 \cap C_2 \\
C_1 \join C_2 \! & \! \defined \! & \! C_1 \cup C_2
\end{eqnarray*}

It is sufficient to establish that $C_1 \cap C_2$ and $C_1 \cup C_2$
are consistent cuts of $G$ which can be easily verified. \qed
\end{proof}

The above theorem is a generalization of the result in lattice theory
that the set of down-sets of a partially ordered set forms a
distributive lattice \cite{DavPri:1990:CUP}. We further prove that the
set of consistent cuts (of a directed graph) does not satisfy any
additional structural property. 
To that end, we need the notion of
{\em join-irreducible} element defined as follows.

\begin{definition}[join-irreducible element \cite{DavPri:1990:CUP}] 
An element of a lattice is {\em join-irreducible}
if ~(1)~it is not the least element of the lattice, and ~(2)~it cannot
be expressed as join of two distinct elements, both different from
itself. Formally, $a \in L$ is join-irreducible if
 
\[
\myexists{x}{}{x < a} \;\; \bigwedge \;\;  \myforall{x, y \in L}{a = x
\sqcup y}{(a = x) \vee (a = y)} 
\] 
\end{definition}

Pictorially, an element of a lattice is join-irreducible if and only
if it has exactly one lower cover, that is, it has exactly one incoming
edge in the corresponding Hasse diagram.
The notion of {\em meet-irreducible element} can be similarly
defined.  It turns out that a distributive lattice is uniquely
characterized by the set of its join-irreducible elements. In
particular, every element of the lattice can be written as join of
some subset of its join-irreducible elements and vice versa. This is
formally captured by the next theorem.

\begin{figure}[h]
\centerline{\resizebox{5.5in}{!}{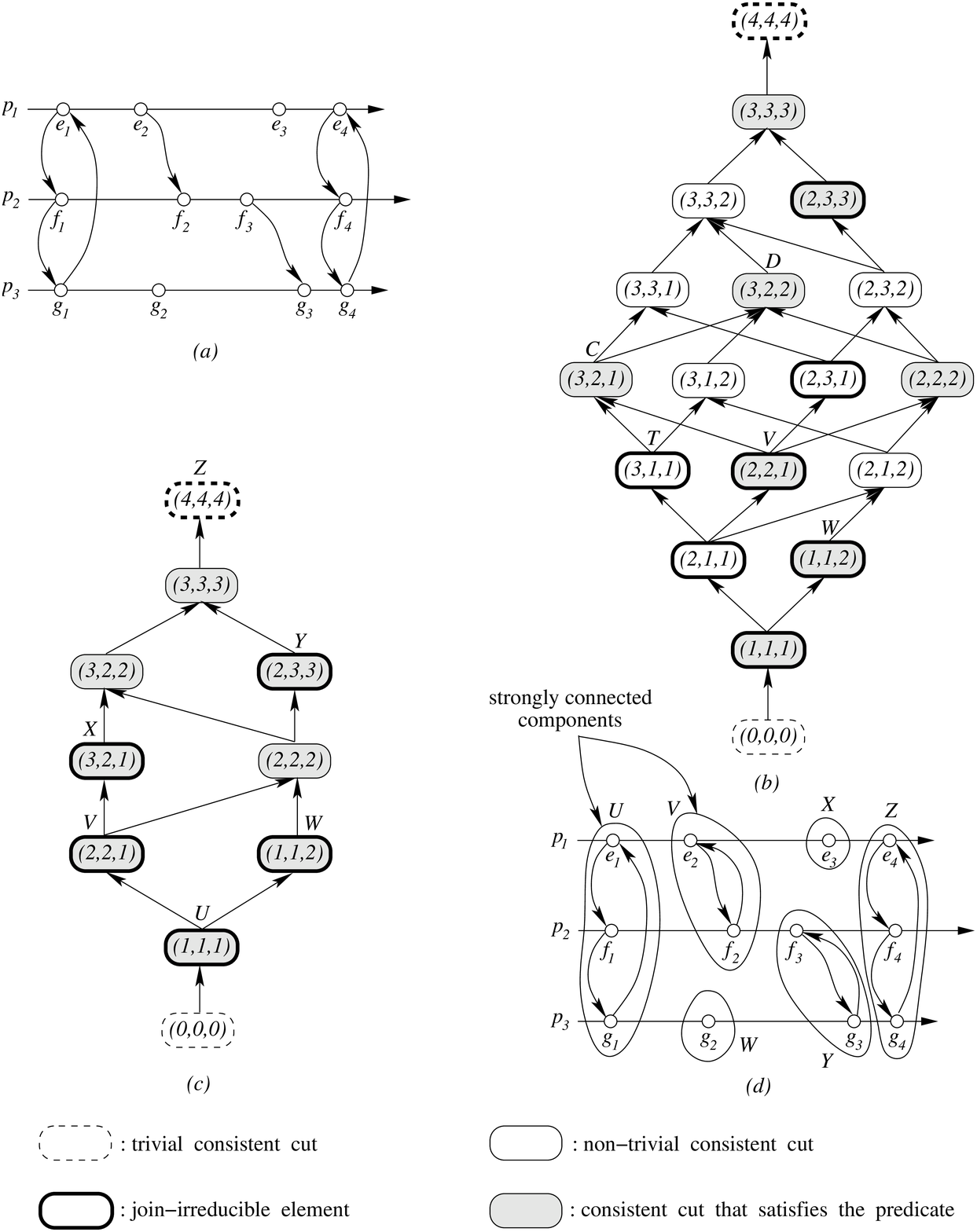}}
\caption{\label{fig:birkhoff} (a)~A computation, (b)~the lattice of
its consistent cuts, (c)~the \mbox{sublattice} of the consistent cuts
that satisfy the \mbox{regular} predicate ``all \mbox{channels} are
empty'', and (d)~the poset induced on the set of join-irreducible
elements of the sublattice.}
\end{figure}

\begin{theorem}{\bf (Birkhoff's Representation Theorem for Finite
Distributive Lattices \linebreak \cite{DavPri:1990:CUP})}~ 
\label{thm:birkhoff}
Let $L$ be a finite distributive lattice and $\ji{L}$ be the set of
its join-irreducible elements. Then the map $f: L \longrightarrow
\cuts{\ji{L}}$ defined by
\[ f(a) = \{ \: x \in \ji{L} \: | \: x \leq a \: \} \]
is an isomorphism of $L$ onto $\cuts{\ji{L}}$. Dually, let $P$ be a
finite poset (partially ordered set). Then the
map  $g: P \longrightarrow \ji{\cuts{P}}$ defined by
\[ g(a) = \{ \: x \in P \: | \: x \leq a \: \} \]
is an isomorphism of $P$ onto $\ji{\cuts{P}}$.
\end{theorem}

Note that the above theorem can also be stated in terms of
meet-irreducible elements.

\begin{example}{\em 
Consider the computation shown in \figref{birkhoff}(a).
\Figref{birkhoff}(b) depicts the lattice of consistent cuts of the
computation.  In the figure, the label of a consistent cut indicates
the number of events that have to be executed on each process to reach
the cut. For example, the label of the consistent cut $C$ is $(3,2,1)$
implying that to reach $C$, three events have to executed on process
$p_1$, two on $p_2$ and one on $p_3$. Mathematically, $C = \{ e_1,
e_2, e_3, f_1, f_2, g_1 \}$.

In \figref{birkhoff}(b), the consistent cuts of the computation
corresponding to the join-irreducible \mbox{elements} of the lattice
have been drawn in thick lines.  There are eight join-irreducible
elements which is same as the number of strongly connected components
of the computation.  Note that the poset induced on the
set of strongly connected components of the computation is isomorphic
to the poset induced on the set of join-irreducible elements of the
lattice. It can be verified that every consistent cut of the
computation can be expressed as the join of some subset of these
join-irreducible elements.  For example, the consistent cut $C$ can be
written as the join of the consistent cuts $T$ and $V$. Moreover, the
join of every subset of these join-irreducible elements is a
consistent cut of the computation. For instance, the join of the
consistent cuts $T$, $V$ and $W$ is given by the consistent cut $D$.
\nseoe
} 
\end{example}

In this paper, we are concerned with only a subset of consistent cuts
and not the entire set of consistent cuts. To that end, the notion of
sublattice of a lattice comes in useful \cite{DavPri:1990:CUP}. Given
a lattice, a subset of its elements forms a {\em sublattice} if the
subset is closed under the meet and join operators of the given
lattice. In our case, the meet and join operators are set intersection
and set union, respectively. Clearly, the set of consistent cuts
satisfying a regular predicate forms a sublattice of the lattice of
consistent cuts.
Finally, we make an important observation regarding a sublattice which
will help us prove the desired result.

\begin{lemma}[\cite{DavPri:1990:CUP}] 
\label{lem:sublattice:distributive}
A sublattice of a distributive lattice is also a distributive lattice.
\end{lemma}

\begin{example}{\em 
In \figref{birkhoff}(b), the consistent cuts for which the regular
predicate ``all channels are empty'' evaluates to true have been shaded. 
\Figref{birkhoff}(c) depicts the poset induced on these \mbox{consistent}
cuts. 
It can be verified that the poset forms a sublattice of the lattice in
\figref{birkhoff}(b). Moreover, the sublattice is, in fact, a
distributive lattice. \eoe
}
\end{example}

We now prove that the slice for a predicate is lean if and only if the
predicate is regular.

\begin{theorem}
\label{thm:regular::lean}
The slice of a computation with respect to a predicate is lean if
and only if the predicate is regular.
\end{theorem}
\begin{proof}
({\em if\/})~ Assume that the predicate, say $b$, is regular. Thus the
set of consistent cuts that satisfy the predicate, denoted by
$\cutsz[b]$, forms a sublattice of the lattice of consistent cuts
(of the computation). From \lemref{sublattice:distributive},
$\cutsz[b]$ is in fact a distributive lattice. Let
$\ji{\cutsz[b]}$ denote the set of join-irreducible elements of
$\cutsz[b]$. From \mbox{Birkhoff's} Representation Theorem,
$\cutsz[b]$ is isomorphic to $\cuts{\ji{\cutsz[b]}}$. Thus
the required slice is given by the poset induced on
$\ji{\cutsz[b]}$ by $\subseteq$. Moreover, every consistent cut
of the slice satisfies the predicate and therefore the slice is lean.

\greaterlinespacing

\noindent ({\em only if\/})~ Assume that the slice of a computation with 
respect to a predicate is lean. From the proof of
\thmref{graph:cuts:lattice}, the set of consistent cuts of the slice 
is closed under set union and set intersection. This in turn implies
that the set of consistent cuts that satisfy the predicate is closed
under set union and set intersection. Thus the predicate is
regular.
\nsqed
\end{proof}

\begin{example}{\em
The sublattice shown in \figref{birkhoff}(c) has exactly six
join-irreducible elements, namely $U$, $V$, $W$, $X$, $Y$ and
$Z$. These elements (or consistent cuts) have been drawn in thick lines.
It can be ascertained that every consistent cut in the sublattice can
be written as the join of some subset of the consistent cuts in
$\mathcal{J} = \{ U, V, W, X, Y, Z \}$. In other words, every
consistent cut of the computation that \mbox{satisfies} the regular
predicate ``all channels are empty'' can be represented as the join of
some subset of the elements in $\mathcal{J}$. Moreover, the join of
every subset of elements in $\mathcal{J}$ yields a consistent cut
contained in the sublattice and hence a cut for which ``all channels
are empty''. The poset induced on the elements of $\mathcal{J}$ by the
relation $\subseteq$ is shown in \figref{birkhoff}(d). (Recall that each
join-irreducible element corresponds to a strongly connected
component, that is, a meta-event.) This poset corresponds to the slice
of the computation shown in \figref{birkhoff}(a) with respect to the
regular predicate ``all channels are empty''. \nseoe
}
\end{example}


\subsection{General Predicate}
\label{sec:establish:general}

To prove that the slice exists even for a predicate that is not a regular
predicate, we define a closure operator, denoted by $reg$, which,
given a computation, converts an arbitrary predicate into a regular
predicate satisfying certain properties.  Given a computation
$\comptn{E}{\hb}$, let $\mathcal{R}(E)$ denote the set of predicates
that are regular with respect to the computation ($\hb$ is implicit).

\begin{definition}[$\mathbf{reg}$]
\noindent Given a predicate $b$, we define {\em $\reglrz{b}$} as the
predicate that satisfies the \linebreak following conditions:

\begin{enumerate}

\item it is regular, that is, $\reglrz{b} \in \mathcal{R}(E)$,

\item it is weaker than $b$, that is, $b \implies \reglrz{b}$,
and

\item it is stronger than any other predicate that  satisfies
(1) and (2), that is, \\ $\myforall{b'}{b' \in \mathcal{R}(E)}{(b
\implies b') \implies (\reglrz{b} \implies b')}$. 

\end{enumerate}

\end{definition}

\begin{figure}[t]
\centerline{\resizebox{4.75in}{!}{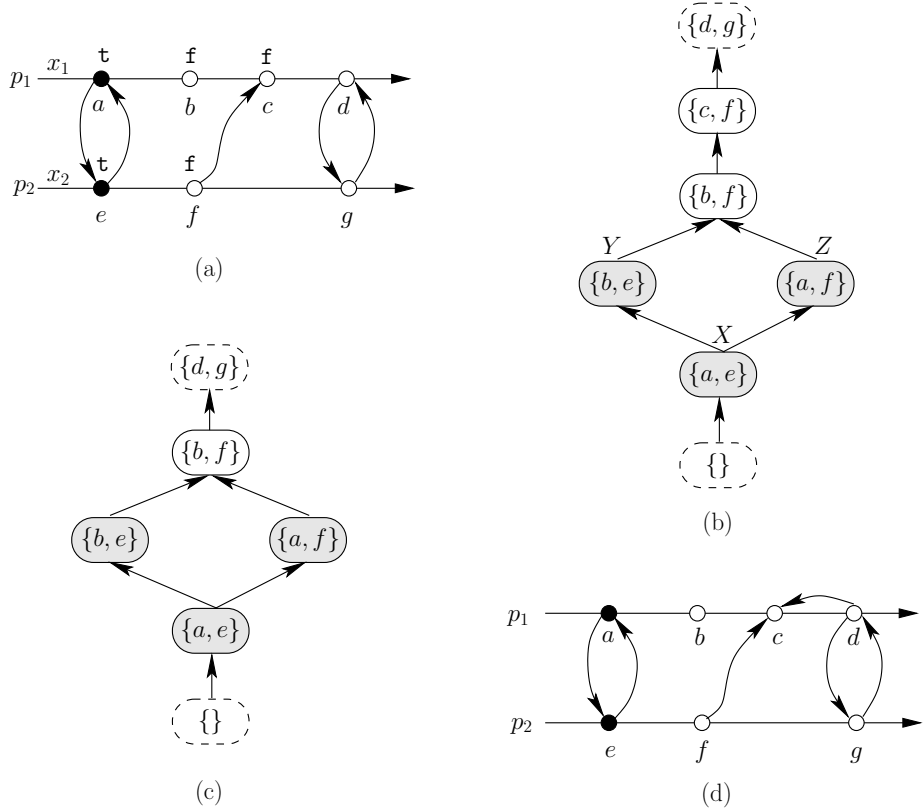}}
\caption{\label{fig:regularize} (a)~A computation, (b)~the lattice of
its consistent cuts, (c)~the sublattice of its consistent cuts
that satisfy $\reglrz{x_1 \vee x_2}$, and (d)~its slice with
\mbox{respect} to $\reglrz{x_1 \vee x_2}$ (and therefore also with
respect to $x_1 \vee x_2$).} 
\end{figure}

Informally, $\reglrz{b}$ is the {\em strongest regular predicate
weaker} than $b$. In general, $\reglrz{b}$ not only depends on the
predicate $b$, but also on the computation under consideration. We
assume the dependence on computation to be implicit and make it
explicit only when necessary. The next theorem establishes that
$\reglrz{b}$ exists for every predicate $b$. Observe that the slice
for $b$ is given by the slice for $\reglrz{b}$. Thus slice exists and
is uniquely defined for all predicates.

\begin{theorem}
\label{thm:reglrz:defined}
Given a predicate $b$, $\reglrz{b}$ exists and is uniquely defined.
\end{theorem}
\begin{proof}
Let $\mathcal{R}_b(E)$ be the set of regular predicates in
$\mathcal{R}(E)$ weaker than $b$.  Observe that $\mathcal{R}_b(E)$ is
non-empty because $\true$ is a regular predicate weaker than $b$ and
therefore contained in $\mathcal{R}_b(E)$.  We set $\reglrz{b}$ to the
conjunction of all predicates in $\mathcal{R}_b(E)$. Formally,

\[ \reglrz{b} \defined \bigwedge_{q \: \in \: \mathcal{R}_b(E)} q \]

It remains to be shown that $\reglrz{b}$ as defined satisfies the
three required conditions.  Now, condition~(1) holds because the class of
regular predicates is closed under conjunction. Condition~(2) holds
because every predicate in $\mathcal{R}_b(E)$ is weaker than $b$ and
hence their conjunction is weaker than $b$. Finally, let $b'$ be a
predicate that satisfies conditions~(1) and (2). Note that $b'
\in \mathcal{R}_b(E)$. Since conjunction of predicates is
stronger than any of its conjunct, $\reglrz{b}$ is stronger than
$b'$. Thus $\reglrz{b}$ satisfies condition~(3). \qed
\end{proof}

Thus, given a computation $\comptn{E}{\hb}$ and a predicate $b$, the
slice of $\comptn{E}{\hb}$ with respect to $b$ can be obtained by
first applying $reg$ operator to $b$ to get $\reglrz{b}$ and then
computing the slice of $\comptn{E}{\hb}$ with respect to $\reglrz{b}$.

\begin{example} {\em 
Consider the computation depicted in \figref{regularize}(a). The
lattice of its consistent cuts is shown in
\figref{regularize}(b). Each consistent cut is labeled with its
frontier. The consistent cuts for which the predicate $x_1 \vee x_2$
evaluates to true have been shaded in the figure. Clearly, the set
of consistent cuts that satisfy $x_1 \vee x_2$ does not form a
sublattice. The smallest sublattice containing the \mbox{subset} is shown
in \figref{regularize}(c); the sublattice corresponds to the predicate
$\reglrz{x_1 \vee x_2}$. The slice for the regular predicate
$\reglrz{x_1 \vee x_2}$ and hence for the predicate $x_1 \vee x_2$ is
portrayed in \figref{regularize}(d). \nseoe
}
\end{example}

\begin{theorem}
\label{thm:reg:closure}
$reg$ is a closure operator. Formally,
\begin{enumerate}
\item $\reglrz{b}$ is weaker than $b$, that is, $b \; \implies \;
\reglrz{b}$, 

\item $reg$ is monotonic, that is, $(b \implies b') \; \implies \;
(\reglrz{b} \implies \reglrz{b'})$, and

\item $reg$ is idempotent, that is, $\reglrz{\reglrz{b}} \: \equiv \:
\reglrz{b}$.

\end{enumerate}
\end{theorem}

From the above theorem it follows that \cite[Theorem
2.21]{DavPri:1990:CUP},

\begin{corollary}
\label{cor:regular:lattice}
$\ang{\mathcal{R}(E);\implies}$ forms a lattice.
\end{corollary}

The meet and join of two regular predicates $b_1$ and $b_2$ is given by
\begin{eqnarray*}
b_1 \meet b_2 \! & \defined & \! b_1 \wedge b_2 \\
b_1 \join b_2 \! & \defined & \! \reglrz{b_1 \vee b_2}
\end{eqnarray*}

The dual notion of $\reglrz{b}$, the weakest regular predicate
stronger than $b$, is also conceivable. However, such a predicate may not
always be unique.

\begin{example} {\em 
In the previous example, three consistent cuts satisfy the predicate
$x_1 \vee x_2$, namely $X$, $Y$ and $Z$, as shown in
\figref{regularize}(b). Two distinct subsets of the set $\mathcal{S} =
\{ X, Y, Z \}$, given by $\{ X, Y \}$ and $\{ X, Z \}$, form maximal
sublattices of $\mathcal{S}$ implying that there is no weakest regular
predicate that is stronger than $x_1 \vee x_2$. \nseoe
}
\end{example}


\section{Representing a Slice}
\label{sec:skeletal}

Any directed graph that is cut-equivalent to a slice constitutes a valid
\mbox{representation} of the slice. However, for computational
purposes, it is preferable to select those graphs to represent a slice
that have fewer edges and can be constructed cheaply. In this section,
we show that every slice can be represented by a directed graph with
$O(|E|)$ vertices and $O(n|E|)$ edges.

Consider a regular predicate $b$ and a computation $\comptn{E}{\hb}$.
Recall that $\cuts{\comptn[b]{E}{\hb}}$ denote the set of consistent
cuts of $\comptn[b]{E}{\hb}$, or, in other words, the set of
consistent cuts of $\comptn{E}{\hb}$ that satisfy $b$.  For reasons of
clarity, we abbreviate $\cuts{\comptn[b]{E}{\hb}}$ by $\cuts[b]{E}$.
From Birkhoff's Representation Theorem, the poset induced on
$\ji{\cuts[b]{E}}$ by the relation $\subseteq$ is cut-equivalent to
the slice $\comptn[b]{E}{\hb}$. It can be proved that
$|\ji{\cuts[b]{E}}|$ is upper-bounded by $|E|$. Therefore the directed
graph corresponding to $\ang{\ji{\cuts[b]{E}}; \subseteq}$ may have
$\Omega(|E|^2)$ edges.

In order to reduce the number of edges, we exploit properties of
join-irreducible elements.
For an event $e$, let $\jvector[b]{e}$ denote the {\em least
consistent cut\/} of $\comptn{E}{\hb}$ that satisfies $b$ and contains
$e$. In case no consistent cut containing $e$ that also satisfies $b$
exists or when $e \in \top$, $\jvector[b]{e}$ is set to $E$---one of
the trivial consistent cuts. Here, we use $E$ as a {\em sentinel\/}
cut. We first show that $\jvector[b]{e}$ is uniquely defined.  Let
$i_e$ be the predicate defined as follows:
\[
C \mbox{ satisfies } i_e \: \defined \: e \in C
\]

It can be proved that $i_e$ is a regular predicate. Next, consider the
predicate $b_e$ defined as the conjunction of $b$ and $i_e$. Since the
class of regular predicates is closed under conjunction, $b_e$ is also
a regular predicate. The consistent cut $\jvector[b]{e}$ can now be
reinterpreted as the least consistent that satisfies $b_e$. Since
$b_e$ is regular, the notion of least consistent cut that satisfies
$b_e$ is uniquely defined, thereby implying that $\jvector[b]{e}$ is
uniquely defined.  For purposes of computing the slice only, we assume
that both trivial consistent cuts satisfy the given regular
predicate. That is, $\{ \emptyset, E \} \subseteq \cuts[b]{E}$.  The
next lemma establishes that $\jvector[b]{e}$ is a join-irreducible
element of $\cuts[b]{E}$.

\begin{lemma} 
\label{lem:least:join}
$\jvector[b]{e}$ is a join-irreducible element of the distributive
lattice $\ang{\cuts[b]{E}; \subseteq}$.
\end{lemma}
\begin{proof}
Suppose $\jvector[b]{e}$ can be expressed as the join (in our case,
set union) of two consistent cuts in $\cuts[b]{E}$, say $C_1$ and
$C_2$. That is, $\jvector[b]{e} = C_1 \cup C_2$, where both $C_1$ and
$C_2$ satisfy $b$. Our obligation is to show that either
$\jvector[b]{e} = C_1$ or $\jvector[b]{e} = C_2$. Since
$\jvector[b]{e}$ contains $e$, either $C_1$ or $C_2$ contains
$e$. Without loss of generality, assume that $e$ belongs to $C_1$. By
definition of set union, $C_1 \subseteq \jvector[b]{e}$. Also,
since $C_1$ is a consistent cut containing $e$ that satisfies $b$,
and $\jvector[b]{e}$ is the {\em least\/} such cut, $\jvector[b]{e}
\subseteq C_1$. Combining the two, $\jvector[b]{e} = C_1$. \qed
\end{proof}

It is possible that $\jvector[b]{e}$s are not all  distinct.
Let $\joins[b]{E}$ denote the set $\{ \jvector[b]{e} \: | \: e \in E
\: \}$. Does $\joins[b]{e}$ capture all join-irreducible elements of
$\cuts[b]{E}$? The following lemma provides the answer.
 
\begin{lemma}
\label{lem:regular:cut|join}
Every consistent cut in $\cuts[b]{E}$ can be expressed as the join of
some subset of consistent cuts in $\joins[b]{E}$.
\end{lemma}
\begin{proof}
Consider a consistent cut $C$ in $\cuts[b]{E}$. Let $D(C)$ be the
consistent cut defined as follows:
\[
D(C) = \bigcup_{e \in C} \jvector[b]{e} 
\]

We prove that $D(C)$ is actually equal to $C$. Since, by definition,
$e \in \jvector[b]{e}$, each event in $C$ is also present in
$D(C)$. Thus $C \subseteq D(C)$. To prove that $D(C) \subseteq C$,
consider an event $e \in C$. Since $C$ is a consistent cut containing
$e$ that satisfies $b$ and $\jvector[b]{e}$ is the {\em least\/} such
cut, $\jvector[b]{e} \subseteq C$. More precisely, for each event $e
\in C$, $\jvector[b]{e} \subseteq C$. This implies that $D(C)
\subseteq C$. \qed 
\end{proof}

From the previous two lemmas, it follows that $\joins[b]{E} =
\ji{\cuts[b]{E}}$. Combining it with Birkhoff's Representation
Theorem, we can deduce that:

\begin{theorem}
\label{thm:j_b:slice}
Given a computation $\comptn{E}{\hb}$ and a regular predicate $b$, the
poset  $\ang{\joins[b]{E};\subseteq}$ is cut-equivalent to
the slice $\comptn[b]{E}{\hb}$.
\end{theorem}


Next, in order to reduce the number of edges, rather than constructing a poset
on the set of join-irreducible elements, we construct a directed graph
with events as vertices. It can be easily verified that:

\begin{observation} 
\label{obs:join:events}
The directed graph $\canonical[b]{E}$ with the set
of vertices as $E$ and an edge from an event $e$ to an event $f$ if
and only if $\jvector[b]{e} \subseteq \jvector[b]{f}$ is
cut-equivalent to the slice $\comptn[b]{E}{\hb}$.
\end{observation}

Whereas the {\em poset representation\/} of a slice is better for
presentation purposes, the {\em graph \mbox{representation}\/} is more suited
for slicing algorithms.
From the way the graph $\canonical[b]{E}$ is constructed, clearly, two
events $e$ and $f$ belong to the same strongly connected component of
$\canonical[b]{E}$ if and only if $\jvector[b]{e} = \jvector[b]{f}$.
As a result, there is a one-to-one correspondence between the strongly
connected components of $\canonical[b]{E}$ and the join-irreducible
elements of $\cuts[b]{E}$.
%

Now, let $\fvector[b]{e}$ be a vector whose $i^{th}$ entry denotes the
earliest event $f$ on process $p_i$ such that $\jvector[b]{e}
\subseteq \jvector[b]{f}$.  Informally, $\fvector[b]{e}[i]$ is the
earliest event on $p_i$ that is reachable from $e$ in the slice
$\comptn[b]{E}{\hb}$.  Using $\fvector[b]{e}$s, we construct a
directed graph we call the {\em skeletal representation} of the slice
and denote it by $\skeletal[b]{E}$. The graph $\skeletal[b]{E}$ has
$E$ as the set of vertices and the following edges:

\begin{enumerate}

\item for each event $e \not\in \top$, there is an
edge from $e$ to $\succ{e}$, and

\item for each event $e$ and process $p_i$, there is an edge from $e$
to $\fvector[b]{e}[i]$.

\end{enumerate}

\begin{example} {\em
Consider the slice depicted in \figref{regularize}(d) of the computation
shown in \figref{regularize}(a) with respect to the predicate $\reglrz{x_1
  \vee x_2}$. Here,
$\jvector[b]{f} = \{ a, e, f \}$ and $\jvector[b]{c} = \{ a, b, c, d,
e, f, g \} = \jvector[b]{d}$. Also, $\fvector[b]{f} = [ c, f ]$ and
$\fvector[b]{c} = [ c, g ] = \fvector[b]{d}$.  \eoe
} 
\end{example} 

To prove that $\skeletal[b]{E}$ faithfully captures the slice
$\comptn[b]{E}{\hb}$, we prove the following two lemmas. The first
lemma establishes that $\jvectorz[b]$ is order-preserving.
 
\begin{lemma}[\mbox{$\jvectorz[b]$} is order-preserving]  
\label{lem:hb:subseteq}
Given events $e$ and $f$, 
$e \hb f \; \implies \; \jvector[b]{e} \subseteq \jvector[b]{f}$.
\end{lemma}
\begin{proof}
Consider $\jvector[b]{f}$. Since $e \hb f$ and $f \in \jvector[b]{f}$,
$e \in \jvector[b]{f}$. Thus $\jvector[b]{f}$ is a consistent cut that
contains $e$ and satisfies $b$. Since $\jvector[b]{e}$ is the {\em
least\/} such cut, $\jvector[b]{e} \subseteq \jvector[b]{f}$. \qed
\end{proof}

The second lemma shows that if $\jvector[b]{e} \subseteq
\jvector[b]{f}$ then there is a path from event $e$ to 
event $f$ in $\skeletal[b]{E}$ and vice versa.

\begin{lemma}
\label{lem:skeletal:subseteq}
Given events $e$ and $f$, 
$\jvector[b]{e} \subseteq \jvector[b]{f} \; \equiv \; (e,f) \in
\paths{\skeletal[b]{E}}$.
\end{lemma}
\begin{proof}
({\em $\implies$})~ Assume that $\jvector[b]{e} \subseteq
\jvector[b]{f}$. Let $\proc{f} = p_i$ and $g =
\fvector[b]{e}[i]$. Since, by definition, $g$ is the earliest event on
$p_i$ such that $\jvector[b]{e} \subseteq \jvector[b]{g}$, $g
\poeq{\hb} f$. This implies that $(g,f) \in \paths{\skeletal[b]{E}}$.
Further, by construction, $(e,g) \in \paths{\skeletal[b]{E}}$. Thus
$(e,f) \in \paths{\skeletal[b]{E}}$. 

\greaterlinespacing

\noindent ({\em $\follows$})~
It suffices to show that for each edge $(u,v)$ in $\skeletal[b]{E}$,
$\jvector[b]{u} \subseteq \jvector[b]{v}$. If $v = \succ{u}$ then
$\jvector[b]{u} \subseteq \jvector[b]{v}$ follows from
\lemref{hb:subseteq}. If $v = \fvector[b]{u}[i]$, where $p_i =
\proc{v}$, then $\jvector[b]{u} \subseteq \jvector[b]{v}$ follows from
the definition of $\fvector[b]{u}$. \qed

\end{proof}

Finally, from \obsref{join:events} and \lemref{skeletal:subseteq}, we
can conclude that:

\begin{theorem}
$\skeletal[b]{E}$ is cut-equivalent to $\comptn[b]{E}{\hb}$.
\end{theorem}

It is easy to see that $\skeletal[b]{E}$ has $O(|E|)$ vertices and
$O(n|E|)$ edges. In the next section we give efficient polynomial-time
algorithms to compute $\jvector[b]{e}$ and $\fvector[b]{e}$ for each
event $e$ when $b$ is a regular  predicate.


\section{Slicing for Regular Predicate}
\label{sec:slicing|regular}

In this section, we discuss our results on slicing with respect to a
regular predicate. They are discussed here separately from our results
on slicing for a general predicate because, as proved in
\secref{establish:regular}, the slice for a regular predicate is lean
and therefore furnishes more information than the slice for a general
predicate. First, we present an efficient $O(n^2|E|)$ algorithm to
compute the slice for a regular predicate. The algorithm can be
optimized for the case when a regular predicate can be decomposed into
a conjunction of clauses, where each clause itself is a
\mbox{$k$-local} regular predicate---a regular predicate that is
also $k$-local---with small $k$. We also provide {\em optimal} algorithms
for special cases of regular predicates such as conjunctive predicates
and certain monotonic channel predicates. Next, we show how a regular
predicate can be monitored under various modalities
\cite{CooMar:1991:WPDD,Gar:2002:Book,MitGar:2000:PODC,StoUnn+:2000:CAV},
specifically $\possiblyz$, $\invariantz$ and $\controllablez$,
using slicing. Finally, we demonstrate that results pertaining to
consistent global checkpoints can be derived as special cases of
slicing.


\subsection{Computing the Slice for Regular Predicate}
\label{sec:regular:efficient}

In this section, given a computation $\comptn{E}{\hb}$ and a regular
predicate $b$, we describe an efficient $O(n^2|E|)$ algorithm to
compute the slice $\comptn[b]{E}{\hb}$. In particular, we construct
$\skeletal[b]{E}$---the skeletal representation of
$\comptn[b]{E}{\hb}$. To that end, it suffices to give an algorithm to
compute $\fvector[b]{e}$ for each event $e$.

Our approach is to first compute $\jvector[b]{e}$ for each event
$e$. Consider the predicate $b_e$ defined in \secref{skeletal}.  Since
$b_e$ is a regular predicate, it is also a linear predicate. (A
predicate is said to be linear if, given two consistent cuts that
satisfy the predicate, the consistent cut given by their set
intersection also satisfies the predicate.) Chase and Garg
\cite{Gar:2002:Book} 
give an efficient algorithm to find the least
consistent cut that satisfies a linear predicate.  Their algorithm is
based on the {\em linearity property} which is defined as follows:

\begin{definition}[linearity property \cite{ChaGar:1998:DC}]
\label{def:linearity}
A predicate satisfies the {\em linearity property\/} if,
given a consistent cut that does not satisfy the predicate, there
exists an event in its frontier, called the {\em forbidden event},
such that there does not exist a consistent cut containing the given
consistent cut that satisfies the \mbox{predicate} and also passes
through the forbidden event. Formally, given a computation
$\comptn{E}{\hb}$, a linear predicate $b$ and a consistent cut $C$,
\[
\nsatisfies{C}{b} \;\: \implies \:\; \myexists{f}{f \in
\frontier{C}}{\myforall{D}{D \: \supseteq \: C}{D \mbox{ satisfies } b
\; \implies \; \succ{f} \in D}}
\]
We denote the forbidden event of $C$ with respect to $b$ by
$\forbiddenC{b}{C}$.
\end{definition}

\begin{figure}[t]
\pseudocode{\> \pushtabs
\underline{Input}: \= (1) a computation $\comptn{E}{\hb}$, ~(2) a
regular predicate $b$, and 
~(3) a process $p_i$ \\[0.5em]
\poptabs 
\> \underline{Output}: $\jvector[b]{e}$ for each event $e$ on $p_i$ \\
\\
\algoline{j_b|initial}
\> $C$ := $\bot$; \\
\algoline{j_b|for|p_i}
\> for each event $e$ on $p_i$ do \>\>\>\>\>\>\>\>\>
\comment{visited in the order given
  by $\po{\hb}$} \\
\algoline{j_b|done}
\>\> $done$ := $\false$; \\
\algoline{j_b|boundary}
\>\> if $C = E$ then $done$ := $\true$; \\
\algoline{j_b|while}
\>\> while not($done$) do \\
\algoline{j_b|if|succf} 
\>\>\> \pushtabs if \= there exist events $f$ and $g$ in  
$\frontier{C}$ \\
\>\>\>\> such that $\succ{f} \hb g$ then \poptabs \>\>\>\>\>\>
\comment{$C$ is not a consistent cut} \\ 
\algoline{j_b|cup|consistent} 
\>\>\>\> $C$ := $C \cup \{ \succ{f} \}$; \>\>\>\>\>\> \comment{advance
beyond $f$} \\
\>\>\> else \>\>\>\>\>\>\> \comment{$C$ is a consistent cut} \\ 
\algoline{j_b|satisfies} 
\>\>\>\> if ($C = E$) or ($C$ \mbox{ satisfies } $b_e$) then $done$ := $\true$;
\\ 
\>\>\>\> else \\
\algoline{j_b|linearity} 
\>\>\>\>\> $f$ := $\forbiddenC{b_e}{C}$; \>\>\>\>\> \comment{invoke the 
linearity property} \\
\algoline{j_b|cup|linearity} 
\>\>\>\>\> $C$ := $C \cup \{ \succ{f} \}$; \>\>\>\>\> \comment{advance beyond $f$} \\
\>\>\>\> endif; \\
\>\>\> endif; \\
\>\> endwhile; \\
\algoline{j_b|assign}
\>\> $\jvector[b]{e}$ := $C$; \\
\> endfor;
}
\caption{\label{fig:j_b} 
The algorithm {\computej} to determine $\jvector[b]{e}$ for each event
$e$ on process $p_i$.}
\end{figure}

\Figref{j_b} describes the algorithm {\computej} to determine
$\jvector[b]{e}$ for each event $e$ on process $p_i$, using the
linearity property. The algorithm scans the computation once from left
to right. Only a single scan is sufficient because, from
\lemref{hb:subseteq}, once we have computed $\jvector[b]{e}$, we do
not need to start all over again to determine $\jvector[b]{\succ{e}}$
but can rather continue on from $\jvector[b]{e}$ itself. The algorithm
basically adds events one-by-one to the cut constructed so far until
either all the events are exhausted or the desired consistent cut is
reached.

The time-complexity analysis of the algorithm {\computej} is as
follows. Each iteration of the while loop at \lineref{j_b|while} has
$O(n)$ time-complexity assuming that the time-complexity of invoking
$\forbidden{b_e}$ at \lineref{j_b|linearity} once is $O(n)$. Moreover,
the while loop is executed at most $O(|E|)$ times because in each
iteration either we succeed in finding the required consistent cut or
we add a new event to $C$. Since there are at most $|E|$ events in the
computation, the while loop cannot be executed more than $O(|E|)$
times. Thus the overall time-complexity of the algorithm {\computej}
is $O(n|E|)$ implying that $\jvector[b]{e}$ for each event $e$
can be computed in $O(n^2|E|)$ time.

\begin{figure}[t]
\begin{center}
\pseudocode{\> \pushtabs
\underline{Input}: \= (1) a computation $\comptn{E}{\hb}$, ~(2)
$\jvector[b]{e}$  for each event $e$, and  
~(3) a process $p_i$ \\[0.5em]
\poptabs 
\> \underline{Output}: $\fvector[b]{e}$ for each event $e$ on $p_i$ \\
\\
\algoline{f_b:j_b|outer}
\> for each process $p_j$ do \\
\algoline{f_b:j_b|f} 
\>\> $f$ := $\bot_j$; \\
\algoline{f_b:j_b|inner} 
\>\> for each event $e$ on $p_i$ do \>\>\>\>\>\>\>\> \comment{visited in
  the order given by $\po{\hb}$} \\
\algoline{f_b:j_b|while} 
\>\>\> while $\jvector[b]{e} \not\subseteq \jvector[b]{f}$ ~do~ $f$
:= $\succ{f}$; endwhile; \\
\algoline{f_b:j_b|gets} 
\>\>\> $\fvector[b]{e}[j] := f$; \\
\>\> endfor; \\
\> endfor; 
}

\caption{\label{fig:f_b:j_b} The algorithm {\computef} to
determine $\fvector[b]{e}$ for each event $e$ on process $p_i$.} 
\end{center}
\end{figure}

Finally, we give an algorithm to compute $\fvector[b]{e}$ for each
event $e$ provided $\jvector[b]{e}$ for each event $e$ is given to
us. We first establish a lemma similar to \lemref{hb:subseteq} for
$\fvectorz[b]$. The lemma allows us to compute the $j^{th}$ entry of
$\fvector[b]{e}$ for each event $e$ on process $p_i$ in a single scan
of the events on process $p_j$ from left to right.

\begin{lemma}[\mbox{$\fvectorz[b]$} is order-preserving]
\label{hb:poeq} 
Given events $e$ and $f$ and a process $p_i$, 
\[e \hb f \: \implies \: \fvector[b]{e}[i] \: \poeq{\hb} \:
\fvector[b]{f}[i]\]
\end{lemma}
\begin{proof} Assume that $e \hb f$. Let $g = \fvector[b]{e}[i]$ and $h =
\fvector[b]{f}[i]$. Note that $\proc{g} = \proc{h} = p_i$.  By
definition of $\fvector[b]{f}$, $\jvector[b]{f} \subseteq
\jvector[b]{h}$.  Since, from \lemref{hb:subseteq}, $\jvector[b]{e}
\subseteq \jvector[b]{f}$, $\jvector[b]{e} \subseteq
\jvector[b]{h}$. Again, by definition of $\fvector[b]{e}$, $g$ is the
earliest event on $p_i$ such that $\jvector[b]{e} \subseteq
\jvector[b]{g}$. Therefore $g \poeq{\hb} h$. \qed
\end{proof}

\begin{figure}[t]
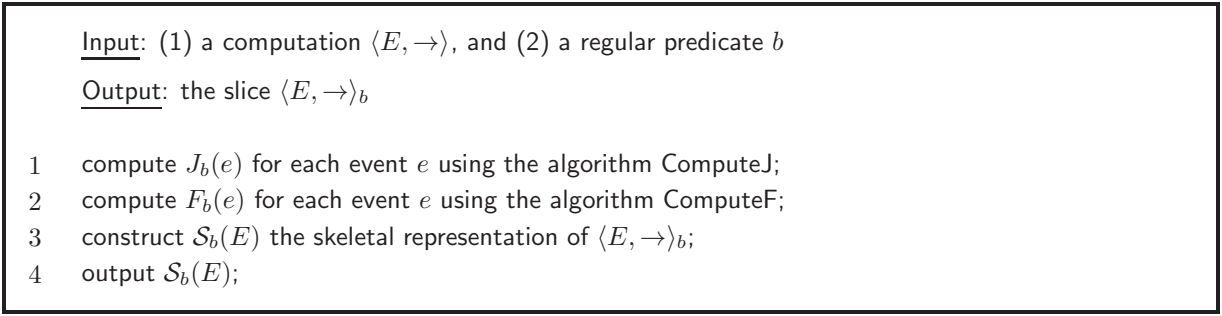

\pseudocode{
\> \underline{Input}: \= (1) a computation
$\comptn{E}{\hb}$, and (2) a regular predicate $b$ \\[0.5em] \>
\underline{Output}: the slice $\comptn[b]{E}{\hb}$ \\ \\
\algoline{slice|j_b} \> compute $\jvector[b]{e}$ for each event $e$
using the algorithm {\computej}; \\ 
\algoline{slice|f_b} \> compute $\fvector[b]{e}$ for each event $e$
using the algorithm {\computef}; \\ 
\algoline{slice|skeletal} \> construct $\skeletal[b]{E}$ the skeletal
representation of $\comptn[b]{E}{\hb}$; \\
\algoline{slice|output}
\> output $\skeletal[b]{E}$; 
}
\caption{\label{fig:slice} 
The algorithm {\computes} to compute the slice for a
regular predicate.}
\end{figure}

\Figref{f_b:j_b} depicts the algorithm {\computef} to determine
$\fvector[b]{e}$ for each event $e$ on process $p_i$. The algorithm is
self-explanatory and its time-complexity analysis is as follows.  Let
$E_j$ denote the set of events on process $p_j$. The outer for loop at
\lineref{f_b:j_b|outer} is executed exactly $n$ times. For $j^{th}$
iteration of the outer for loop, the while loop at
\lineref{f_b:j_b|while} is executed at most $O(|E_i| + |E_j|)$
times. Each iteration of the while loop has $O(1)$ time-complexity
because whether $\jvector[b]{e} \subseteq \jvector[b]{f}$ can be
ascertained by performing only a single comparison, namely testing
whether $e$ is contained $\jvector[b]{f}$. More precisely, 
$\jvector[b]{e} \subseteq \jvector[b]{f}$ if and only if $e \in
\jvector[b]{f}$. The reason is as follows. Since $e \in
\jvector[b]{e}$, if $\jvector[b]{e} \subseteq \jvector[b]{f}$, then $e
\in \jvector[b]{f}$. Also, if $e \in \jvector[b]{f}$, then
$\jvector[b]{f}$ is a consistent cut that contains $e$ and satisfies
$b$. Since $\jvector[b]{e}$ is the {\em least\/} such cut,
$\jvector[b]{e} \subseteq \jvector[b]{f}$. Combining the two, we
obtain the desired equivalence.
The overall time-complexity of the algorithm {\computef} is,
therefore, $O(n|E_i| + |E|)$. Summing up over all processes,
$\fvector[b]{e}$ for each event $e$ can be determined in $O(n|E|)$
time.  The overall algorithm is summarized in
\figref{slice}.

{


\subsection{Optimizing for the Special Case: Computing the Slice
for Decomposable \mbox{Regular} Predicate}
\label{sec:regular:decomposable}

In this section, we explore the possibility of a faster algorithm for
the case when a regular predicate can be expressed as a
conjunction of clauses where each clause is again a regular predicate
but depends on variables of only a small number of processes.
For example, consider the regular predicate ``counters on all
processes are approximately synchronized'', denoted by $b_{sync}$,
which can be expressed formally as:
\[
b_{sync} \defined \bigwedge_{1 \leq i,j \leq n} (|counter_i -
counter_j| \leq \triangle_{ij})
\]
where each $counter_i$ is a monotonically \mbox{non-decreasing}
variable on process $p_i$. 
In this example, each clause depends on variables of at most two
processes and is therefore 2-local.  Using the algorithm discussed in
this section, it is possible to compute the slice for $b_{sync}$ in
$O(n|E|)$ time---a factor of $n$ faster than using the algorithm
\computes.  We describe the algorithm in two steps. In the first step,
we give a fast algorithm to compute the slice for each clause. In the
second step, we describe how to combine slices for all clauses
together in an efficient manner to obtain the slice for the given regular predicate.

\subsubsection{Step 1}

Consider a computation $\comptn{E}{\hb}$ and a $k$-local regular predicate $b$.
Let ${Q}$ denote the subset of processes whose variables $b$ depends on.
{\em Without loss of generality, assume that $\hb$ is a transitive
relation.}  We denote the projection of $E$ on $Q$ by
$\projectE{E}{{Q}}$ and that of $\hb$ on ${Q} \times {Q}$ by
$\projectHB{\hb}{{Q}}$. Thus the projection of the computation
$\comptn{E}{\hb}$ on ${Q}$ is given by $\projectC{E}{\hb}{{Q}}$.

We first show that the slice $\comptn[b]{E}{\hb}$ of the computation
$\comptn{E}{\hb}$ can be recovered exactly from the slice
$\projectC[b]{E}{\hb}{{Q}}$ of the projected computation
$\projectC{E}{\hb}{{Q}}$.  To that end, we extend the definition of
$\fvector[b]{e}$ and define $\projectFF[b]{e}{{Q}}{F}$ to be a vector
whose $i^{th}$ entry represents the earliest event on process $p_i$
that is \mbox{reachable} from $e$ in the slice
$\projectC[b]{E}{\hb}{{Q}}$. Thus $\fvector[b]{e} =
\projectFF[b]{e}{P}{F}$, where $P$ is the entire set of processes,
$\projectFF{e}{{Q}}{F} = \projectFF[\true]{e}{{Q}}{F}$ and
$\fvector{e} = \fvector[\true]{e}$.  We next define $\kvector[b]{e}$
as follows:

\[ \kvector[b]{e}[i] = \left\{
\begin{array}{l@{\quad:\quad}l}
\projectFF[b]{e}{{Q}}{F}[i] & (e \in \projectE{E}{{Q}})
\wedge (p_i \in {Q}) \\ 
\fvector{e}[i] & \mbox{otherwise}
\end{array} \right. \]

We claim that it suffices to know $\kvector[b]{e}$ for each event $e$
to be able to compute the slice $\comptn[b]{E}{\hb}$.  
We build a graph $\hgraph[b]{E}$ in a similar fashion as the skeletal
representation $\skeletal[b]{E}$ of $\comptn[b]{E}{\hb}$ except that
we use $\kvectorz[b]$ instead of $\fvectorz[b]$ in its construction.
The next lemma proves that every path in $\hgraph[b]{E}$ is also a
path in $\skeletal[b]{E}$.

\begin{figure}[t]
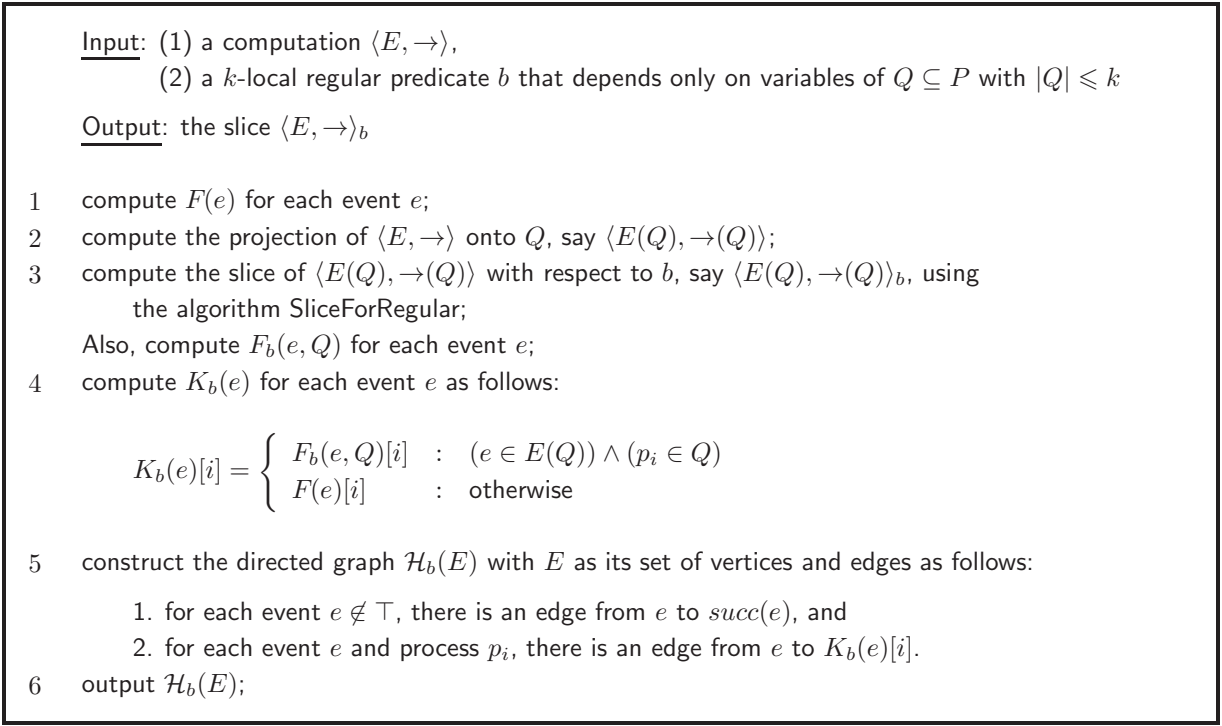

\pseudocode{\> \pushtabs
\underline{Input}: \= (1) a computation $\comptn{E}{\hb}$, \\ 
\>\> (2) a $k$-local regular predicate $b$ that depends only on
variables of $Q \subseteq P$ with $|Q| \leq k$
\\[0.5em] 
\poptabs 
\> \underline{Output}: the slice $\comptn[b]{E}{\hb}$ \\
\\
\algoline{project|F}
\> compute $\fvector{e}$ for each event $e$; \\
\algoline{project|project}
\> compute the projection of $\comptn{E}{\hb}$ onto $Q$, say
$\projectC{E}{\hb}{{Q}}$; \\ 
\algoline{project|slice}
\> compute the slice of $\projectC{E}{\hb}{{Q}}$ with respect to $b$, say 
$\projectC[b]{E}{\hb}{{Q}}$, using \\ 
\>\> the algorithm {\computes}; \\
\> Also, compute $\projectFF[b]{e}{{Q}}{F}$ for each event $e$; \\
\algoline{project|K}
\> compute $\kvector[b]{e}$ for each event $e$ as follows:  \\

\>\> \begin{minipage}[c]{20ex}
\[ \kvector[b]{e}[i] = \left\{
\begin{array}{l@{\quad:\quad}l}
\projectFF[b]{e}{{Q}}{F}[i] & (e \in \projectE{E}{{Q}})
\wedge (p_i \in {Q}) \\ 
\fvector{e}[i] & \mbox{otherwise}
\end{array} \right. \]
\end{minipage} \\[1.25em]

\algoline{project|graph}
\> construct the directed graph $\hgraph[b]{E}$ with $E$ as its set of
vertices and  edges as follows: \\[0.5em]

\>\> 1. for each event $e \not\in \top$, there is an edge from $e$ to
$\succ{e}$, and \\ 

\>\> 2. for each event $e$ and process $p_i$, there is an edge from $e$
to $\kvector[b]{e}[i]$. \\

\algoline{project|output}
\> output $\hgraph[b]{E}$; 

}
\caption{\label{fig:project} The algorithm {\computeklr} to compute
the slice for a $k$-local regular predicate.}
\end{figure}

\begin{lemma}
For each event $e$ and process $p_i$, $\fvector[b]{e}[i] \: \poeq{\hb}
\: \kvector[b]{e}[i]$.
\end{lemma}
\begin{proof}
Every consistent cut of the slice $\comptn[b]{E}{\hb}$ is a
consistent cut of the computation $\comptn{E}{\hb}$ as well.
Therefore, by \lemref{graph:paths:cuts}, every path in
$\comptn{E}{\hb}$ is also a path in $\comptn[b]{E}{\hb}$. This in turn
implies that, for each event $e$ and process $p_i$, $\fvector[b]{e}[i]
\poeq{\hb} \fvector{e}[i]$. Our obligation is to prove that
$\fvector[b]{e}[i] \: \poeq{\hb} \: \projectFF[b]{e}{{Q}}{F}[i]$ when
$e \in \projectE{E}{{Q}}$ and $p_i \in {Q}$.

Consider an event $e \in \projectE{E}{{Q}}$ and process $p_i \in {Q}$. For
convenience, let $f = \projectFF[b]{e}{{Q}}{F}[i]$. Let $C$ be the {\em
least\/} consistent cut of the slice $\comptn[b]{E}{\hb}$ that
contains $f$. Clearly, $C$ is also a consistent cut of the computation
$\comptn{E}{\hb}$. We have,
\begin{formal}
\reason{}{definition of projection}
\statement{$\projectE{C}{{Q}}$ is a consistent cut of
  $\projectC{E}{\hb}{{Q}}$} 
\reason{$\equiv$}{predicate calculus}
\statement{($\projectE{C}{{Q}}$ is a consistent cut of
$\projectC{E}{\hb}{{Q}}$) $\wedge \Big( (C = \top) \vee (C \neq \top) \Big)$}  
\reason{$\implies$}{in case $C\neq \top$, by definition, $C$  
satisfies $b$ and $b$ depends only on variables of $Q$}
\statement{$\projectE{C}{{Q}}$ is a consistent cut of
$\projectC[b]{E}{\hb}{{Q}}$} 
\reason{$\implies$}{$f \in C$, $\proc{f} = p_i$, and $p_i \in Q$}
\statement{($f \in \projectE{C}{{Q}}$) $\wedge$ ($\projectE{C}{{Q}}$
is a consistent cut of  $\projectC[b]{E}{\hb}{{Q}}$)}
\reason{$\implies$}{using definition of $f$, which is
$\projectFF[b]{e}{{Q}}{F}[i]$} 
\raw{\> $\begin{array}{@{}l@{}} (f \in \projectE{C}{{Q}}) \wedge (\mbox{there
is a path from } e \mbox{ to } f \mbox{ in }
\projectC[b]{E}{\hb}{{Q}}) \wedge \\ (\projectE{C}{{Q}} \mbox{ is a
  consistent cut of } \projectC[b]{E}{\hb}{{Q}}) \end{array}$ \\}
\reason{$\implies$}{definition of consistent cut}
\statement{$(f \in \projectE{C}{{Q}}) \wedge (e \in
  \projectE{C}{{Q}})$}
\reason{$\implies$}{$\{ e, f \} \subseteq \projectE{C}{{Q}}$ implies
  $\{ e, f \} \subseteq C$}
\statement{$(f \in C) \wedge (e \in C)$} 
\reason{$\equiv$}{definition of $C$}
\statement{there is a path from $e$ to $f$ in $\comptn[b]{E}{\hb}$}
\reason{$\equiv$}{definition of $\fvector[b]{e}[i]$}
\conclusion{$\fvector[b]{e}[i] \poeq{\hb} f$}
\end{formal}

Thus $\fvector[b]{e}[i] \poeq{\hb} \projectFF[b]{e}{{Q}}{F}[i]$.
\qed
\end{proof}

We now prove the converse, that is, every path in $\skeletal[b]{E}$ is
also a path in $\hgraph[b]{E}$. To that end, by virtue of
\lemref{graph:paths:cuts}, it suffices to show that every consistent
cut of $\hgraph[b]{E}$ is also a consistent cut of $\skeletal[b]{E}$
or, equivalently, every consistent cut of $\hgraph[b]{E}$ satisfies
$b$.
 
\begin{lemma}
\label{lem:project:every|satisfies}
Every (non-trivial) consistent cut of $\hgraph[b]{E}$ satisfies $b$.
\end{lemma}

The proof is in the appendix. Finally, the previous two lemmas can be
combined to give the following theorem:

\begin{theorem}
\label{thm:klocal|regular:slice}
$\hgraph[b]{E}$ is cut-equivalent to $\skeletal[b]{E}$.
\end{theorem}

Observe that the two graphs $\hgraph[b]{E}$ and $\skeletal[b]{E}$ may
actually be different. However, \thmref{klocal|regular:slice} ensures
that the two will be cut-equivalent, thereby implying that
$\hgraph[b]{E}$ captures the slice faithfully.  \Figref{project}
describes the algorithm {\computeklr} for computing the slice for a
$k$-local regular predicate. We assume that the computation is given
to us as $n$ queues of events---one for each process. Further, the
Fidge/Mattern's timestamp $ts(e)$ for each event $e$ is also available
to us, using which $\jvector{e}$ can be computed easily.
The algorithm {\computef} can be used to $\fvector{e}$ for each event
$e$ in $O(n|E|)$. The projection of the computation on $Q$ can then be
computed at \lineref{project|project} in a straightforward
fashion---by simply ignoring events on other processes.  The slice of
the projected computation can be computed at \lineref{project|slice}
in $O(|{Q}|^2|\projectE{E}{{Q}}|)$ time. The vector $\kvector[b]{e}$
for each event $e$ can be determined at \lineref{project|K} in
$O(n|E|)$ time. Finally, the graph $\hgraph[b]{E}$ can be constructed
at \lineref{project|graph} in $O(n|E|)$ time. Thus the overall
time-complexity of the algorithm is $O(|{Q}|^2|\projectE{E}{{Q}}| +
n|E|)$. If $|Q|$ is small, say at most $\sqrt{n}$, then the
time-complexity of the algorithm is $O(n|E|)$---a factor of $n$ faster
than computing the slice directly using the algorithm {\computes}.

A natural question to ask is: ``Can this technique of taking a
projection of a computation on a subset of processes, then computing
the slice of the projection and finally mapping the slice back to
the original set of processes be used for a non-regular predicate as
well?'' The answer is no in general as the following example
demonstrates. 

\begin{figure}[t]
\centerline{\includegraphics[width=5.5in]{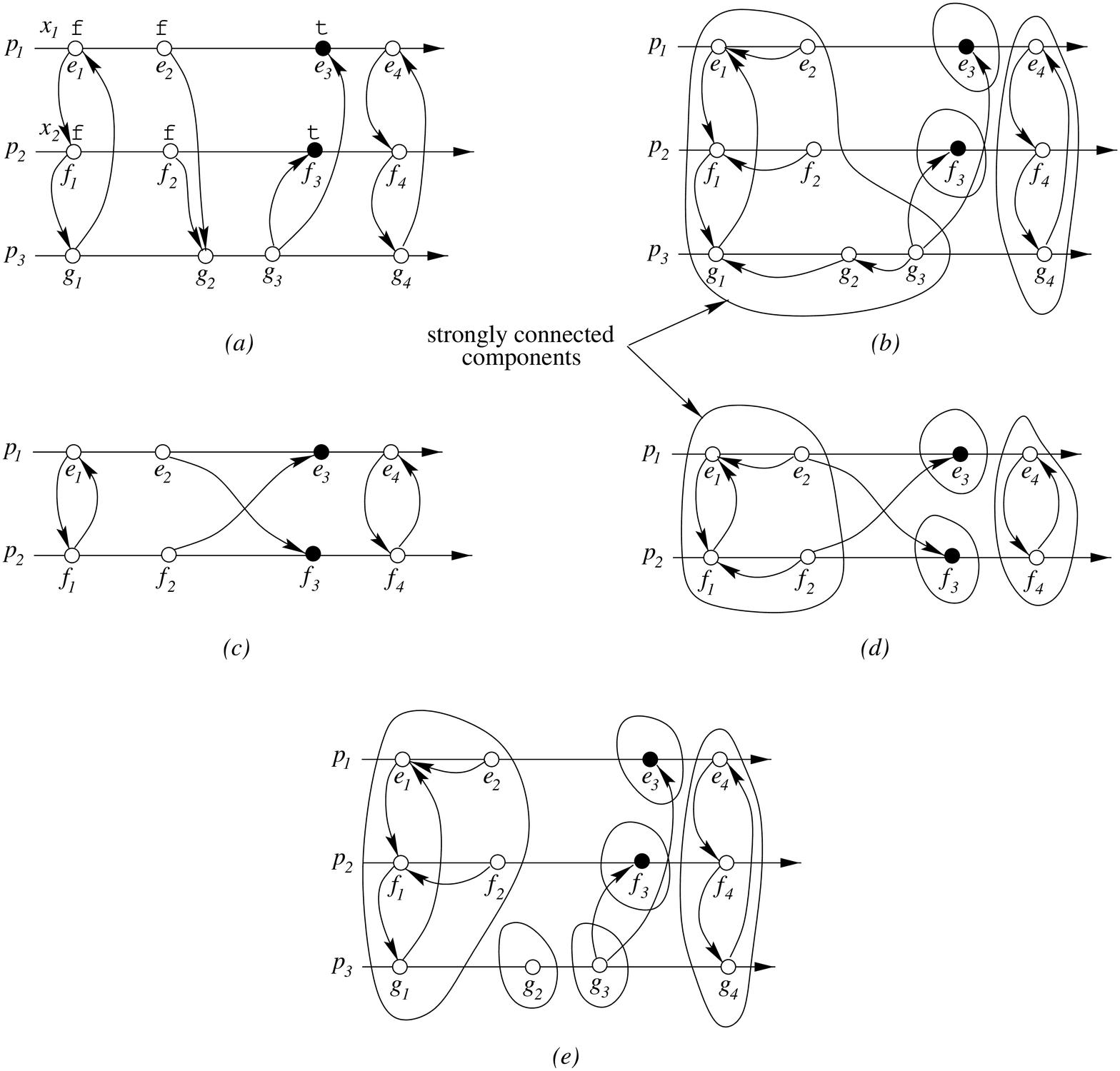}}
\caption{\label{fig:projection-counter} (a) A computation, (b) its
slice with respect to the predicate $x_1 \vee x_2$, (c) its projection
on processes $p_1$ and $p_2$, (d) the slice of the projected
computation with respect to the predicate $x_1 \vee x_2$, and (e) the
slice computed in (d) mapped to the original set of
processes.}
\end{figure}

\begin{example}{\em
Consider the computation shown in \figref{projection-counter}(a)
involving three processes $p_1$, $p_2$ and $p_3$. Let $x_1$ and $x_2$
be boolean variables on processes $p_1$ and $p_2$, respectively. In
the figure, the solid events, namely $e_3$ and $f_3$, satisfy the
respective boolean variable. The slice of the computation for the
(non-regular) predicate $x_1 \vee x_2$ is depicted in
\figref{projection-counter}(b).
\Figref{projection-counter}(c) displays the projection of the
computation on processes on which the predicate $x_1 \vee x_2$
depends, namely $p_1$ and $p_2$. The slice of the projected
computation is shown in \figref{projection-counter}(d) and its mapping
back to the original set of processes is depicted in
\figref{projection-counter}(e). As it can be seen, the slice
shown in \figref{projection-counter}(e) computed
using the algorithm {\computeklr} 
is different from the actual slice
shown in \figref{projection-counter}(b). For instance, events $g_2$ and $g_3$
belong to the same meta-event in the actual slice but not in the slice
computed using the algorithm {\computeklr}. The reason for this
difference is as follows. Since the predicate $x_1 \vee x_2$ is
non-regular, the slice of the projected computation shown in
\figref{projection-counter}(d) contains the consistent cut $X = \{
e_1, e_2, f_1, f_2 \}$ which does not satisfy $x_1 \vee x_2$ but has
to be included anyway so as to complete the sublattice. Now, on
mapping this slice back to the original set of processes, the
resulting slice depicted in \figref{projection-counter}(e) will
contain {\em all} consistent cuts of the original computation whose
projection on $\{ p_1, p_2\}$ is $X$. There are three such consistent
cuts, namely $X \cup \{g_1\}$, $X \cup \{ g_1, g_2 \}$ and $X \cup \{
g_1, g_2, g_3 \}$. However, only one of these consistent cuts, given
by $X \cup \{ g_1, g_2, g_3 \}$, is required to complete the
sublattice for the actual slice. \eoe
}
\end{example}

It can be verified that the algorithm {\computeklr} when used for
computing the slice for a non-regular predicate, in general, produces a
slice bigger than the actual slice. Thus it yields  a fast way to
compute an approximate slice for a non-regular predicate ({\em e.g.},
linear predicate).

\subsubsection{Step 2}

Now, consider a decomposable regular predicate $b$ expressible as
conjunction of $k$-local regular predicates $b^{(j)}$, where $1 \leq j
\leq m$. Let $\supportQ[j]$ denote the subset of processes whose
variable(s) the $j^{th}$ clause $b^{(j)}$ depends on. For a process
$p_i$, we define $clauses_i$ as the set of those clauses that depend
on some variable of $p_i$, that is, $clauses_i \defined \{ \: b^{(j)}
\: | \: p_i \in \supportQ[j] \: \}$. Also, let $s = \max\limits_{1
\leq i \leq n} |clauses_i|$. For example, for the regular predicate
$b_{sync}$, $k = 2$ and $s = n$.

To obtain the slice with respect to $b$, we can proceed as follows. We
first compute the slice for each clause using the algorithm
{\computeklr}. This will give us $\kvector[b^{(j)}]{e}$ for each
clause $b^{(j)}$ and event $e$. Then, for each event $e$ and process
$p_i$, we simply set $\kvector[b]{e}[i]$ to the earliest event 
in the set $\{ \kvector[b^{(1)}]{e}[i],
\kvector[b^{(2)}]{e}[i], \ldots, \kvector[b^{(m)}]{e}[i] \}$.
However, this approach has time-complexity of $O((n m + k^2 s)|E|)$.

To reduce the time-complexity, after computing the slice
$\projectC[b^{(j)}]{E}{\hb}{\supportQ[j]}$ for each clause $b^{(j)}$,
we compute $\kvectorz[b]$ directly without first computing
$\kvectorz[b^{(j)}]$ for each clause $b^{(j)}$. The algorithm is
shown in \figref{K}.
%
Intuitively, among all the slices for the clauses belonging to
$clauses_i$, $\kvector[b]{e}[i]$ is the earliest event on $p_i$ that is
reachable from $e$.
Formally,
\[ \kvector[b]{e}[i] = \min_{b^{(j)} \in clauses_i} 
\fvector[b^{(j)}]{e,\supportQ[j]}[i]  \]

\begin{figure}[t]
\pseudocode{

for each event $e \in E$ do \\
\> $\kvector[b]{e}$ := $\fvector{e}$; \\
endfor; \\
\\
for each conjunct $b^{(j)}$ do \\
\> for each event $e \in \projectE{E}{\supportQ[j]}$ do \\
\>\> for each process $p_i \in \supportQ[j]$ do \\[0.4em]
\>\>\> $\kvector[b]{e}[i]$ := $\min \{ \: \kvector[b]{e}[i], \:
\projectFF[b^{(j)}]{e}{\supportQ[j]}{F}[i] \: \};$ \\[0.4em] 
\>\> endfor; \\
\> endfor; \\
endfor;

}
\caption{\label{fig:K} Computing $\kvector[b]{e}$ for each event $e$.}
\end{figure}



It can be verified that the graph $\hgraph[b]{E}$ then constructed 
using $\kvector[b]{e}$ for each event $e$---in a similar fashion as in
Step 1---is actually \mbox{cut-equivalent} to the slice
$\comptn[b]{E}{\hb}$. The proof is similar to that in Step 1 and
has been omitted.
The overall time-complexity of the algorithm is
given by:

\begin{formal}
\statement{$O(n|E|)  + \sum\limits_{j=1}^{m}
O(|\supportQ[j]|^2 |\projectE{E}{\supportQ[j]}|)$}
\reason{$=$}{each $b^{(j)}$ is a $k$-local predicate,  therefore
$|\supportQ[j]| \leq k$} 
\statement{$O(n|E|) + O(k^2 \sum\limits_{j=1}^{m}
|\projectE{E}{\supportQ[j]}|)$}
\reason{$=$}{simplifying}
\conclusion{$O(n|E| + k^2 s \: |E|) = O((n + k^2 s) |E|)$}
\end{formal}

In case $k$ is $O(1)$ and $s$ is $O(n)$, as is the case with $b_{sync}$,
the overall time-complexity is $O(n|E|)$, which is a factor of
$n$ less than computing the slice directly using the algorithm
{\computes}.
}


\subsection{Optimal Algorithms for Special Cases}
\label{sec:regular:conjunctive}

For special cases of regular predicates, namely conjunctive predicates
and certain monotonic channel predicates, it is possible to give an
$O(|E|)$ optimal algorithm for computing the slice. We only present
the slicing algorithm for the class of conjunctive predicates
here. The slicing algorithm for the class of monotonic channel
predicates is given in the appendix.


\subsubsection{Computing the Slice for Conjunctive Predicate}
\label{sec:conjunctive}

Consider a computation $\comptn{E}{\hb}$ and a conjunctive predicate
$b$. The first step is to partition events on each process into true
events and false events. Having done that, we then construct a graph
$\hgraph[b]{E}$ with vertices as the events in $E$ and the following edges:

\begin{enumerate}

\item from an event, that is not a final event, to its successor,

\item from a send event to the corresponding receive event, and

\item from the successor of a false event to the false event.

\end{enumerate}

\begin{figure}[t]
\centerline{\resizebox{5.5in}{!}{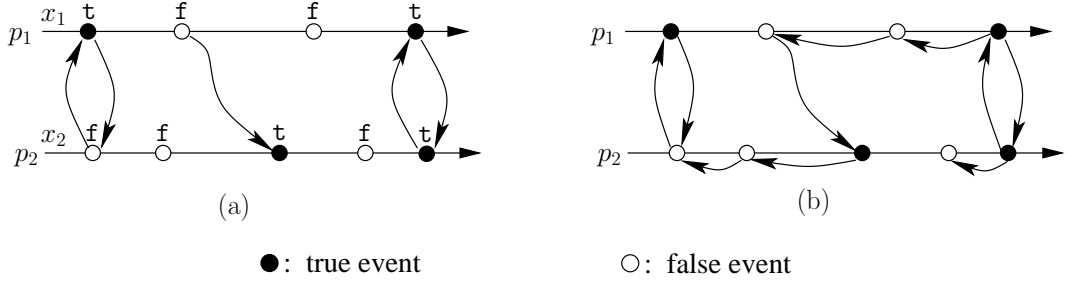}}
\caption{\label{fig:conjunctive} (a) A computation, and (b) its slice
with respect to the conjunctive predicate $x_1 \wedge x_2$.}
\end{figure}


For the purpose of building the graph, we assume that all final events
are true events. Thus every false event has a successor. The first two
types of edges ensure that the Lamport's happened-before relation
\cite{Lam:1978:CACM} is contained in $\paths{\hgraph[b]{E}}$.
Consider the computation depicted in \figref{conjunctive}(a) and the
conjunctive predicate $x_1 \wedge x_2$. The corresponding graph
constructed as described is shown in
\figref{conjunctive}(b).
We now establish that the above-mentioned edges are sufficient to
eliminate all those consistent cuts of the computation that do not satisfy
the conjunctive predicate.

\begin{lemma}
\label{lem:conj:cut:satisfies}
Every (non-trivial) consistent cut of $\hgraph[b]{E}$ satisfies $b$.
\end{lemma}
\begin{proof}
It is sufficient to prove that no consistent cut of $\hgraph[b]{E}$ contains
a false event in its frontier. Consider a consistent cut $C$ of
$\hgraph[b]{E}$. Assume, on the contrary, that $C$ contains a false event,
say $e$, in its frontier. Since every false event has a successor, by
construction, there is an edge from the successor of $e$, say $f$, to
$e$. Therefore $f$ also belongs to $C$. This contradicts the fact that
$e$ is the last event on its process to be contained in $C$.  \qed
\end{proof}

We next show that the above constructed graph retains all consistent
cuts of the computation that satisfy the conjunctive predicate.

\begin{lemma}
\label{conj:cut|satisfies:cut}
Every consistent cut of $\comptn{E}{\hb}$ that \mbox{satisfies} $b$ is a
consistent cut of $\hgraph[b]{E}$.
\end{lemma}
\begin{proof}
Consider a consistent cut $C$ of $\comptn{E}{\hb}$ that satisfies
$b$. Assume, on the contrary, that $C$ is not a consistent cut of
$\hgraph[b]{E}$. Thus there exist events $e$ and $f$ such that there is an
edge from $e$ to $f$ in $\hgraph[b]{E}$, $f$ belongs to $C$ but $e$ does not.
Since $C$ is a consistent cut of $\comptn{E}{\hb}$, the edge from $e$
to $f$ could only of type~(3). (The other two types of edges are
present in $\comptn{E}{\hb}$ as well.) Equivalently, $e$ and $f$ occur on the
same process, $e$ is the successor of $f$, and $f$ is a false
event. Again, since $f$ is contained in $C$ but its successor $e$ is
not, $f$ belongs to the frontier of $C$. However, $C$ satisfies $b$
and hence cannot contain any false event in its frontier. \qed
\end{proof}

From the previous two lemmas, it follows that:

\begin{theorem}
$\hgraph[b]{E}$ is cut-equivalent to $\comptn[b]{E}{\hb}$.
\end{theorem}

It is easy to see that the graph $\hgraph[b]{E}$ has $O(|E|)$ vertices,
$O(|E|)$ edges (at most three edges per event assuming that an event
that is not local either sends at most one message or receives at most
one message but not both) and can be built in $O(|E|)$ time. Thus the
algorithm has $O(|E|)$ overall time-complexity. It also gives us an
$O(|E|)$ algorithm to evaluate $\possibly{b}$ when $b$ is a
conjunctive predicate.


\subsection{Applications of Slicing}
\label{sec:regular:applications}

In this section, we show that slicing can be used to solve other
problems in distributed systems.

\subsubsection{Monitoring Regular Predicate under Various Modalities}

A predicate can be monitored under four modalities, namely $\possiblyz$,
$\definitelyz$, $\invariantz$ and $\controllablez$
\cite{CooMar:1991:WPDD,Gar:2002:Book,MitGar:2000:PODC,StoUnn+:2000:CAV}.
A predicate is {\em possibly\/} true in a computation if there exists a
consistent cut of the computation that satisfies the predicate. On the
other hand, a predicate {\em definitely\/} holds in a computation if
it eventually becomes true in all runs of the computation (a {\em
run\/} is a path in the lattice of consistent cuts starting from the
initial consistent cut and ending at the final consistent cut). The
modalities $\invariantz$ and $\controllablez$ are duals of the modalities
$\possiblyz$ and $\definitelyz$, respectively. Monitoring has
applications in the areas of testing and debugging and software fault
tolerance of distributed programs.

We show how to monitor a regular predicate under $\possibly{b}$,
$\invariant{b}$ and $\controllable{b}$ modalities using
slicing.
Given a directed graph $G$, let $\scc{G}$ denote
the number of strongly connected components of $G$.

\begin{theorem}
\label{thm:monitor}
A regular predicate is
\begin{enumerate}

\item 
{\bf possibly} true in a computation if and only if the slice of the
computation with respect to the predicate has at least one non-trivial
consistent cut, that is, it has at least two strongly connected
components.  Formally,
$\possibly{b} \; \equiv \; \scc{\comptn[b]{E}{\hb}}
\geq 2$.


\item 
{\bf invariant} in a computation if and only if the slice of the
computation with respect to the predicate is cut-equivalent to the
computation. Formally,
$\invariant{b} \; \equiv \;
\comptn[b]{E}{\hb} \myequiv \comptn{E}{\hb}$.


\item 
{\bf controllable} in a computation if and only if the slice of the
computation with respect to the predicate has the same number of
strongly connected components as the computation. Formally,
$\controllable{b} \;\; \equiv \;\; \scc{\comptn[b]{E}{\hb}} =
\scc{\comptn{E}{\hb}}$.


\end{enumerate}
\end{theorem}

The proof of the theorem can be found in the appendix. We do not yet
know how to monitor a regular predicate under $\definitelyz$
modality.

\subsubsection{Zig-Zag Consistency Theorem: A Special Case of Slicing}

We now show how slicing relates to some of the well-known results in
checkpointing. Consider a conjunctive predicate such that the local
predicate for an event on a process is true if and only if the event
corresponds to a local checkpoint. It can be verified that there is a
{\em zigzag path\/} \cite{NetXu:1995:TPDS,Wan:1997:TC} from a local
checkpoint $c$ to a local checkpoint $c'$ in a computation if and only
if there is a path from $\succ{c}$, if it exists, to $c'$ in the
corresponding slice---which can be ascertained by comparing
$\jvector[b]{\succ{c}}$ and $\jvector[b]{c'}$. An alternative
formulation of the consistency theorem in \cite{NetXu:1995:TPDS} can
thus be obtained as follows:

\begin{theorem}
A set of local checkpoints can belong to the same consistent global
snapshot if and only if the local checkpoints in the set are mutually
consistent (\mbox{including} with itself) in the slice.
\end{theorem}

Moreover, the R-graph (rollback-dependency graph) \cite{Wan:1997:TC}
is path-equivalent to the slice when each contiguous sequence of false
events on a process is merged with the nearest true event that occurs
later on the process. The minimum consistent global checkpoint that
contains a set of local checkpoints \cite{Wan:1997:TC} can be computed
by taking the set union of $\jvectorz[b]$'s for each local checkpoint in the
set. The maximum consistent global checkpoint can be similarly
obtained by using the dual of $\jvectorz[b]$.


\section{Slicing for General Predicate}
\label{sec:slicing|general}

In this section, we describe our results on slicing for general
predicates. We first prove that it is in general NP-hard to compute
the slice for an arbitrary predicate.  Nonetheless, polynomial-time
algorithms can be developed for certain special classes of predicates.
In particular, we provide efficient algorithm to compute the slice for
a \mbox{linear} predicate and its dual---a post-linear predicate
\cite{Gar:2002:Book}.  
We next present an efficient algorithm for
composing two slices efficiently; composition can be done with respect
to meet or join operator as explained later. Slice composition can be
used to compute the slice for a predicate in DNF (disjunctive normal
form).  We further provide three more applications of
composition. First, we demonstrate how composition can be employed to
compute the slice for a {\em co-regular predicate}---complement of a
regular predicate---in polynomial-time.
Second, using composition, we derive a polynomial-time algorithm to
the compute the slice for a {\em $k$-local predicate\/} for constant
$k$.
%
Lastly, we employ slice composition to compute an {\em approximate
slice}---in polynomial-time---for a predicate composed from linear
predicates, post-linear predicates, co-regular predicates and
$k$-local predicates, for constant $k$, using $\wedge$ and $\vee$
operators.


\subsection{NP-Hardness Result}
\label{sec:general:hardness}

It is evident from the definition of slice that the following is true:

\begin{observation}
The necessary and sufficient condition for the slice of a
\mbox{computation} with respect to a predicate to be non-empty is that
there exists a \mbox{consistent} cut of the computation that satisfies
the predicate.
\end{observation}

However, finding out whether some consistent cut of the computation
satisfies a predicate is an NP-complete problem
\cite{Gar:2002:Book}. 
Thus it is in general NP-complete to
determine whether the slice for a predicate is non-empty. This further
implies that computing the slice for an arbitrary predicate is an
NP-hard problem. From the results of \cite{MitGar:2001:ICDCS}, it
follows that this is the case even when the predicate is a singular
2-CNF (conjunctive normal form) predicate.
 

\subsection{Computing the Slice for Linear Predicate and its Dual}
\label{sec:slice|linear}

Recall that a predicate is linear if given two consistent cuts that
satisfy the predicate, the cut given by their set intersection also
satisfies the predicate 
\cite{Gar:2002:Book}.  A post-linear predicate can be defined dually
\cite{Gar:2002:Book}.
In this section, we prove that the slicing algorithm {\computes} for a
regular predicate described in \secref{regular:efficient} can be used
for a linear predicate as well.  For a post-linear predicate, however,
a slightly different version of the algorithm based on the notion of
meet-irreducible element will be applicable. The proof is given in the
appendix.


\subsection{Composing Two Slices}
\label{sec:composing}

Given two slices, slice composition can be used to
either compute the smallest slice that contains all consistent cuts
common to both slices---composing with respect to meet---or compute
the smallest slice that contains consistent cuts of both
slices---composing with respect to join.  In other words, given slices
$\comptn[b_1]{E}{\hb}$ and $\comptn[b_2]{E}{\hb}$, where $b_1$ and
$b_2$ are regular predicates, composition can be used to compute the
slice $\comptn[b]{E}{\hb}$, where $b$ is either $b_1 \!  \meet b_2 =
b_1 \! \wedge b_2$ or $b_1 \! \join b_2 = \reglrz{b_1 \!  \vee
b_2}$. Slice composition enables us to compute the actual slice for an
arbitrary boolean expression of local predicates---by rewriting it in
DNF---although it may require exponential time in the worst case.

\subsubsection{Composing with respect to Meet: $\mathbf{b \equiv b_1
\meet b_2 \equiv b_1 \wedge b_2}$}

In this case, the slice $\comptn[b]{E}{\hb}$ contains a consistent cut
of $\comptn{E}{\hb}$ if and only if the cut satisfies $b_1$ as well as
$b_2$.  Given an event $e$, let $\fvector[\min]{e}$ denote the vector
obtained by taking componentwise minimum of $\fvector[b_1]{e}$ and
$\fvector[b_2]{e}$. We first prove that no component of
$\fvector[\min]{e}$ is less than (or occurs before) the
corresponding component of $\fvector[b]{e}$.

\begin{lemma} 
\label{lem:b:min}
For each event $e$ and process $p_i$,
$\fvector[b]{e}[i] \; \poeq{\hb} \; \fvector[\min]{e}[i]$.
\end{lemma}
\begin{proof}
For convenience, let $f = \fvector[b_1]{e}[i]$. Let $C$ be the {\em
least\/} consistent cut of the slice $\comptn[b]{E}{\hb}$ that
contains $f$. Clearly, $C$ is also a consistent cut of the computation
$\comptn{E}{\hb}$. We have,
\begin{formal}
\statement{($C$ is a consistent cut of
$\comptn{E}{\hb}$) $\wedge \Big( (C = \top) \vee (C \neq \top) \Big)$}  
\reason{$\implies$}{in case $C\neq \top$, by definition, $C$  
satisfies $b$ and therefore satisfies $b_1$ as well} 
\statement{$C$ is a consistent cut of $\comptn[b_1]{E}{\hb}$} 
\reason{$\implies$}{by definition, $C$ contains $f$}
\statement{$(f \in C) \wedge (C \mbox{ is a consistent cut of } \comptn[b_1]{E}{\hb})$}
\reason{$\implies$}{using definition of $f$, which is
$\fvector[b_1]{e}[i]$}
\statement{$(f \in C) \wedge (\mbox{there
is a path from } e \mbox{ to } f \mbox{ in }
\comptn[b_1]{E}{\hb}) \wedge (C \mbox{ is a
  consistent cut of } \comptn[b_1]{E}{\hb})$}
\reason{$\implies$}{definition of consistent cut}
\statement{$(f \in C) \wedge (e \in C)$}
\reason{$\equiv$}{by definition, $C$ is the {\em least\/} consistent
  cut of $\comptn[b]{E}{\hb}$ that contains $f$}
\statement{there is a path from $e$ to $f$ in $\comptn[b]{E}{\hb}$}
\reason{$\equiv$}{definition of $\fvector[b]{e}[i]$}
\conclusion{$\fvector[b]{e}[i] \poeq{\hb} f$}
\end{formal}

Therefore $\fvector[b]{e}[i] \poeq{\hb}
\fvector[b_1]{e}[i]$. Likewise, $\fvector[b]{e}[i] \poeq{\hb}
\fvector[b_2]{e}[i]$.
\qed
\end{proof}

We now construct a directed graph $\skeletal[\min]{E}$ that is similar
to $\skeletal[b]{E}$ except that we use $F_{\min}$ instead of $\fvectorz[b]$ in
its construction. The following theorem proves that
$\skeletal[\min]{E}$ is in fact cut-equivalent to $\skeletal[b]{E}$.

\begin{theorem}
\label{thm:min:cuteqv:b}
$\skeletal[\min]{E}$ is cut-equivalent to $\skeletal[b]{E}$.
\end{theorem}
\begin{proof} We have,
\begin{formal}
\reason{}{definition of $F_{\min}$}
\statement{$\Big( \paths{\skeletal[b_1]{E}} \: \subseteq \:
\paths{\skeletal[\min]{E}} \Big) \; \bigwedge \; \Big(
\paths{\skeletal[b_2]{E}} \: \subseteq \: \paths{\skeletal[\min]{E}}
\Big)$}  
\reason{$\equiv$}{using \lemref{graph:paths:cuts}}
\statement{$\Big( \cuts{\skeletal[\min]{E}} \: \subseteq \:
\cuts{\skeletal[b_1]{E}} \Big) \; \bigwedge \; \Big(
\cuts{\skeletal[\min]{E}} \: \subseteq \: \cuts{\skeletal[b_2]{E}}
\Big)$}  
\reason{$\equiv$}{set calculus}
\statement{$\cuts{\skeletal[\min]{E}} \; \subseteq \:
\Big( \cuts{\skeletal[b_1]{E}} \: \cap \: \cuts{\skeletal[b_2]{E}}
\Big)$}  
\reason{$\equiv$}{$b \equiv b_1 \wedge b_2$}
\conclusion{$\cuts{\skeletal[\min]{E}} \; \subseteq \; 
\cuts{\skeletal[b]{E}}$}
\end{formal}

Also, we have,

\begin{formal}
\reason{}{using \lemref{b:min}}
\statement{$\paths{\skeletal[\min]{E}} \; \subseteq \; 
\paths{\skeletal[b]{E}}$}
\reason{$\equiv$}{using \lemref{graph:paths:cuts}}
\conclusion{$\cuts{\skeletal[b]{E}} \; \subseteq \; 
\cuts{\skeletal[\min]{E}}$}
\end{formal}

Thus $\cuts{\skeletal[\min]{E}} \; = \;
\cuts{\skeletal[b]{E}}$.  \qed 
\end{proof} 

Roughly speaking, the aforementioned algorithm computes the union of
the sets of edges of each slice.  Note that, in general,
$\fvector[b]{e}[i]$ need not be same as $\fvector[\min]{e}[i]$.  This
algorithm can be generalized to conjunction of an arbitrary number of
regular predicates.

\subsubsection{Composing with respect to Join: $\mathbf{b \equiv b_1
\join b_2 \equiv \reglrz{b_1 \vee b_2}}$}

In this case, the slice $\comptn[b]{E}{\hb}$ contains a consistent cut
of $\comptn{E}{\hb}$ if the cut satisfies either $b_1$ or $b_2$.
Given an event $e$, let $\fvector[\max]{e}$ denote the vector obtained
by taking componentwise maximum of $\fvector[b_1]{e}$ and
$\fvector[b_2]{e}$. We first prove that no component of
$\fvector[b]{e}$ is less than (or occurs before) the corresponding
component of $\fvector[\max]{e}$.

\begin{lemma} 
\label{lem:max:b}
For each event $e$ and process $p_i$,
$\fvector[\max]{e}[i] \; \poeq{\hb} \; \fvector[b]{e}[i]$.
\end{lemma}

The proof of \lemref{max:b} is similar to that of \lemref{b:min} and
therefore has been omitted. We now construct a directed graph
$\skeletal[\max]{E}$ that is similar to $\skeletal[b]{E}$ except that
we use $F_{\max}$ instead of $\fvectorz[b]$ in its construction. The following
theorem proves that $\skeletal[\max]{E}$ is in fact cut-equivalent to
$\skeletal[b]{E}$.

\begin{theorem}
\label{thm:max:cuteqv:b}
$\skeletal[\max]{E}$ is cut-equivalent to $\skeletal[b]{E}$.
\end{theorem}

Again, the proof of \thmref{max:cuteqv:b} is similar to that of
\thmref{min:cuteqv:b} and hence has been omitted.
Intuitively, the above-mentioned algorithm computes the \mbox{intersection}
of the sets of edges of each slice. In this case, in contrast to the
\mbox{former} case, $\fvector[b]{e}[i]$ is identical to
$\fvector[\max]{e}[i]$.  The reason is as follows. Recall that
$\fvector[b]{e}[i]$ is the earliest event on $p_i$ that is reachable
from $e$ in $\comptn[b]{E}{\hb}$. From
\thmref{max:cuteqv:b}, at least $\fvector[\max]{e}[i]$ is reachable
from $e$ in $\comptn[b]{E}{\hb}$. Thus $\fvector[b]{e}[i] \poeq{\hb}
\fvector[\max]{e}[i]$. Combining it with \lemref{max:b}, we obtain,

\begin{observation}
\label{obs:max::b}
For each event $e$ and process $p_i$,
$\fvector[b]{e}[i] \; =  \; \fvector[\max]{e}[i]$.
\end{observation}

This algorithm can be generalized to disjunction of
an arbitrary number of regular predicates.


\subsection{Computing the Slice for Co-Regular Predicate}
\label{sec:co-regular:algorithm}

Given a regular predicate, we give an algorithm to compute the slice
of a \mbox{computation} with respect to its negation---a co-regular
predicate. In particular, we express the negation
as disjunction of polynomial number of regular predicates. The slice
can then be computed by composing together slices for each disjunct.

Consider a computation $\comptn{E}{\hb}$ and a regular predicate $b$.
For convenience, let $\hb_b$ be the edge relation for the slice
$\comptn[b]{E}{\hb}$. Without loss of generality, assume that both
$\hb$ and $\hb_b$ are transitive relations. Our objective is to find a
property that {\em distinguishes} the consistent cuts that belong to
the slice from the consistent cuts that do not. Consider events $e$
and $f$ such that $e \not\hb f$ but $e \hb_b f$. Then, clearly, a
consistent cut that contains $f$ but does not contain $e$ cannot
belong to the slice. On the other hand, every consistent cut of the
slice that contains $f$ also contains $e$. This motivates us to define
a predicate $\prevents{f}{e}$ as follows:
\[ C \mbox{ ~satisfies~ } \prevents{f}{e} \defined (f \in C) \: \wedge
\: (e \not\in C) \] 

We now prove that the predicate $\prevents{f}{e}$ is a regular
predicate. Specifically, we establish that $\prevents{f}{e}$ is a
conjunctive predicate.

\begin{lemma}
$\prevents{f}{e}$ is a conjunctive predicate.
\end{lemma}
\begin{proof}
Let $\proc{e} = p_i$ and $\proc{f} = p_j$. We define a local predicate
$l_i(e)$ to be true for an event $g$ on process $p_i$ if $g
\po{\hb} e$. Similarly, we define a local predicate $m_j(f)$ to be
true for an event $h$ on process $p_j$ if $f \poeq{\hb}
h$. Clearly, $\prevents{f}{e}$ is equivalent to $l_i(e) \wedge
m_j(f)$. \qed
\end{proof}

It turns out that every consistent cut that does not belong to the
slice satisfies $\prevents{f}{e}$ for some pair of events $(e,f)$ such
that $(e \not\hb f) \wedge (e \hb_b f)$ holds.  Formally,

\begin{theorem} 
\label{thm:co-regular:disjunction}
Let $C$ be a consistent cut of $\comptn{E}{\hb}$. Then,
\[C \mbox{ ~satisfies~ } \neg b \; \equiv \; \myexists{e, f}{(e
\nhb f) \wedge (e \hb_b f)}{C \mbox{ ~satisfies~ } \prevents{f}{e}}\]
\end{theorem} 
\begin{proof} We have,
\begin{formal}
\statement{$C$ satisfies $\neg b$}
\reason{$\equiv$}{$b$ is a regular predicate}
\statement{$\neg \Big( C \in \cuts{\comptn[b]{E}{\hb}} \Big)$}
\reason{$\equiv$}{definition of $\cuts{\comptn[b]{E}{\hb}}$}
\statement{$\neg \myforall{e,f}{e \hb_b f}{f \in C \implies e \in C}$}
\reason{$\equiv$}{predicate calculus}
\statement{$\myexists{e,f}{e \hb_b f}{(f \in C) \wedge (e \not\in
C)}$}
\reason{$\equiv$}{definition of $\prevents{f}{e}$}
\statement{$\myexists{e,f}{e \hb_b f}{C \mbox{ ~satisfies~ }
\prevents{f}{e}}$}  
\reason{$\equiv$}{predicate calculus}
\statement{$\myexists{e,f}{(e \hb_b f) \: \bigwedge \: \Big((e \hb f)
\: \vee \: (e \nhb f)\Big)}{C \mbox{ ~satisfies~ } \prevents{f}{e}}$}
\reason{$\equiv$}{$e \hb f$ ~implies~ $e \hb_b f$}
\statement{$\myexists{e,f}{(e \hb f) \: \bigvee \: \Big((e \hb_b f)
\: \wedge \: (e \nhb f)\Big)}{C \mbox{ ~satisfies~ } \prevents{f}{e}}$}
\reason{$\equiv$}{$\mbox{since } C \mbox{ is a consistent cut of }
\comptn{E}{\hb}, \mbox{ } C \mbox{ satisfies } \prevents{f}{e}
\mbox{ implies } e \nhb f$}
\conclusion{$\myexists{e,f}{(e \hb_b f) \: \wedge \: (e \nhb
f)}{C \mbox{ satisfies } \prevents{f}{e}}$}
\end{formal}
 
This establishes the theorem.  \qed
\end{proof}

\Thmref{co-regular:disjunction} can also be derived using the results
in lattice theory \cite{Riv:1974:AMS}.
We now give the time-complexity of the algorithm. We start by making
the following observation.

\begin{observation}
\label{obs:co-regular:weakest}
Let $e$, $f$ and $g$ be events such that $f \hb g$. Then,
\[\prevents{g}{e} \: \implies \: \prevents{f}{e}\]
\end{observation}

\halflinespacing

Let $\kvector[b]{e}$ denote the vector whose $i^{th}$ entry denote the
earliest event $f$ on \mbox{process} $p_i$, if it exists, such that
$(e \nhb f) \wedge (e \hb_b f)$ holds. \Obsref{co-regular:weakest}
implies that $\prevents{\kvector[b]{e}[i]}{e}$, whenever
$\kvector[b]{e}[i]$ exists, is the weakest predicate among all
predicates $\prevents{f}{e}$, where $\proc{f} = p_i$ and $(e \nhb f)
\wedge (e \hb_b f)$. Thus we can ignore all other events on $p_i$ for
the purpose of computing the slice for a co-regular predicate. More
precisely, \thmref{co-regular:disjunction} can be restated as:

\begin{theorem}
\label{thm:co-regular:disjunction:new}
Let $C$ be a consistent cut of $\comptn{E}{\hb}$. Then,
\[C \mbox{ ~satisfies~ } \neg b \; \equiv \; \myexists{e,
  p_i}{\kvector[b]{e}[i] \mbox{ exists}}{C
\mbox{ ~satisfies~ } \prevents{\kvector[b]{e}[i]}{e}} \]  
\end{theorem} 

\halflinespacing

It turns out that $\kvector[b]{e}[i]$ and $\fvector[b]{e}[i]$ are
closely related. 

\begin{observation}
\label{obs:k_b:f_b}
$\kvector[b]{e}[i]$ exists if and only if $e \nhb
\fvector[b]{e}[i]$. Moreover, whenever $\kvector[b]{e}[i]$ exists it
is identical to $\fvector[b]{e}[i]$.
\end{observation}

\halflinespacing

\Thmref{co-regular:disjunction:new} implies that the number of
disjuncts in the predicate equivalent to the negation of a regular
predicate is at most $O(n|E|)$. Further, using \obsref{k_b:f_b}, these
disjuncts can be determined in $O(n^2|E|)$ time using the algorithms
{\computej} and {\computef} discussed in
\secref{regular:efficient}. The slice with respect to each disjunct
can be computed in $O(|E|)$ time using the slicing algorithm for
conjunctive predicate. Moreover, for a disjunct $b^{(i)}$,
$\jvector[b^{(i)}]{e}$ for each event $e$ can be computed in $O(n|E|)$
time from its slice $\comptn[b^{(i)}]{E}{\hb}$ (by topologically
sorting the strongly connected components). Using
$\jvectorz[b^{(i)}]$, it is possible to determine
$\fvector[b^{(i)}]{e}$ for each event $e$ in $O(n|E|)$ time using the
algorithm {\computef}. Finally, these slices can be composed together
to produce the slice for a \mbox{co-regular} predicate in $O(n|E|
\times n|E|) = O(n^2|E|^2)$ time. This is because, given an event $e$,
computing each entry of $\fvector[b']{e}$, where $b' = \reglrz{\neg
b}$, using \obsref{max::b} requires $O(n|E|)$ time. Thus the overall
time-complexity of the algorithm is $O(n^2|E| + n^2|E|^2) = O(n^2
|E|^2)$.
 

\subsection{Computing the Slice for $\mathbf{k}$-Local Predicate for
Constant $\mathbf{k}$ }
\label{sec:klocal}  

In case the predicate is regular, we can simply use the algorithm
{\computeklr} to compute the slice in $O(n|E|)$ time.
However, if the predicate is not regular, then the slice produced will
only be an approximate one.  To compute the slice for a $k$-local
predicate, which is not regular, we use the technique developed by
Stoller and Schneider \cite{StoSch:1995:WDAG}. For a given
computation, their technique can be used to transform a $k$-local
predicate into a predicate in $k$-DNF (disjunctive normal form) with
at most $m^{k - 1}$ clauses, where $m$ is the maximum number of events
on a process. For example, consider the predicate $x_1 \neq x_2$. Let
$V$ denote the set of values that $x_1$ can take in the given
computation.  Then $x_1 \neq x_2$ can be rewritten as:
\[
x_1 \neq  x_2  \; \equiv \; \bigvee_{v \in V} \Big( (x_1 = v) \wedge
(x_2 \neq v) \Big)
\]

Note that $|V| \leq m$.  Thus the resultant predicate, in the above
case, consists of at most $m$ clauses where each clause is a conjunctive
predicate 
\cite{Gar:2002:Book}).  In general, the resultant $k$-DNF predicate
will consist of at most $m^{k-1}$ clauses.  To compute the slice for
each clause, we use the optimal $O(|E|)$ algorithm given in
\secref{conjunctive}. We then compose these slices together with
respect to disjunction to obtain the slice for the given $k$-local
predicate.  The overall time-complexity of the algorithm is given by
$O(n m^{k-1} |E|)$.


\subsection{Computing an Approximate Slice}
\label{sec:general:approximate}



Even though it is, in general, NP-hard to compute the slice for an
arbitrary predicate, it is still possible to compute an {\em
approximate slice\/} in many cases. The slice is ``approximate'' in
the sense that it is bigger than the actual slice for the
predicate. Nonetheless, it still contains all consistent cuts of the
computation that satisfy the predicate.  In many cases, the
approximate slice that we obtain is much smaller than the computation
itself and therefore can be used to prune the search-space for many
intractable problems such as monitoring predicates under various
modalities.

In particular, using slice composition and slicing algorithms for
various classes of predicates, it is possible to compute an
approximate slice in an efficient manner for a large class of
predicates---namely those derived from linear predicates, post-linear
predicates, regular predicates, \mbox{co-regular} predicates, and
$k$-local predicates for constant $k$ using $\wedge$ and $\vee$
operators.

To compute an approximate slice for such a predicate, we first
construct the parse tree for the corresponding boolean expression; all
predicates occupy leaf nodes whereas all operators occupy non-leaf
nodes.  We then recursively compute the slice by starting with
leaf-nodes and moving up, level by level, until we reach the root.
For a leaf node, we use the slicing algorithm appropriate for the
predicate contained in the node.  For example, if the leaf node
contains to a linear predicate, we use the algorithm described in
\secref{slice|linear}.
For a non-leaf node, we use the suitable composition algorithm
depending on the operator.

\begin{example} {\em
Suppose we wish to compute an approximate slice for the predicate
$(x_1 \vee x_2) \wedge (x_3 \vee x_4)$, where each $x_i$ is a regular
predicate. First, we compute slices for regular predicates $x_1$,
$x_2$, $x_3$ and $x_4$. Next, we compose the first two and the last
two slices together with respect to join to obtain slices for the
clauses $x_1 \vee x_2$ and $x_3 \vee x_4$, respectively.  Finally, we
compose the slices for both clauses together with respect to meet. The
slice obtained will contain all consistent cuts that satisfy the
predicate $(x_1 \vee x_2) \wedge (x_3 \vee x_4)$.
\nseoe
}
\end{example}

\punt{

\begin{theorem}
\label{thm:monitor}
Given a computation $\comptn{E}{\hb}$ and predicates $b$ and $b'$ with
$b \implies b'$, 

\begin{eqnarray*}
\satisfies{\comptn{E}{\hb}}{\possibly{b}} & \equiv &
\satisfies{\comptn[b']{E}{\hb}}{\possibly{b}} \\
\satisfies{\comptn{E}{\hb}}{\invariant{b}} & \equiv &
(\comptn{E}{\hb} \myequiv \comptn[b']{E}{\hb}) \wedge \\
& & (\satisfies{\comptn[b']{E}{\hb}}{\invariant{b}}) \\
\satisfies{\comptn{E}{\hb}}{\controllable{b}} & \equiv & 
(\scc{\comptn{E}{\hb}} = \scc{\comptn[b']{E}{\hb}}) \wedge \\
& & \satisfies{\comptn[b']{E}{\hb}}{\controllable{b}} \\
\satisfies{\comptn{E}{\hb}}{\definitely{b}} & \equiv & 
\Big(\scc{\comptn{E}{\hb}}\neq \scc{\comptn[\neg b']{E}{\hb}}\Big)
\vee \\ & & \satisfies{\comptn[\neg b']{E}{\hb}}{\definitely{b}} 
\end{eqnarray*}

\end{theorem}
}

\subsection{Experimental Evaluation}
\label{sec:general:experimental}

In this section, we evaluate the effectiveness of slicing in pruning
the search-space for detecting a predicate under $possibly$ modality.
We compare our approach with that of Stoller, Unnikrishnan and Liu
\cite{StoUnn+:2000:CAV}, which is based on {\em partial-order methods}
\cite{God:1996:SV}. Intuitively, when searching the state-space, at
each consistent cut, partial-order methods allow only a small subset
of enabled transitions to be explored.  In particular, we use
partial-order methods employing both persistent and sleep sets for
comparison. We consider two examples that were also used by Stoller,
Unnikrishnan and Liu to evaluate their approach
\cite{StoUnn+:2000:CAV}.

The first example, called {\em primary-secondary}, concerns an algorithm 
designed to ensure that the system always contains a pair of processes
acting together as primary and secondary.  The invariant for the
algorithm requires that there is a pair of processes $p_i$ and $p_j$
such that (1)~$p_i$ is acting as a primary and correctly thinks that
$p_j$ is its secondary, and (2)~$p_j$ is acting as a secondary and
correctly thinks that $p_i$ is its primary.  Both the primary and the
secondary may choose new processes as their successor at any time; the
algorithm must ensure that the invariant is never falsified. A global
fault, therefore, corresponds to the complement of the invariant which
can be expressed as:
\[
\neg I_{ps} = \bigwedge_{i,j \in [1 \ldots n], \: i \neq j} \Big( \neg
isPrimary_i \vee \neg isSecondary_j \vee (secondary_i \neq  p_j) \vee
(primary_j \neq p_i) \Big)
\]

Note that $\neg I_{ps}$ is a predicate in CNF where each clause is a
disjunction of two local predicates. An approximate slice for $\neg
I_{ps}$ can be computed in $O(n^3|E|)$ time.
%
%
In the second example, called {\em database partitioning}, a database
is partitioned among processes $p_2$ through $p_n$, while process
$p_1$ assigns tasks to these processes based on the current
partition. A process $p_i$, $i \in [2 \ldots n]$, can suggest a new
partition at any time by setting variable $change_i$ to true and
broadcasting a message containing the proposed partition. An invariant
that should be maintained is: if no process is changing the partition,
then all processes agree on the partition. Its complement,
corresponding to a global fault, can be expressed as:
\[
\neg I_{db} = \neg change_2 \wedge \neg change_3 \wedge \cdots \wedge \neg
change_n \wedge \Big( \bigvee_{i, j \in [1 \ldots n], \: i \neq j}
(partition_i \neq partition_j) \Big)\]

Note that the first $n-1$ clauses of $\neg I_{db}$ are local
predicates and the last clause, say $LC$, is a disjunction of
\mbox{2-local} predicates. Thus, using the technique described in
\secref{klocal}, $LC$ can be rewritten as a predicate in DNF with
$O(n|E|)$ clauses. To reduce the number of clauses, we proceed as
follows.  Let $V$ denote the set of values that $partition_1$ assumes
in the given computation. Then it can be verified that $LC$
is logically equivalent to:
\[ 
\bigvee_{v \in V} \Big( (partition_1 = v) \wedge 
\Big( (partition_2 \neq v) \vee (partition_3 \neq v) \vee \cdots \vee
 (partition_n \neq v) \Big) \Big)\]

\begin{table}[t] 
\renewcommand{\arraystretch}{1.05}
\begin{center}
\begin{tabular}{||c||c|c||c|c||c|c||c|c||} \hline
& \multicolumn{4}{|c||}{\rule[-3mm]{0mm}{8mm} No Faults} &
\multicolumn{4}{|c||}{One Injected Fault} \\ \hline
\multicolumn{1}{||c||}{Number of}
& \multicolumn{2}{|c||}{\rule[-3mm]{0mm}{8mm} Partial-Order}
& \multicolumn{2}{|c||}{Computation} 
& \multicolumn{2}{|c||}{\rule[-3mm]{0mm}{8mm} Partial-Order} 
& \multicolumn{2}{|c||}{Computation} \\ [-0.8em]
\multicolumn{1}{||c||}{Processes}
& \multicolumn{2}{|c||}{\rule[-3mm]{0mm}{8mm} Methods} 
& \multicolumn{2}{|c||}{Slicing} 
& \multicolumn{2}{|c||}{\rule[-3mm]{0mm}{8mm} Methods} &
\multicolumn{2}{|c||}{Slicing} \\ \hline
$n$ & $T$ & $M$ & $T$ & $M$ & $T$ & $M$ & $T$ & $M$ \\ \hline
6 & 69 & 0.62 & 356 & 1.21 & 46 & 0.41 & 366 & 1.38 \\ \hline
7 & 163 & 1.11 & 609 & 1.34 & 110 & 0.81 & 584 &  1.41 \\ \hline
8 & 367 & 2.06 & 901 & 1.54 & 312 & 1.79 & 908 &  1.61 \\ \hline
9 & 832 & 4.37 & 1243 & 1.70 & 586 & 3.05 & 1207 &  1.77 \\ \hline
10 & 1516 & 7.26 & 1734 & 1.81 & 1115 & 5.54 & 1700 & 2.00 \\ \hline
11 & 2992$^*$ & 13.14$^*$ & 2147 & 1.93 & 2087$^*$ & 9.50$^*$ & 2128 &
2.27 \\ \hline 
12 & 4997$^*$ & 21.56$^*$ & 2849 & 2.16 & 3510$^*$ & 14.13$^*$ & 2765
& 2.43 \\ \hline 
\multicolumn{9}{l}{\rule[-3mm]{0mm}{8mm} $n$: number of processes \quad
$T$: amount of time spent (in s)} \\
\multicolumn{9}{l}{$M$: amount of memory used (in MB)} \\
\multicolumn{9}{l}{*: does not include the cases in which the
  technique runs out of memory} \\
\end{tabular}
\end{center} 
\vspace*{-1em}
\caption{\label{tab:dataps} Primary-Secondary example with the
  number of events on a process bounded by $90$.}
\end{table}
 
This decreases the number of clauses, when $LC$ is rewritten in a form
that can be used to compute a slice, to $O(n|V|)$. Note that $|V|$ is
bounded by the number of events on the first process, and therefore we
expect $n|V|$ to be $O(|E|)$. We use the simulator implemented in Java
by Stoller, Unnikrishnan and Liu to generate computations of these
protocols.  Message latencies and other delays (e.g., how long to wait
before looking for a new successor) are selected randomly using the
distribution $1 + \exp(x)$, where $\exp(x)$ is the exponential
distribution with mean $x$.  Further details of the two protocols and
the simulator can be found elsewhere \cite{StoUnn+:2000:CAV}.
We consider two different scenarios: {\em fault-free} and {\em faulty}. The
simulator always produces fault-free computations. A
faulty computation is generated by randomly injecting faults into a
fault-free computation.
%
Note that in the first (fault-free) scenario, we know {\em a priori}
that the computation does not contain a faulty consistent cut. We
cannot, however, assume the availability of such knowledge in
general. Thus it is important to study the behaviour of the two
predicate detection techniques in the fault-free scenario as well.  We
implement the algorithm for slicing a computation in Java.  We compare
the two predicate detection techniques with respect to two metrics:
amount of time spent and amount of memory used. In the case of the
former technique, both metrics also include the overhead of computing
the slice. We run our experiments on a machine with Pentium 4
processor operating at 1.8GHz clock frequency and 512MB of physical
memory.

\begin{figure}[t]
\begin{center}

\begin{minipage}{3.2in}
\centerline{\includegraphics[width=2.75in]{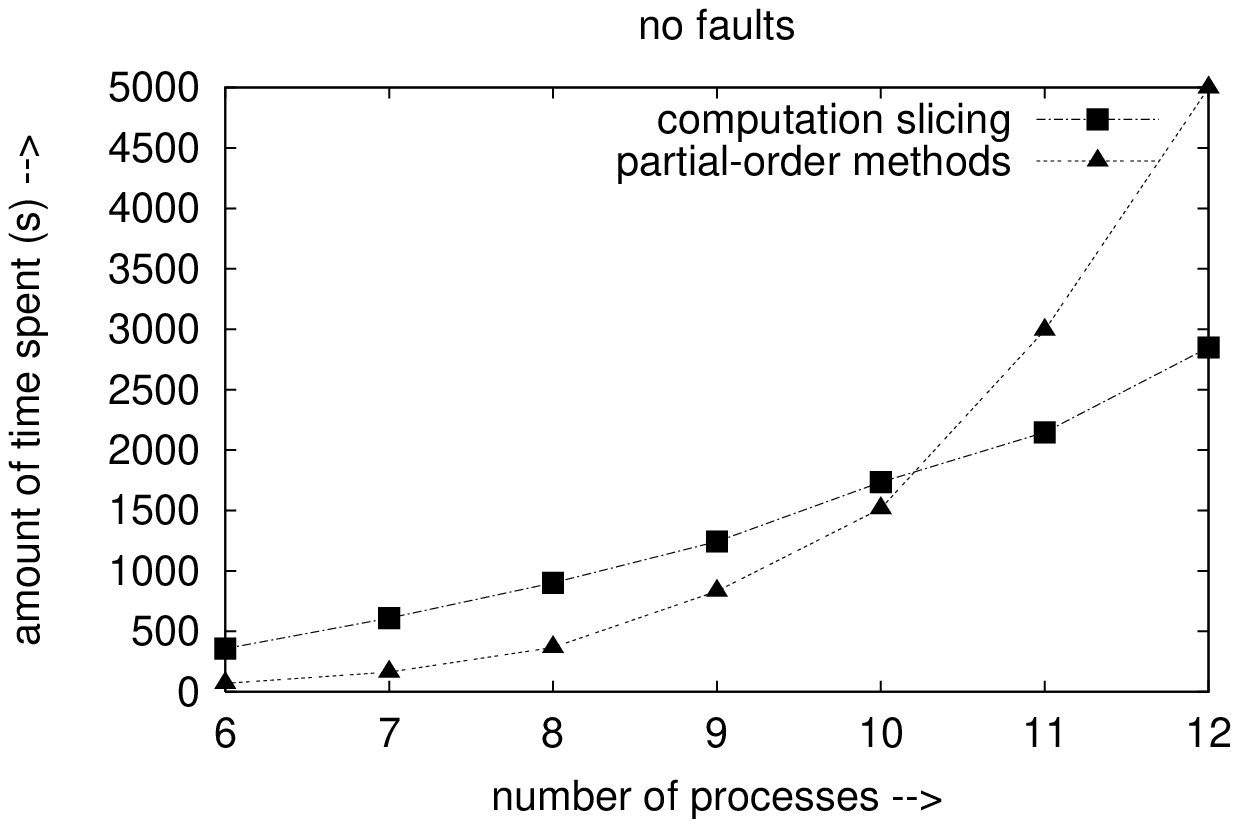}}
\end{minipage}
\begin{minipage}{3.2in}
\centerline{\includegraphics[width=2.75in]{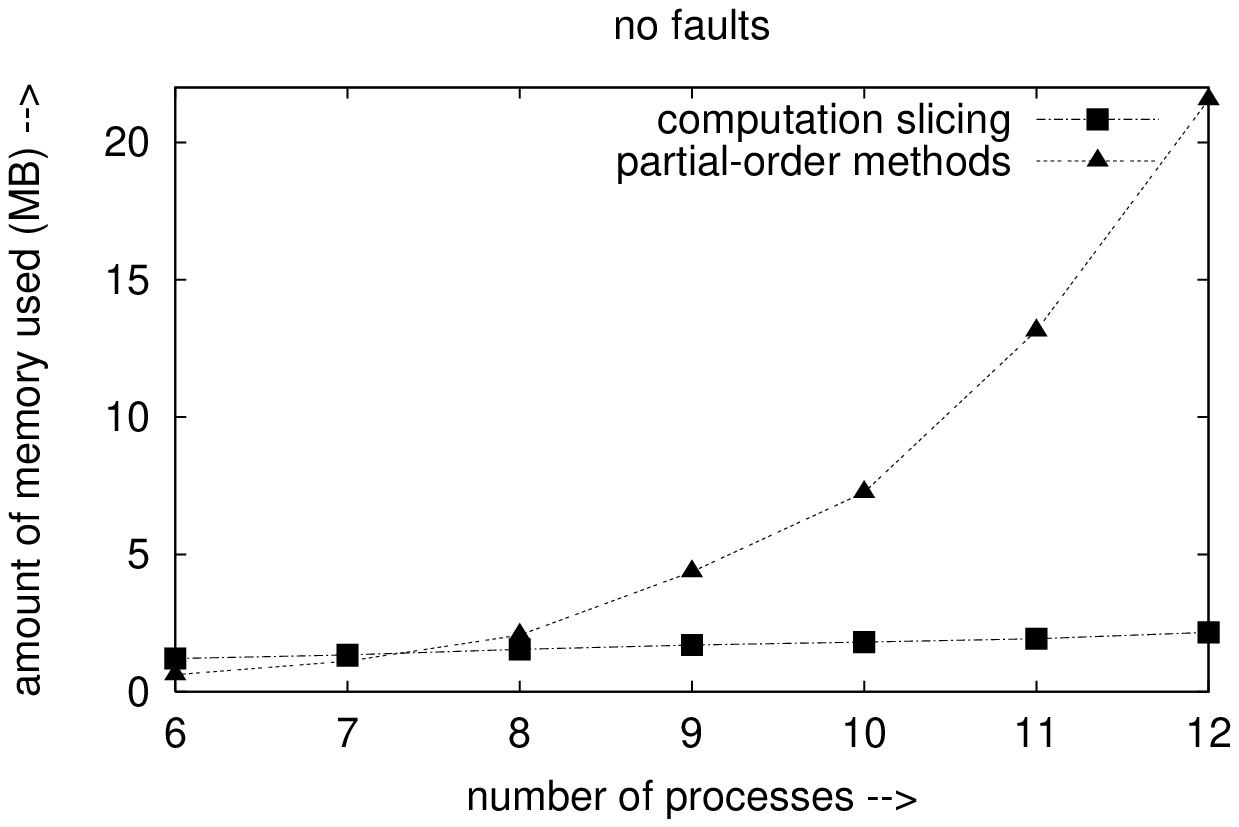}} 
\end{minipage} 

\centerline{(a)} 

\vspace*{0.5em}

\begin{minipage}{3.2in}
\centerline{\includegraphics[width=2.75in]{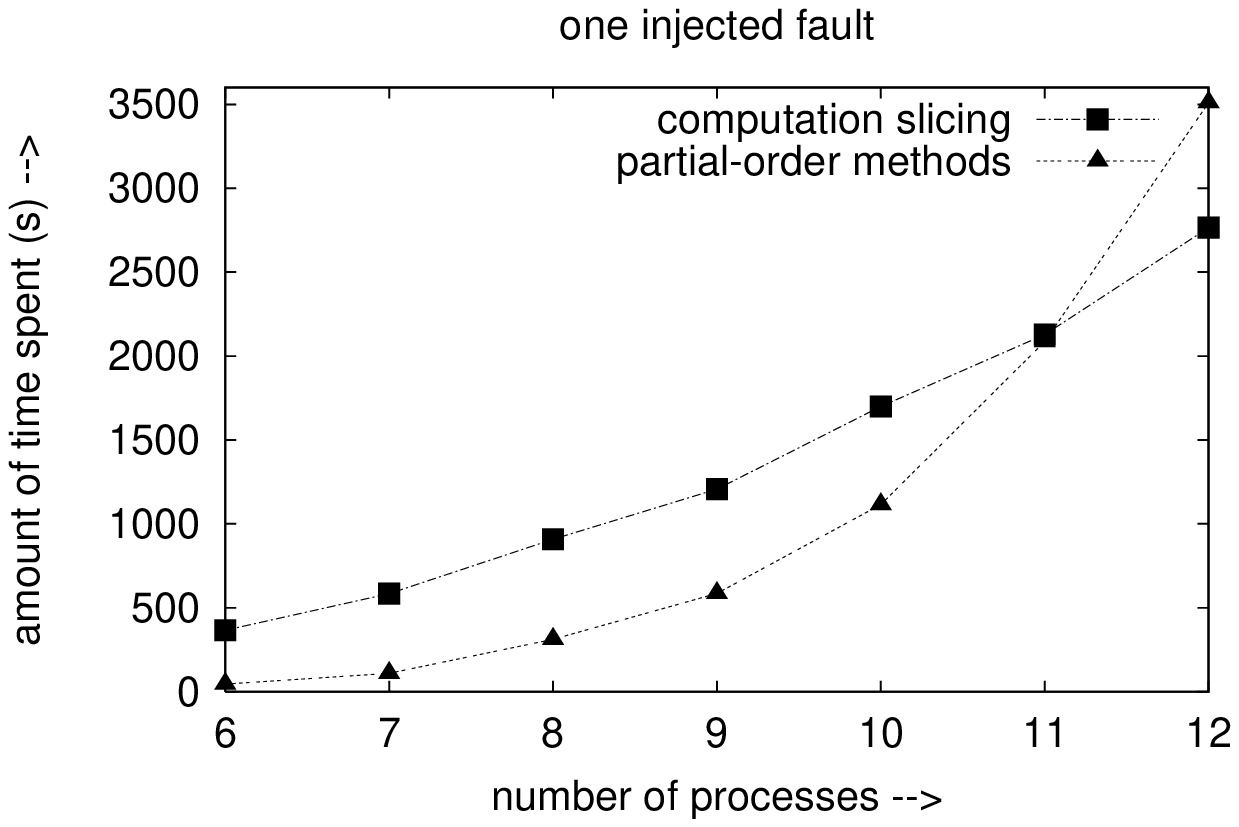}}
\end{minipage}
\begin{minipage}{3.2in}
\centerline{\includegraphics[width=2.75in]{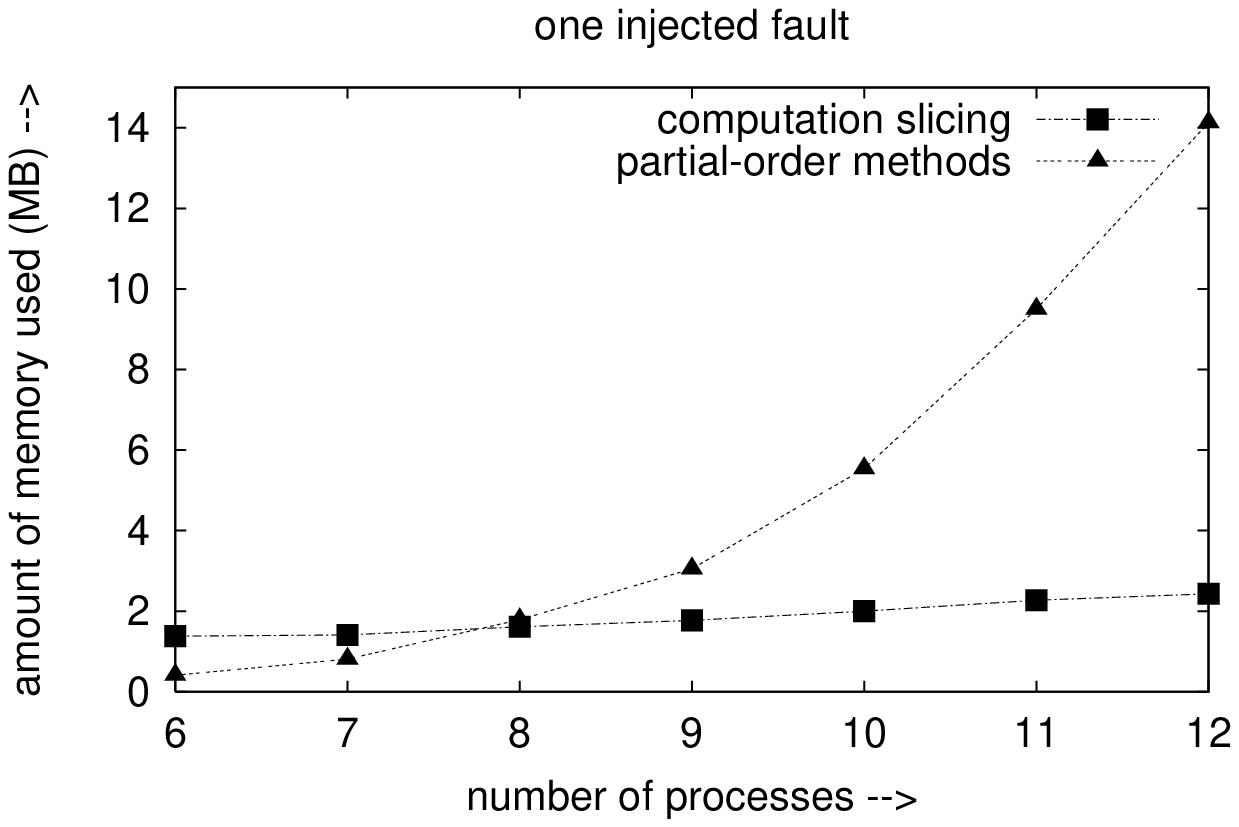}}
\end{minipage} 

\centerline{(b)}

\end{center}

\vspace*{-1.0em}
\caption{\label{fig:dataps} Primary-Secondary example with the number of
events on a process bounded by $90$ for (a)~no faults and
(b)~one injected fault.}

\end{figure}

%
%
For primary-secondary example, the simulator is run until the number
of events on some process reaches 90. The measurements averaged over
300 computations are displayed in \tabref{dataps}.
With computation slicing, for fault-free \mbox{computations}, the
slice is always empty. As the number of processes is increased from 6
to 12, the amount of time spent increases from 356s to 2,849s, whereas
the amount of memory used increases from 1.21M to 2.16M.  On the other
hand, with partial-order methods, they increase, almost exponentially,
from 69s to 4,997s and 0.62M to 21.56M, respectively.
Even on injecting a fault, the slice stays quite small. After
computing the slice, in our experiments, we only need to examine at
the most 13 consistent cuts to locate a faulty consistent cut, if
any. The amount of time spent and the amount of memory used, with
computation slicing, increase from 366s to 2,765s and 1.38M to 2.43M,
respectively, as the number of processes is increased from 6 to 12.
However, with partial-order methods, they again increase almost
exponentially from 46s to 3,510s and 0.41M to 14.13M, respectively.
Clearly, with slicing, both time and space \mbox{complexities}
for detecting a global fault, if it exists, in primary-secondary
example are polynomial in input size for the specified range of
parameters.  In contrast, with partial-order methods, they are
exponential in input size.  
\Figref{dataps}(a) and \figref{dataps}(b) plot the variation in the
two metrics with the number of processes for the two approaches.

\begin{table}[t]
\renewcommand{\arraystretch}{1.1}
\begin{center}
\begin{tabular}{||c||c|c||c|c||c|c||c|c||} \hline
& \multicolumn{4}{|c||}{\rule[-3mm]{0mm}{8mm} No Faults} &
\multicolumn{4}{|c||}{One Injected Fault} \\ \hline
\multicolumn{1}{||c||}{Number of}
& \multicolumn{2}{|c||}{\rule[-3mm]{0mm}{8mm} Partial-Order}
& \multicolumn{2}{|c||}{Computation} 
& \multicolumn{2}{|c||}{\rule[-3mm]{0mm}{8mm} Partial-Order} 
& \multicolumn{2}{|c||}{Computation} \\ [-0.8em]
\multicolumn{1}{||c||}{Processes}
& \multicolumn{2}{|c||}{\rule[-3mm]{0mm}{8mm} Methods} 
& \multicolumn{2}{|c||}{Slicing} 
& \multicolumn{2}{|c||}{\rule[-3mm]{0mm}{8mm} Methods} &
\multicolumn{2}{|c||}{Slicing} \\ \hline
$n$ & $T$ & $M$ & $T$ & $M$ & $T$ & $M$ & $T$ & $M$ \\ \hline
4  & 0.05 & 0.07 & 0.24 & 1.06 & 0.03 & 0.05 & 0.24 & 0.95 \\ \hline
5  & 0.05 & 0.09 & 0.34 & 1.13 & 0.03 & 0.08 & 0.36 & 0.99 \\ \hline
6  & 0.05 & 0.13 & 0.50 & 1.22 & 0.03 & 0.10 & 0.48 & 1.13 \\ \hline
7  & 0.05 & 0.22 & 0.59 & 1.33 & 0.04 & 0.16 & 0.62 & 1.25 \\ \hline
8  & 0.07 & 0.31 & 0.76 & 1.41 & 0.04 & 0.23 & 0.73 & 1.57 \\ \hline
9  & 0.07$^*$ & 0.36$^*$ & 0.89 & 1.56 & 0.05 & 0.31 & 0.92 & 1.69 \\
\hline 
10 & 0.08$^*$ & 0.40$^*$ & 1.09 & 1
.80 & 0.05$^*$ & 0.42$^*$ & 1.07 &
1.80 \\ \hline 
\multicolumn{9}{l}{\rule[-3mm]{0mm}{8mm} $n$: number of processes \quad
$T$: amount of time spent (in s)} \\
\multicolumn{9}{l}{$M$: amount of memory used
(in MB)} \\
\multicolumn{9}{l}{*: does not include the cases in which the
  technique runs out of memory} \\
\end{tabular}
\end{center} 
\vspace*{-1em}
\caption{\label{tab:datadb} Database partitioning example with the
  number of events on a process bounded by $80$.}
\end{table}

The worst-case performance of the partial-order methods approach is
quite bad. With 12 processes in the system and the limit on the memory
set to 100MB, the approach runs out of memory in approximately 6\% of
the cases. In around two-thirds of such cases, the computation
actually contains a consistent cut that does not satisfy the
invariant.
It may be noted that we do not include the above-mentioned cases in
computing the average amount of time spent and memory used. Including
them will only make the average performance of the partial-order
methods approach worse. Further, the performance of the partial-order
methods approach appears to be very sensitive to the location of the
fault, in particular, whether it occurs earlier during the search or
much later or perhaps does not occur at all. Consequently, the
variation or standard deviation in the two metrics is very large. This
has implications when predicate detection is employed for achieving
software fault tolerance. Specifically, it becomes hard to provision
resources (in our case, memory) when using partial-order methods
approach. If too little memory is reserved, then, in many cases, the
predicate detection algorithm will not be able to run successfully to
completion. On the other hand, if too much memory is reserved, the
memory utilization will be sub-optimal.

\begin{figure}[t]
\begin{center}

\begin{minipage}{3.2in}
\centerline{\includegraphics[width=2.75in]{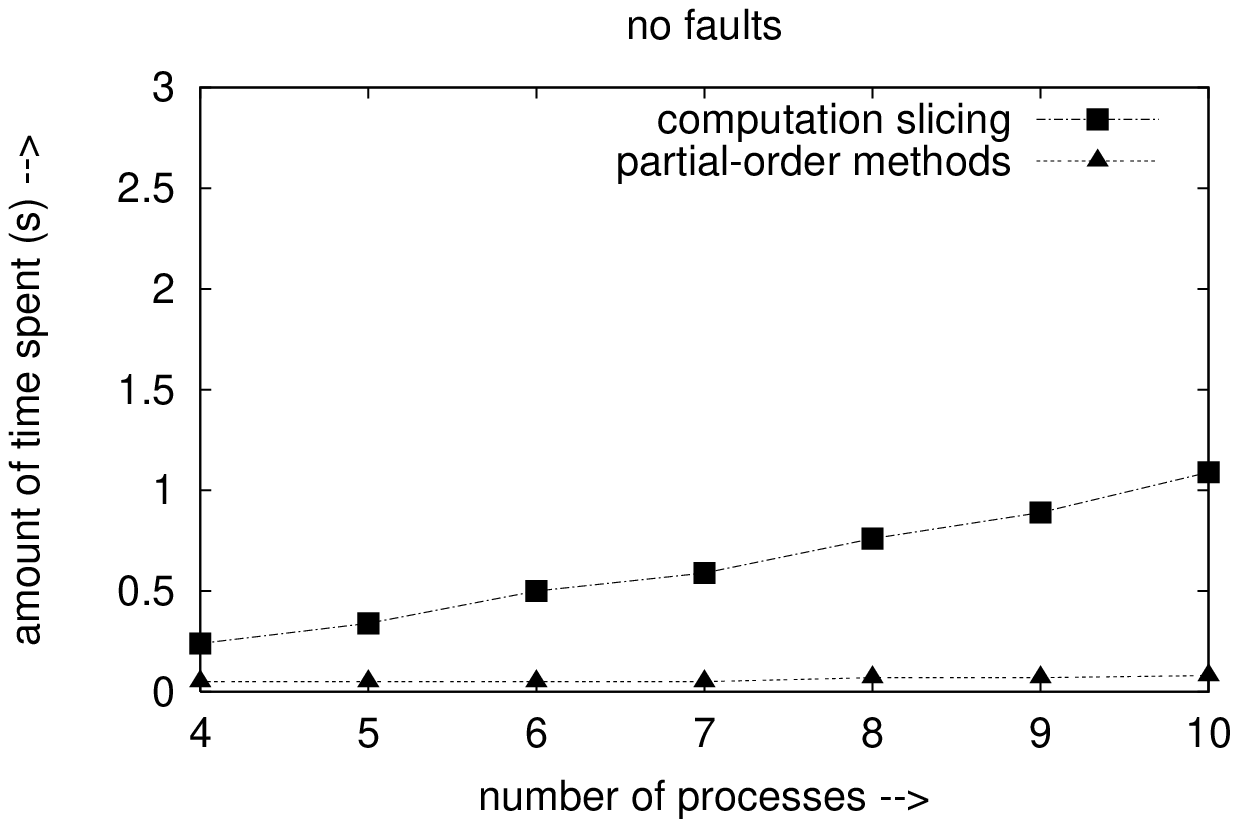}}
\end{minipage}
\begin{minipage}{3.2in}
\centerline{\includegraphics[width=2.75in]{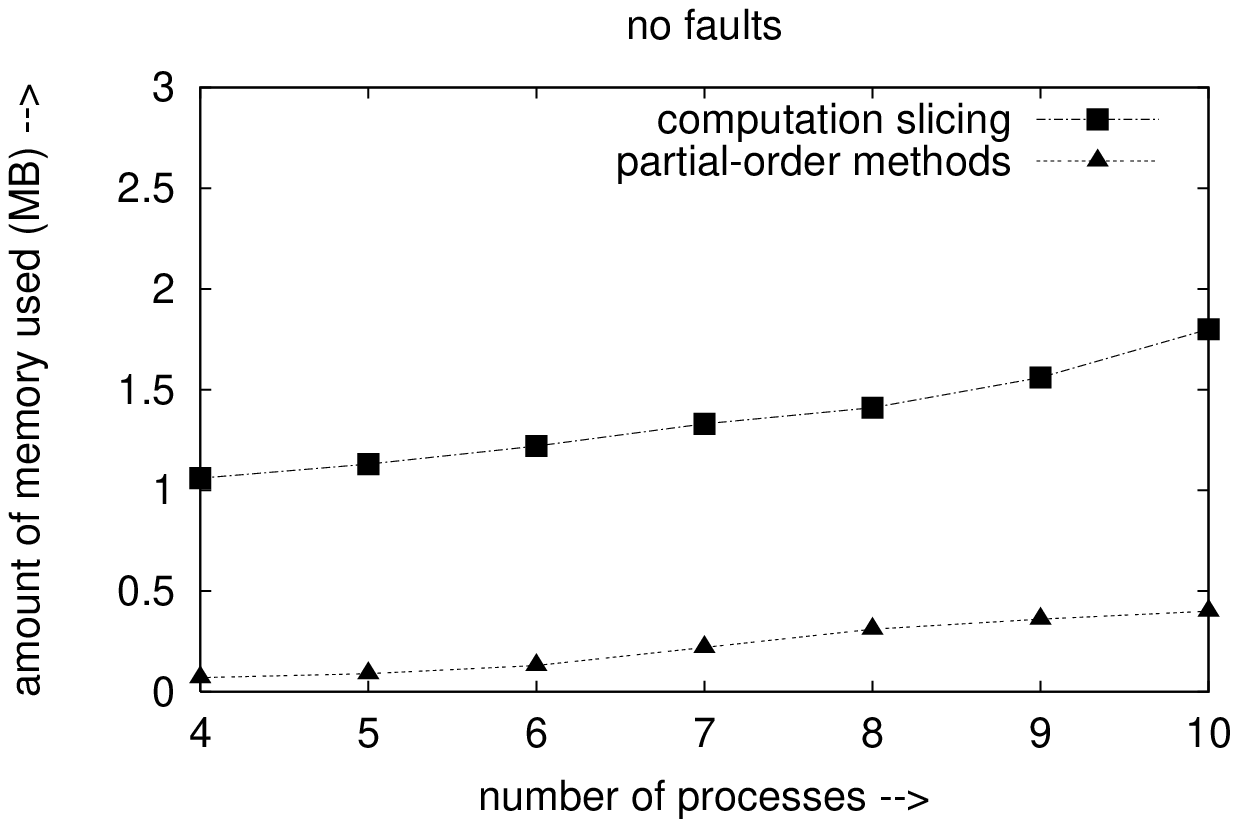}} 
\end{minipage} 

\centerline{(a)} 

\vspace*{0.5em}

\begin{minipage}{3.2in}
\centerline{\includegraphics[width=2.75in]{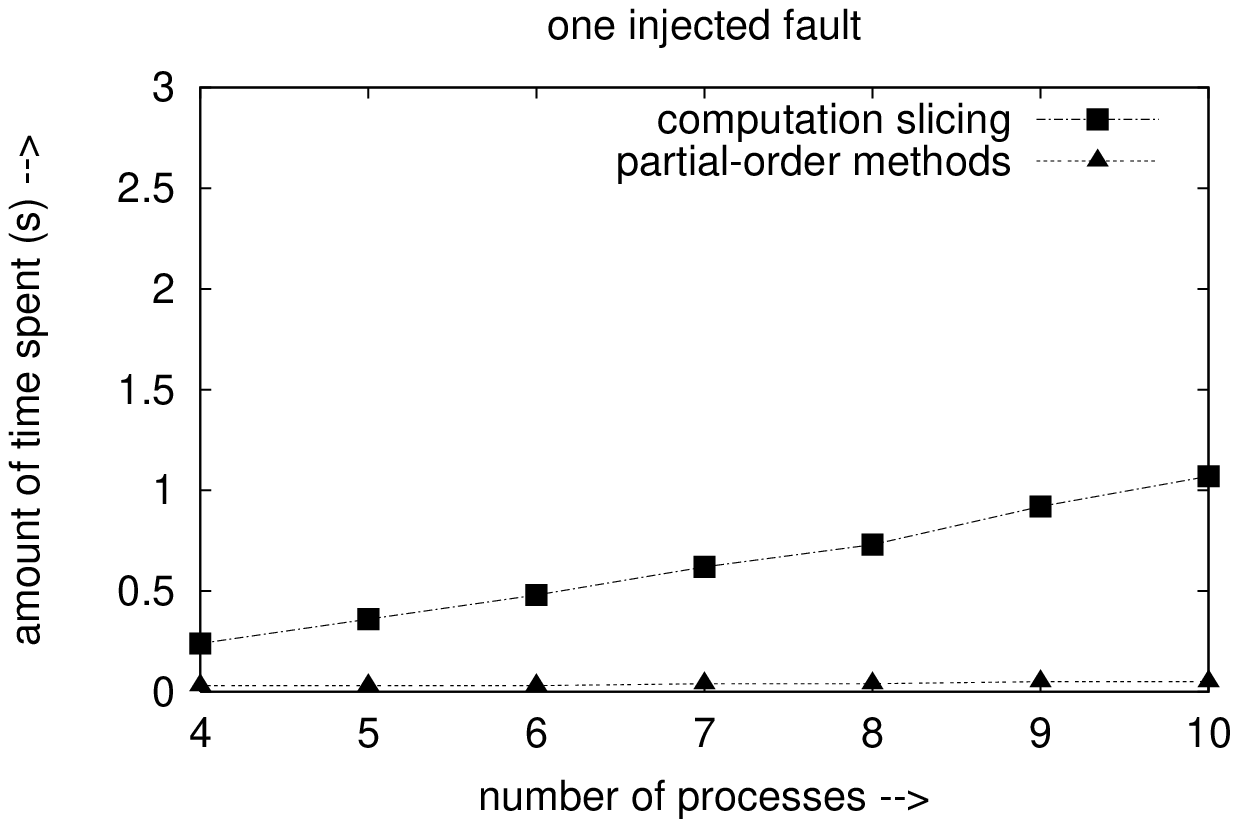}}
\end{minipage}
\begin{minipage}{3.2in}
\centerline{\includegraphics[width=2.75in]{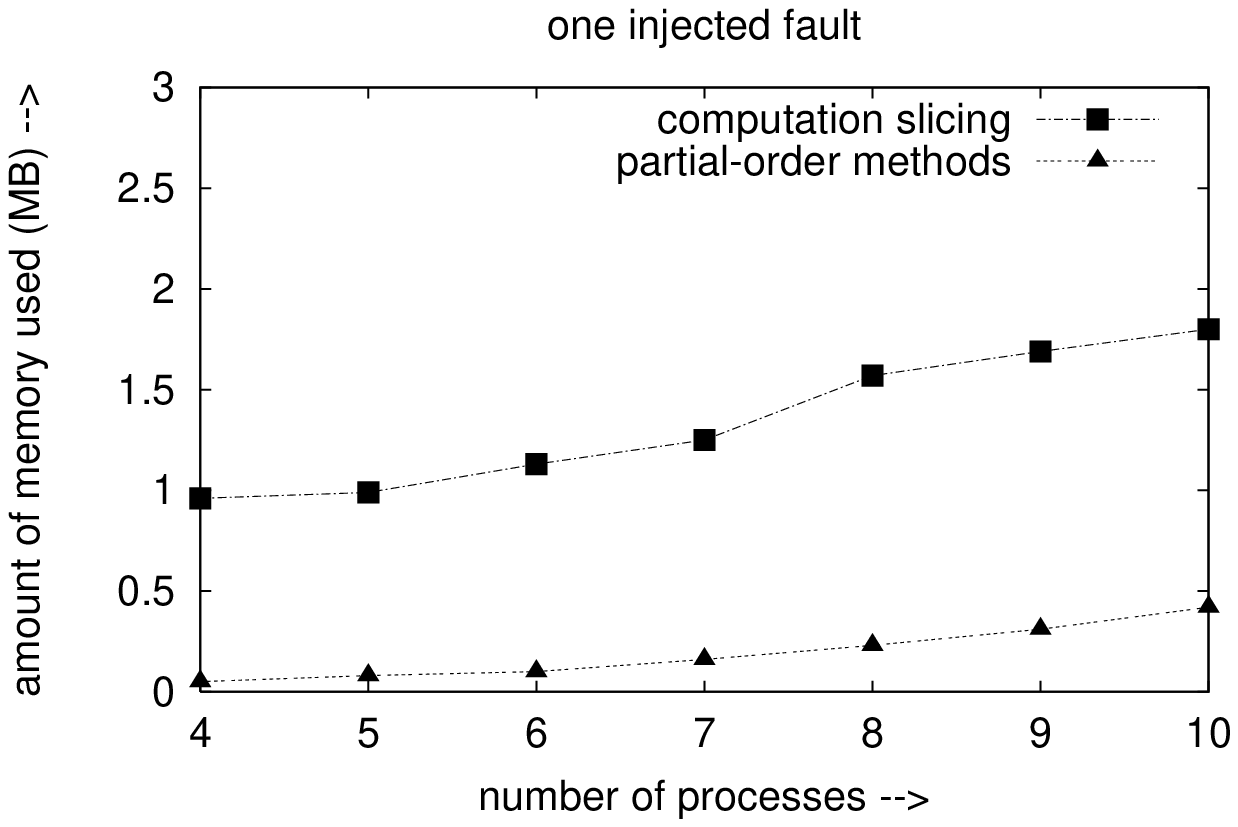}}
\end{minipage} 

\centerline{(b)}

\end{center}

\vspace*{-1.0em}
\caption{\label{fig:datadb}  Database partitioning example with the number
of events on a process bounded by $80$ for (c)~no faults and
(d)~one injected fault.}
\end{figure}

For database partitioning example, the simulator is run until the
number of events on some process reaches 80. The measurements averaged
over 300 computations are shown in \tabref{datadb}.
\mbox{\Figref{datadb}(c)} and \figref{datadb}(d) plot the variation in the
two metrics with the number of processes for the two approaches. 
As it can be seen, the average performance of partial-order methods is
much better than computation slicing. This is because substantial
overhead is incurred in computing the slice. The slice itself is quite
small. Specifically, for the fault-free scenario, the slice is always
empty. On the other hand, for the faulty scenario, only at most 4
transitions need to be explored after computing the slice to locate a
faulty consistent cut, if any.

Even for database partitioning example, for 10 processes, the
partial-order methods approach runs out of memory in a small
fraction---approximately 1\%---of the cases. Therefore the worst-case
performance of computation slicing is better than partial-order
methods.  To get the best of both worlds, predicate detection can be
first done using the partial-order methods approach. In case it turns
out that the approach is using too much memory, say more than $cn|E|$
for some small constant $c$, and still has not terminated, it can be
aborted and the computation slicing approach can then be used for
predicate detection.


\section{Discussion}
\label{sec:discussion}

In this paper, we introduce the notion of computation slice and prove
its usefulness in evaluating global properties in distributed
computations. We provide efficient polynomial-time algorithms for
computing the slice for several useful classes of predicates. For many
other classes predicates for which it is otherwise provably
NP-complete to compute the actual slice, we present efficient
heuristic algorithms for computing an approximate slice.  Our
experimental results demonstrate that slicing can lead to an
exponential improvement over existing techniques in terms of time and
space for intractable problems such as predicate detection.

Recently, we have been able to prove that exists a polynomial-time
algorithm for detecting a predicate if and only if there exists a
polynomial-time algorithm for computing its slice.
At first glance it may seem that we are not any better off than we
were before. After all, predicate detection is ``equivalent'' to
computation slicing. Then, how can slicing be used to improve the
complexity of predicate detection? The answer is in affirmative;
slicing can indeed be used to facilitate predicate detection as
illustrated by the following example.
Consider a predicate $b$ that is a  conjunction of two clauses $b_1$
and $b_2$. Now, assume that $b_1$ can be detected efficiently but
$b_2$ has no structural property that can be exploited for efficient
detection. To detect $b$, without computation slicing, we are forced
to use techniques
\cite{CooMar:1991:WPDD,AlaVen:2001:TSE,StoUnn+:2000:CAV} which do not take
advantage of the fact that $b_1$ can be detected efficiently. With
computation slicing, however, we can first compute the slice for
$b_1$. If only a small fraction of consistent cuts satisfy $b_1$, then,
instead of detecting $b$ in the computation, it is much more
efficient detect $b$ in the slice. Therefore by spending only
polynomial amount of time in computing the slice we can throw away
exponential number of consistent cuts, thereby obtaining an
exponential speedup overall.
Consequently, the equivalence result, rather than diminishing the
benefits of computation slicing, actually enhances them by
significantly enlarging the class of predicates for which the slice
can be computed efficiently.


{

Although in this paper our focus is on distributed systems, slicing
has applications in other areas as well, such as combinatorics
\cite{Gar:2002:FSTTCS}.
A combinatorial problem usually requires counting, enumerating or
ascertaining the existence of structures that satisfy a given
property.  We cast the combinatorial problem as a distributed
computation such that there is a bijection between the combinatorial
structures satisfying a property $b$ and the global states (or
consistent cuts) that satisfy a property equivalent to $b$. We then
apply results in slicing a computation with respect to a predicate to
obtain a slice consisting of only those global states that satisfy
$b$.  This gives us an efficient algorithm to count, enumerate or
detect structures that satisfy $b$ when the total set of structures is
large but the set of structures satisfying $b$ is small.

For example, consider the following problem in combinatorics: {\em
Count the number of subsets of size $k$ of the set $\{ 1, 2, \ldots, n
\}$ (hereafter denoted by $[n]$) which do not contain any consecutive numbers (for
given values of $n$ and $k$)}. To solve this problem, we first come up
with a distributed computation such that there is a one-to-one
correspondence between global states and subsets of size $k$.
\Figref{kset}(a) depicts a computation such that all subsets of $[n]$
of size $k$ are its global states.  There are $k$ processes in this
computation and each process executes exactly $n - k$ events. By the
structure of the computation, if, in a global state, process $p_i$ has
executed $j$ events, then process $p_{i+1}$ must have also executed at
least $j$ events.  The correspondence between subsets of $[n]$ and
global states can be understood as follows. If a process $p_{i}$ has
executed $m$ events in a global state, then the element $m+i$ belongs
to the corresponding subset.  Thus process $p_1$ chooses a number from
$1 \ldots (n - k + 1)$ (because there are $n - k$ events); process $p_2$
chooses the next larger number and so on.  \Figref{kset}(b) gives an
example of the computation for subsets of size $3$ of the set
$[6]$. The global state shown corresponds to the subset $\{1,3,4\}$.

\begin{figure}[t]
\centerline{\resizebox{4.5in}{!}{\input{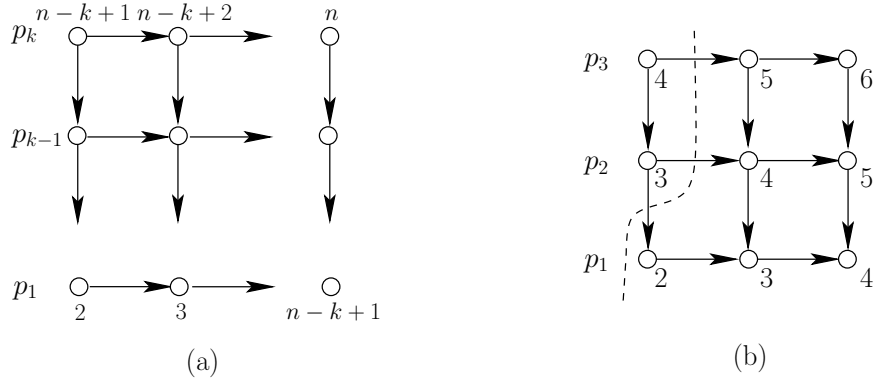}}}
\caption{\label{fig:kset} (a) Computation for subsets of $[n]$ of size
  $k$, and (b) example when $n = 6$ and $k=3$.}
\end{figure}

Now we define predicate $b$ to be ``the global state does not contain
any consecutive numbers''. For the computation we have constructed, it
can be easily verified that the predicate $b$ is regular. Therefore
one can mechanically and efficiently compute the slice of the
computation with respect to $b$. \Figref{kset2} shows the slice which
includes precisely such subsets.
Clearly, if the event labeled $m$ on process $p_i$ has been executed,
then the event labeled $m+2$ on process $p_{i+1}$ should also have
been executed.  This can be accomplished by adding the dotted arrows
to the computation as depicted in the figure.
%
By collapsing all strongly connected components and by removing the
transitively implied edges, we obtain a graph that is isomorphic to
the graph shown in \figref{kset}(a), with $k$ processes and in which each
process executes $n-k-(k-1)$ events. Therefore the total number of
such sets is ${n-k+1} \choose {k}$.
\cite{Gar:2002:FSTTCS} gives several other applications of slicing for
analyzing problems in integer partitions, set families, and set of
permutations.

\begin{figure}[t]
\centerline{\resizebox{3.25in}{!}{\input{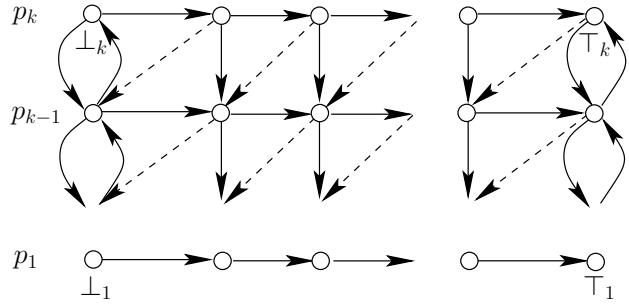}}}
\caption{\label{fig:kset2} Slice with respect to the predicate ``the global state
does not contain any consecutive numbers''.}
\end{figure}


}


At present, all our algorithms for computing a slice are centralized
in nature. They assume that there is a designated process that is
responsible for collecting all the events that have been generated and
constructing a trace using them.  Slicing algorithms use this trace to
compute the slice. While the centralized approach is quite adequate
for applications such as testing and debugging, for other applications
including software fault tolerance, a more distributed approach is
desirable.
Also, currently, algorithms for computing a slice and therefore for
detecting a predicate work in an off-line manner. 
%
%
%
To detect a software fault in a more timely manner, however, it is
desirable and sometimes essential that its slice be computed and
analyzed for any possible fault in an incremental fashion.  As the
execution of the system progresses and more and more events in the
trace become available, the current slice is updated to reflect the
newly generated events.
In the future, we plan to develop slicing algorithms that are
incremental and more distributed in nature.


{

\raggedright


}

\pagebreak

\begin{appendix}

\section{Omitted Proofs}

\begin{proof}[ for \thmref{regular:closed}]
We have to prove that if $b_1$ and $b_2$ are regular predicates then
so is $b_1 \wedge \: b_2$.  Consider consistent cuts $C_1$ and $C_2$
that satisfy $b_1 \wedge \: b_2$. By semantics of conjunction, both
$C_1$ and $C_2$ satisfy $b_1$ as well as $b_2$. Since $b_1$ and $b_2$
are regular predicates, $C_1 \cap C_2$
satisfies $b_1$ and $b_2$. Again, by semantics of conjunction, $C_1
\cap C_2$ satisfies $b_1 \wedge \: b_2$. Likewise, $C_1 \cup C_2$
satisfies $b_1 \wedge \: b_2$. Thus $b_1
\wedge \: b_2$ is a regular predicate. \qed
\end{proof}

~

\begin{proof}[ of \thmref{reg:closure}]({\em $\reglrz{b}$ is weaker
than $b$})~ Follows from the definition. 
\greaterlinespacing

\noindent 
({\em $reg$ is monotonic})~ Since $\reglrz{b'}$ is weaker than $b'$,
it is also weaker than $b$.  That is, $\reglrz{b'}$ is a regular
predicate weaker than $b$.  By definition, $\reglrz{b}$ is the
strongest regular predicate weaker than $b$. Therefore $\reglrz{b}$ is
stronger than $\reglrz{b'}$ or, in other words, $\reglrz{b} \implies
\reglrz{b'}$.

\greaterlinespacing

\noindent
({\em $reg$ is idempotent})~ Follows from the fact that $\reglrz{b}$ is a
regular predicate and is weaker than $\reglrz{b}$. \qed
\end{proof}

~

\begin{proof}[ for \lemref{project:every|satisfies}]
%
It suffices to prove that if $C$ is a consistent cut of
$\hgraph[b]{E}$, then $\projectE{C}{{Q}}$ is a consistent cut of
$\projectC[b]{E}{\hb}{{Q}}$. We prove the contrapositive. We have,
\begin{formal}
\statement{$\projectE{C}{{Q}}$ is not a consistent cut of
  $\projectC[b]{E}{\hb}{{Q}}$}
\reason{$\implies$}{definition of consistent cut}
\statement{$\myexists{e,f \in \projectE{E}{{Q}}}{\mbox{there is a path from } e \mbox{ to } f
  \mbox{ in } \projectC[b]{E}{\hb}{{Q}}}{(f \in \projectE{C}{{Q}})
  \wedge (e \not\in \projectE{C}{{Q}}}$} 
\reason{$\implies$}{using definition of $\projectFF[b]{e}{{Q}}{F}[i]$
  where $p_i = \proc{f}$}
\statement{$\myexists{e,f \in
  \projectE{E}{{Q}}}{\projectFF[b]{e}{{Q}}{F}[i] \poeq{\hb} f}{(f \in \projectE{C}{{Q}})
  \wedge (e \not\in \projectE{C}{{Q}}}$}
\reason{$\implies$}{using definition of $\kvector[b]{e}[i]$}
\statement{$\myexists{e,f \in
  \projectE{E}{{Q}}}{\kvector[b]{e}[i] \poeq{\hb} f}{(f \in \projectE{C}{{Q}})
  \wedge (e \not\in \projectE{C}{{Q}}}$}
\reason{$\implies$}{using definition of $\hgraph[b]{E}$}
\statement{$\myexists{e,f \in
  \projectE{E}{{Q}}}{\mbox{there is a path from } e \mbox{ to } f
  \mbox{ in } \hgraph[b]{E}}{(f \in \projectE{C}{{Q}})
  \wedge (e \not\in \projectE{C}{{Q}}}$}
\reason{$\implies$}{$f \in \projectE{C}{{Q}} \implies f \in C$ and $(e \not\in
  \projectE{C}{{Q}}) \wedge (e \in \projectE{E}{{Q}}) \implies e \not\in C$}
\statement{$\myexists{e,f \in E}{\mbox{there is a path from } e \mbox{ to } f
  \mbox{ in } \hgraph[b]{E}}{(f \in C) \wedge (e \not\in C)}$}
\reason{$\implies$}{definition of consistent cut}
\conclusion{$C$ is not a consistent cut $\hgraph[b]{E}$}
\end{formal}

This establishes the lemma.
\qed
\end{proof}  

~

\begin{proof}[ of \thmref{monitor}]
The first two propositions are easy to verify. We only prove the last
proposition.  As for the last proposition, it can be verified that a
regular predicate is controllable in a computation if and only if
there exists a path from the initial to the final consistent cut in
the lattice (of consistent cuts) such that every consistent cut along
the path satisfies the predicate \cite{TarGar:1998:SPDP}.  Note that
the path from the initial to the final consistent cut actually
corresponds to a longest chain in the lattice of consistent
cuts. For a lattice $L$, let $\height{L}$ denote the length of a
longest chain in $L$. Therefore if $b$ is controllable in
$\comptn{E}{\hb}$, then a longest chain in $\cuts{E}$ is contained
in $\cuts[b]{E}$ as well and vice versa. This implies that
$\height{\cuts{E}} \leq \height{\cuts[b]{E}}$. However, $\cuts[b]{E}
\subseteq \cuts{E}$ implying that $\height{\cuts[b]{E}} \leq
\height{\cuts{E}}$. Therefore we have:
\[ \controllable{b} \;\; \equiv \;\; \height{\cuts{E}} = \height{\cuts[b]{E}} \]

For a finite distributive lattice $L$, the length of its longest chain
is equal to the number of its join-irreducible elements
\cite{DavPri:1990:CUP}. In other words,  $\height{L} = \ji{L}$.
Also, as observed before in \secref{skeletal}, for a directed graph, the
number of join-irreducible elements of the lattice generated by its
set of consistent cuts---including trivial consistent cuts---is same
as the number of its strongly connected components. As a result,
$\height{\cuts{E}} = \ji{\cuts{E}} = \scc{\comptn{E}{\hb}}$ and
$\height{\cuts[b]{E}} = \ji{\cuts[b]{E}} =
\scc{\comptn[b]{E}{\hb}}$. \nsqed
\end{proof}

\pagebreak 

\section{Computing the Slice for Monotonic Channel Predicate}

We present an optimal algorithm to compute the slice with respect to
monotonic \mbox{channel} predicates such as:

\begin{itemize}

\item $\bigwedge\limits_{i,j \in [1..n]} \: (\mbox{at most } k_{ij}
\mbox{ messages in transit from process } p_i \mbox{ to process }
p_j), \mbox{ and}$ 

\item $\bigwedge\limits_{i,j \in [1..n]} \: (\mbox{at least } k_{ij}
\mbox{ messages in transit}  \mbox{ from process } p_i \mbox{ to
process } p_j)$ 

\end{itemize}

We only provide the slicing algorithm for the first predicate
here. The slicing algorithm for the second predicate is very similar
and has been omitted. Let $\snd{i}{j}{x}$ denote the send event on
$p_i$ corresponding to the send of the $x^{th}$ message to
$p_j$. Similarly, let $\rcv{i}{j}{k}$ denote the receive event
on $p_i$ corresponding to the receive of the $x^{th}$ message
from $p_j$. (Note that the $x^{th}$ message sent by $p_i$ to $p_j$ may
be different from the $x^{th}$ message received by $p_j$ from $p_i$
because we do not assume that channels are FIFO.)

Consider a computation $\comptn{E}{\hb}$ and a monotonic channel
predicate $b$ of the form in the first example. As in the case of
conjunctive predicate, we construct a graph $\hgraph[b]{E}$ with
vertices as the events in $E$ and the following edges:

\begin{enumerate}

\item from an event, that is not a final event, to its successor,

\item from a send event to the corresponding receive event, and

\item from a receive event $\rcv{j}{i}{x}$ to the send event
$\snd{i}{j}{x + k_{ij}}$, if it exists.

\end{enumerate}

\begin{figure}[t]
\centerline{\resizebox{5.5in}{!}{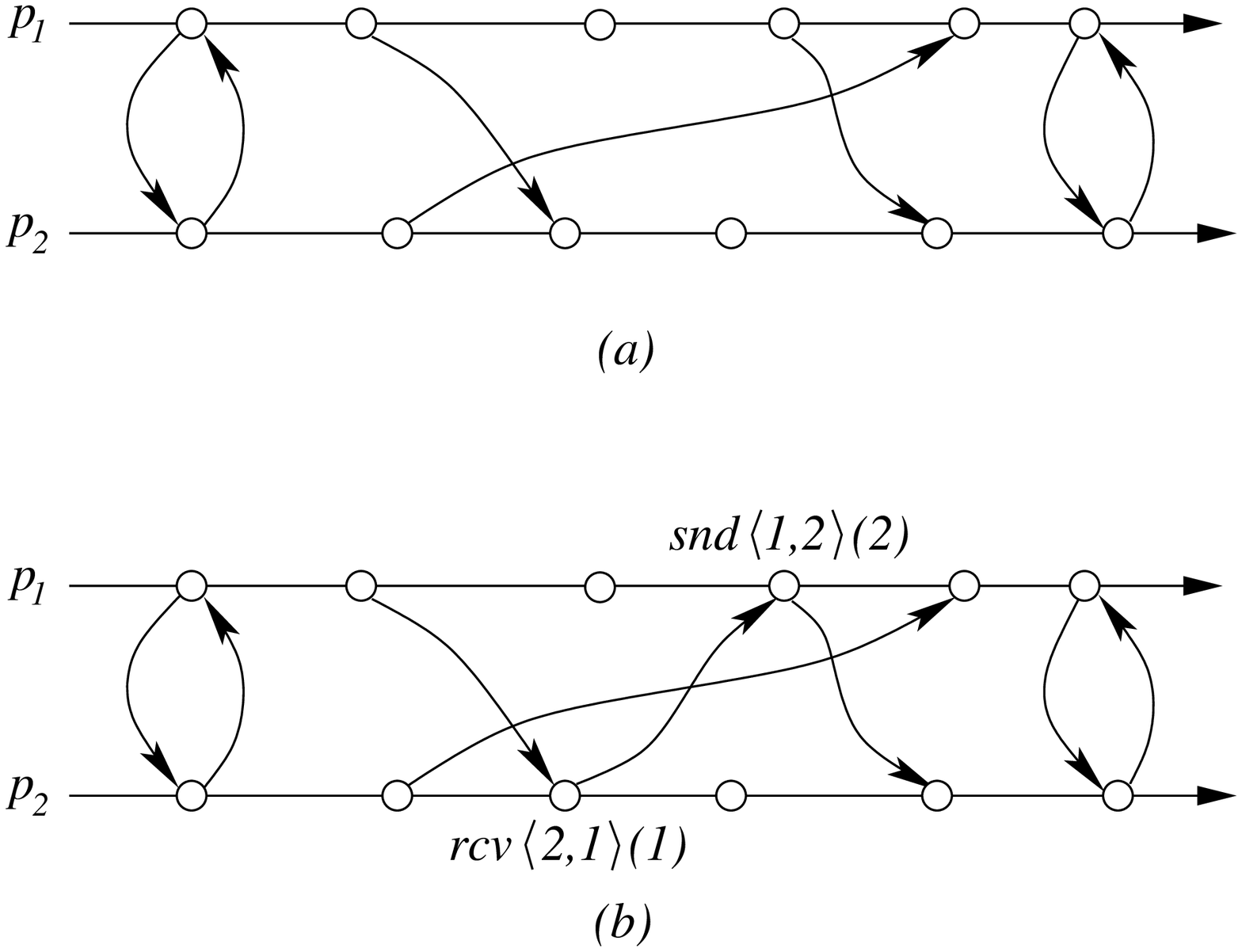}}
\caption{\label{fig:channel} (a) A computation, and (b) its slice with
respect to the monotonic channel predicate ``at most one message in
transit in any channel''.} 
\end{figure}

As before, the first two types of edges ensure that the Lamport's
happened-before relation \cite{Lam:1978:CACM} is contained in
$\paths{\hgraph[b]{E}}$.  For an example, consider the
computation shown in \mbox{\figref{channel}(a)} and the monotonic
channel predicate ``at most one message is in transit in any
channel''. Here, $k_{12} = k_{21} = 1$. The corresponding graph
constructed, as described above, is depicted in \mbox{\figref{channel}(b)}.
We first establish that the graph contains a consistent cut of the
computation only if the cut satisfies the channel predicate.

\begin{lemma}
\label{lem:chan:cut:satisfies}
Every (non-trivial) consistent cut of $\hgraph[b]{E}$ satisfies $b$.
\end{lemma}
\begin{proof}
Consider a consistent cut $C$ of $\hgraph[b]{E}$ and processes $p_i$
and $p_j$. Let $\snd{i}{j}{x}$ be the send event corresponding to the
last message sent by $p_i$ to $p_j$ such that $\snd{i}{j}{x} \in
C$. Since $C$ is a \mbox{consistent} cut of $\hgraph[b]{E}$ and there is an
edge from $\rcv{j}{i}{x - k_{ij}}$ to $\snd{i}{j}{x}$, $\rcv{j}{i}{x -
k_{ij}}$ also belongs to $C$. This implies that there are at most
$k_{ij}$ messages in transit from $p_i$ to $p_j$. \qed
\end{proof}

We next show that the graph retains all consistent cuts of the
computation that satisfy the channel predicate.

\begin{lemma}
\label{lem:chan:cut|satisfies:cut}
Every consistent cut of $\comptn{E}{\hb}$ that satisfies $b$ is a
consistent cut of $\hgraph[b]{E}$.
\end{lemma}
\begin{proof}
Consider a consistent cut $C$ of $\comptn{E}{\hb}$ that satisfies
$b$. Assume, on the contrary, that $C$ is not a consistent cut of
$\hgraph[b]{E}$. Thus there exist events $e$ and $f$ such that there
is a path from $e$ to $f$ in $\hgraph[b]{E}$, $f$ belongs to $C$ but
$e$ does not. Since $C$ is a consistent cut of $\comptn{E}{\hb}$, the
edge from $e$ to $f$ could only of type~(3). (The other two types of
edges are present in $\comptn{E}{\hb}$ as well.) Let $e$ be
$\rcv{j}{i}{x}$ and $f$ be $\snd{i}{j}{x + k_{ij}}$. Since $C$
satisfies $b$, $\rcv{j}{i}{x}$ belongs to $C$. In other words, $e$
belongs to $C$---a contradiction. \qed
\end{proof}

From the previous two lemmas, it follows that:

\begin{theorem}
$\hgraph[b]{E}$ is cut-equivalent to $\comptn[b]{E}{\hb}$.
\end{theorem}

It is easy to see that the graph $\hgraph[b]{E}$ has $O(|E|)$ vertices,
$O(|E|)$ edges (at most three edges per event assuming that an event
that is not local either sends at most one message or receives at most
one message but not both) and can be built in $O(|E|)$ time. Thus the
algorithm has $O(|E|)$ overall time-complexity.

\pagebreak

\section{Computing the Slice for Linear Predicate: Proof of Correctness}

Consider a computation $\comptn{E}{\hb}$ and a linear predicate $b$.
First, we extend the definition of $\jvector[b]{e}$ for an event $e$
and a regular predicate $b$ to the case when $b$ is a linear
predicate. It can be easily verified that $\jvector[b]{e}$ is uniquely
defined for each event $e$ even when $b$ is a linear predicate.
Now, consider the directed graph $\canonical[b]{E}$ with vertices as
events in $E$ and an edge from an event $e$ to an event $f$ if and
only if $\jvector[b]{e} \subseteq \jvector[b]{f}$. We establish that
the directed graph $\canonical[b]{E}$ is cut-equivalent to the slice
$\comptn[b]{E}{\hb}$.
It suffices to prove that $\cuts{\canonical[b]{E}}$ is the {\em
smallest\/} sublattice of $\cuts{E}$ that contains $\cuts[b]{E}$.  To
that end, the following lemma comes in useful. The lemma basically
states that, for each event $e$, $\jvector[b]{e}$ is the {\em least
consistent cut\/} of $\canonical[b]{E}$ that contains $e$. (Note that
$\jvector[b]{e} \subseteq \jvector[b]{f}$ is equivalent to saying that
there is an path from $e$ to $f$ in $\canonical[b]{E}$.)

\begin{lemma} Given events $e$ and $f$, 
$e \in \jvector[b]{f} \; \equiv \; \jvector[b]{e} \subseteq
\jvector[b]{f}$.
\label{lem:in|subseteq}
\end{lemma}
\begin{proof}
({\em $\implies$})~ 
Assume that $e \in \jvector[b]{f}$. Let $C = \jvector[b]{e} \cap
\jvector[b]{f}$. Since $e \in \jvector[b]{e}$, $e \in C$. Note that
$\jvector[b]{e}$ and $\jvector[b]{f}$ are consistent cuts of
$\comptn{E}{\hb}$. Moreover, both of them satisfy $b$. Since $b$ is a
linear predicate, their conjunction, given by $C$, also satisfies
$b$. This implies that $C$ is a consistent cut of $\comptn{E}{\hb}$
which contains $e$ and satisfies $b$. However, $\jvector[b]{e}$ is the
{\em least} such cut. Therefore $\jvector[b]{e} \subseteq C$ or
$\jvector[b]{e} \subseteq \jvector[b]{e} \cap \jvector[b]{f}$. This
implies that $\jvector[b]{e} = \jvector[b]{e} \cap
\jvector[b]{f}$. Equivalently, $\jvector[b]{e} \subseteq
\jvector[b]{f}$.  
\greaterlinespacing

\noindent
({\em $\follows$})~ Assume that $\jvector[b]{e} \subseteq
\jvector[b]{f}$. Since $e \in \jvector[b]{E}$, trivially, $e \in \jvector[b]{f}$.
\qed
\end{proof}

Again, as before, let $\joins[b]{E} = \{ \: \jvector[b]{e} \: | \: e \in E \:
\}$. Using \lemref{in|subseteq}, the following theorem can be proved
in a similar fashion as \lemref{least:join} and
\lemref{regular:cut|join}.

\begin{theorem}
\label{thm:cuts:join}
$\cuts{\canonical[b]{E}}$ forms a distributive lattice under
$\subseteq$. Further, the set of join-irreducible elements of
$\cuts{\canonical[b]{E}}$ is given by $\joins[b]{E}$.
\end{theorem}

The next lemma demonstrates that $\cuts{\canonical[b]{E}}$ contains at
least $\cuts[b]{E}$.

\begin{lemma}
\label{lem:any:join}
Every consistent cut in $\cuts[b]{E}$ can be written as the join of
some subset of elements in $\joins[b]{E}$.
\end{lemma}

The proof of the above lemma is similar to the proof of
\lemref{regular:cut|join} and therefore has been omitted. Observe
that, for every event $e$, by definition, either $\jvector[b]{e}$
satisfies $b$ or is same as $E$. In either case, $\jvector[b]{e} \in
\cuts[b]{E}$. Therefore we have,

\begin{observation}
$\joins[b]{E} \subseteq \cuts[b]{E}$.
\end{observation}

Finally, the next theorem establishes that $\cuts{\canonical[b]{E}}$
is indeed the smallest sublattice of $\cuts{E}$ that contains
all consistent cuts satisfying $b$.

\begin{theorem} 
\label{thm:sublattice:contains}
Any sublattice of $\cuts{E}$ that contains $\cuts[b]{E}$ also
contains $\cuts{\canonical[b]{E}}$.
\end{theorem}
\begin{proof}
Consider a sublattice $\mathcal{D}$ of $\cuts{E}$ such that
$\mathcal{D}$ contains $\cuts[b]{E}$. Also, consider a consistent cut
$C$ of $\cuts{\canonical[b]{E}}$. From Birkhoff's Representation
Theorem and \thmref{cuts:join}, $C$ can be expressed as the join of
some subset of elements in $\joins[b]{E}$. Since $\joins[b]{E}
\subseteq \cuts[b]{E}$ and $\cuts[b]{E} \subseteq \mathcal{D}$,
$\joins[b]{E} \subseteq \mathcal{D}$. This implies that $C$ can be
written as the join of some subset of elements in
$\mathcal{D}$. However, $\mathcal{D}$ is a sublattice and thus closed
under set union. Therefore $C \in \mathcal{D}$.
\qed
\end{proof}

The directed graph $\canonical[b]{E}$ has $|E|$ vertices and can have
as many as $\Omega(|E|^2)$ edges. However, by constructing
$\skeletal[b]{E}$, the skeletal representation of
$\comptn[b]{E}{\hb}$, instead of $\canonical[b]{E}$, the number of
edges and the time-complexity can be reduced to $O(n|E|)$ and
$O(n^2|E|)$, respectively.




\end{appendix}

\end{document}